\documentclass[12pt]{article}
\usepackage{axodraw,amsmath,amssymb,amsfonts,color,graphicx,cite,color,feynarts,soul} 
\input paperdef

\newcommand{\decayH}{\Stopz \to \Sbote H^+}
\newcommand{\decaySbzH}{\Stopz \to \Sbotz H^+}
\newcommand{\decaySbiH}{\Stopz \to \Sboti H^+}
\newcommand{\decaySbiW}{\Stopz \to \Sboti W^+}
\newcommand{\decaySbeW}{\Stopz \to \Sbote W^+}
\newcommand{\decaySbzW}{\Stopz \to \Sbotz W^+}
\newcommand{\SE}{S1}
\newcommand{\SZ}{S2}

\graphicspath{{figs/}}

\evensidemargin \oddsidemargin
\allowdisplaybreaks

\begin{document}
\thispagestyle{empty}

\def\thefootnote{\fnsymbol{footnote}}

\begin{flushright}
KA-TP-17-2010 \\
SFB/CPP-10-55
\end{flushright}

\vspace{0.5cm}

\begin{center}

{\large\sc {\bf Proposals for Bottom Quark/Squark Renormalization \\[.5em]
in the Complex MSSM}}

\vspace{0.4cm}

\vspace{1cm}

{\sc
S.~Heinemeyer$^{1}$%
\footnote{email: Sven.Heinemeyer@cern.ch}%
, H.~Rzehak$^{2}$%
\footnote{email: hr@particle.uni-karlsruhe.de}%
~and C.~Schappacher$^{2}$%
\footnote{email: cs@particle.uni-karlsruhe.de}%
}

\vspace*{.7cm}

{\sl
$^1$Instituto de F\'isica de Cantabria (CSIC-UC), Santander, Spain

\vspace*{0.1cm}

$^2$Institut f\"ur Theoretische Physik, 
Karlsruhe Institute of Technology, \\
D--76128 Karlsruhe, Germany
}

\end{center}

\vspace*{0.1cm}

\begin{abstract}
\noindent
We present a consistent renormalization of the top and bottom
quark/squark sector of the MSSM with complex parameters
(cMSSM). 
Various renormalization schemes are defined, analyzed analytically and
tested numerically in the decays $\Stopz \to \Sboti\, H^+/W^+$ ($i = 1,2$).
No scheme is found that produces numerically acceptable results
over all the cMSSM parameter space, where problems occur mostly already
for real parameters.
Two schemes are identified that show the most robust behavior.
A numerical analysis of the four partial stop
decay widths is performed in our
``preferred'' scheme, ``$\mb,\,\Ab$~\DRbar''.
The full one-loop corrections to the corresponding partial
decay widths are evaluated including hard QED and QCD radiation. 
We find mostly modest corrections at the one-loop level.
\end{abstract}

\def\thefootnote{\arabic{footnote}}
\setcounter{page}{0}
\setcounter{footnote}{0}

\newpage


\section{Introduction}

One of the main tasks of the LHC is to search for Supersymmetry
(SUSY)~\cite{mssm}. 
The Minimal Supersymmetric Standard Model (MSSM) predicts two scalar
partners for all Standard Model (SM) fermions as well as fermionic
partners to all SM bosons. 
Of particular interest are the scalar partners of the heavy SM
quarks, the scalar top quarks, $\Stopi$ ($i = 1,2$) and scalar bottom
quarks $\Sbotj$ ($j = 1,2$) due to their large Yukawa couplings. 
A scalar top quark $\Stopi$ has many possible
decay modes, depending on the mass patterns of the SUSY particles. 
Among  those decay modes are the decays to a scalar bottom quark, 
$\Sbotj$, and a charged Higgs boson, $H^+$, or
  $W$~boson, $W^+$,  
\begin{align}
\label{stsbH}
&\Stopi \to \Sbotj H^+ \quad (i,j = 1,2)~, \\
\label{stsbW}
&\Stopi \to \Sbotj W^+ \quad (i,j = 1,2)~.
\end{align}
If these channels are kinematically allowed they can even be dominant if
(most of) the other decay modes are kinematically forbidden.
Consequently, these processes can constitute a large part of
the total stop decay width, and, in case of decays to a Higgs boson, they
can serve as a source of charged Higgs bosons in cascade decays at the LHC.
                      
For a precise prediction of the partial decay widths corresponding to
\refeq{stsbH} and \refeq{stsbW}, at least the one-loop level contributions
have to be taken into account.
This in turn requires a renormalization of the relevant sectors,
especially a simultaneous renormalization of the top and bottom
quark/squark sector. 
Due to the $SU(2)_L$ invariance of the left-handed scalar top and
bottom quarks, these two sectors cannot be treated independently.
Within the framework of the MSSM with complex parameters (cMSSM) 
we analyze various bottom quark/squark sector renormalization
schemes, while we 
apply a commonly used on-shell renormalization scheme for the
top quark/squark sector throughout all the investigations.
Special attention is payed to ``perturbativity'', i.e.\ the loop
corrections should not be enhanced by large counterterm contributions
resulting from an inappropriate renormalization scheme.
This turns out to be a constraint that is very difficult to fulfill over
the whole cMSSM parameter range, where it is especially 
difficult to achieve this simultaneously for small and large values of
$\tan \beta$.

Higher-order corrections to scalar fermion decays have been evaluated in
various analysis over the last
decade. The simultaneous renormalization 
of the top and the bottom quark/squark sector was taken
into account only in a relatively small subset.
In \citeres{squark_q_V_als,stopsbot_phi_als} stop and sbottom decays,
including the ones to charged Higgs and SM
gauge bosons, have been 
evaluated at \order{\als} within the MSSM with real parameters
(rMSSM). The numerical investigation 
was restricted to relatively low $\tb$ values. These
calculations are implemented in the program {\tt SDECAY}~\cite{sdecay}.
A similar analysis in
\citere{sbot_stop_Hpm_altb} included electroweak one-loop corrections,
where again only relatively low $\tb$ values were considered. 
The decays of Higgs bosons to scalar fermions, including the charged
Higgs decays, at the full one-loop level within the rMSSM was presented in
\citeres{A_sferm_sferm_full,H_sferm_sferm_full}, indicating very
large one-loop corrections 
for large $\tb$. An effective Lagrangian approach in the rMSSM 
for these types of decays was given in \citere{squark_eff}, with a
numerical analysis for $\tb = 5$. 

The renormalization of the top and bottom quark/squark
sector has been analyzed also in the context of other calculations 
in the past. A comparison of different renormalization schemes within 
the rMSSM was performed in \citeres{dissHR,mhiggsFDalbals}, focusing on
large $\tb$. One of the renormalization schemes considered therein had
been used before within the calculation of the two-loop bottom
quark/squark contributions to the neutral Higgs boson masses~\cite{sbotrenold}
which are important for large $\tb$~values. Within the cMSSM a
renormalization was presented in \citere{dissTF}, however without an
analysis of its practicability. In \citeres{dissHR,mhcMSSM2L} the top and
bottom quark/squark sector was renormalized within the cMSSM, but only
the QCD part needed for the presented calculation was considered. Thus
no complete top and bottom quark/squark sector renormalization has been
performed within the cMSSM. Recently a renormalization of nearly all
sectors of the rMSSM appeared~\cite{FawziRen}. In this analysis,
however, the main focus has been on gauge parameter independence.

Complex phases, as assumed here in the cMSSM, can be relevant for
collider observables and possibly extracted from experimental data. 
Scalar top quark branching ratios at a linear collider are discussed in 
\citere{Bartl:2003pd}. Concerning LHC measurements, triple products
involving the decay of scalar 
top or bottom quarks are analyzed in
\citeres{Bartl:2004jr,Ellis:2008hq,Deppisch:2009nj,Deppisch:2010nc}.
Finally, rate asymmetries are examined in \citere{Eberl:2009xe}.
Depending on assumptions about the LHC performance it might be
possible to extract information on the phases of $M_1$, $\At$ and $\Ab$
at the LHC.

In this paper we analyze the renormalization of the full top
and bottom quark/squark sector in the cMSSM. We show analytically 
(and numerically) why
certain renormalization schemes 
fail for specific parts of the parameter space. Finally, we explore the
one-loop effects 
for the decays (\ref{stsbH},\ref{stsbW}) for important parts of the
cMSSM parameter space in the favored
renormalization scheme. We present numerical results showing the size
of the one-loop corrections, especially including small and large $\tb$.
The evaluation of the partial decay widths of the scalar top quarks are
being implemented into the Fortran code 
{\tt FeynHiggs}~\cite{feynhiggs,mhiggslong,mhiggsAEC,mhcMSSMlong}. 
A numerical analysis of {\em all} scalar top quark decay modes,
involving a renormalization of {\em all} relevant sectors will be 
presented elsewhere~\cite{Stop2decay}.


\section{The generic structure of the quark/squark sector}
\label{sec:generic}

The decay channels (\ref{stsbH},\ref{stsbW}) are calculated at the full
one-loop level (including hard QED and QCD radiation). This requires the
renormalization of several sectors of the cMSSM as discussed below. 
The sectors not discussed in detail are renormalized as follows:
\begin{itemize}
\item 
The gauge and Higgs sector renormalization has been performed following
\citere{mhcMSSMlong}. The gauge boson masses, $M_W$ and $M_Z$, as well
as the mass of the 
charged Higgs boson, $M_H^\pm$, has been defined on-shell while the sine squared
of the weak mixing
angle, $s_{\mathrm{w}}^2$, is defined via the gauge boson masses,
    $s_{\mathrm{w}}^2 = 1 - M_W^2/M_Z^2$. The $Z$ factors for 
the $W$~boson field are also determined within an on-shell scheme while
the $Z$ factors of the charged Higgs boson field are given by a linear
combination of the $\overline{\text{DR}}$ $Z$ factors of the Higgs
doublets (see \citere{mhcMSSMlong}). An additional finite $Z$ factor is
introduced to fulfill on-shell conditions for the external charged
$H^\pm$~field. 
 $\tan \beta$ is defined as $\overline{\text{DR}}$
parameter. 
\item
The Higgs mixing parameter $\mu$ has been renormalized via an
on-shell (OS) procedure for the neutralino and chargino
sector~\cite{dissTF,diplTF}. 
\item
For the renormalization of the electromagnetic charge we require that the
renormalized $ee\ga$-vertex in the Thomson limit is not changed by
higher order corrections with respect to the corresponding tree-level
vertex~\cite{denner}. 
\end{itemize}
A detailed description of our renormalization of all sectors will be
given in \citere{Stop2decay}.

\bigskip
In the following we focus on the top and bottom quark/squark sector.
The bilinear part of the  Lagrangian with top and bottom squark fields,
$\Stop$ and $\Sbot$, 
\begin{align}
\cL_{\Stop/\Sbot\text{ mass}} &= - \begin{pmatrix}
{{\tilde{t}}_{L}}^{\dagger}, {{\tilde{t}}_{R}}^{\dagger} \end{pmatrix}
\matr{M}_{\tilde{t}}\begin{pmatrix}{\tilde{t}}_{L}\\{\tilde{t}}_{R}
\end{pmatrix} - \begin{pmatrix} {{\tilde{b}}_{L}}^{\dagger},
{{\tilde{b}}_{R}}^{\dagger} \end{pmatrix}
\matr{M}_{\tilde{b}}\begin{pmatrix}{\tilde{b}}_{L}\\{\tilde{b}}_{R} 
\end{pmatrix}~,
\end{align}
contains the stop and sbottom mass matrices
$\matr{M}_{\tilde{t}}$ and $\matr{M}_{\tilde{b}}$,
given by 
\begin{align}\label{Sfermionmassenmatrix}
\matr{M}_{\tilde{q}} &= \begin{pmatrix}  
 M_{\tilde Q_L}^2 + m_q^2 + M_Z^2 c_{2 \beta} (T_q^3 - Q_q \sw^2) & 
 m_q \Xq^* \\[.2em]
 m_q \Xq &
 M_{\tilde{q}_R}^2 + m_q^2 +M_Z^2 c_{2 \beta} Q_q \sw^2
\end{pmatrix} 
\end{align}
with
\begin{align}\label{kappa}
\Xq &= \Aq - \mu^*\kappa~, \qquad \kappa = \{\cot\beta, \tan\beta\} 
        \quad {\rm for} \quad q = \{t, b\}
~.
\end{align}
$M_{\tilde Q_L}^2$ and $M_{\tilde{q}_R}^2$ are the soft SUSY-breaking mass
parameters. $m_q$ is the mass of the corresponding quark.
$Q_{{q}}$ and $T_q^3$ denote the charge and the isospin of $q$, and
$A_q$ is the trilinear soft SUSY-breaking parameter.
The mass matrix can be diagonalized with the help of a unitary
 transformation ${\matr{U}}_{\tilde{q}}$, 
\begin{align}\label{transformationkompl}
\matr{D}_{\tilde{q}} &= 
\matr{U}_{\tilde{q}}\, \matr{M}_{\tilde{q}} \, 
{\matr{U}}_{\tilde{q}}^\dagger = 
\begin{pmatrix} \msqe^2 & 0 \\ 0 & \msqz^2 \end{pmatrix}~, \qquad
{\matr{U}}_{\tilde{q}}= 
\begin{pmatrix} U_{\tilde{q}_{11}}  & U_{\tilde{q}_{12}} \\  
                U_{\tilde{q}_{21}} & U_{\tilde{q}_{22}}  \end{pmatrix}~. 
\end{align}
The scalar quark masses, $\msqe$ and $\msqz$, will always be mass
ordered, i.e.\  
$m_{\tilde{q}_1} \le m_{\tilde{q}_2}$:
\begin{align}
m_{\tilde{q}_{1,2}}^2 &= \edz \KL M_{\tilde{Q}_L}^2 + M_{\tilde{q}_R}^2 \KR 
       + m_q^2 + \edz T_q^3 \MZ^2 c_{2\be} \non \\
&\quad \mp \edz \sqrt{\KKL M_{\tilde{Q}_L}^2 - M_{{\tilde{q}_R}}^2 
       + \MZ^2 c_{2\be} (T_q^3 - 2 Q_q \sw^2) \KKR^2 + 4 m_q^2 |\Xq|^2}~.
\label{MSbot}
\end{align}

\smallskip
The parameter renormalization can be performed as follows, 
\begin{align}
\matr{M}_{\sq} &\to \matr{M}_{\sq} + \de\matr{M}_{\sq}
\end{align}
which means that the parameters in the mass matrix $\matr{M}_{\sq}$ 
are replaced by the renormalized parameters and a counterterm. After the
expansion $\de\matr{M}_{\sq}$ contains the counterterm part,
\begin{align}\label{proc1a}
\de\matr{M}_{\sq_{11}} &= \de M_{\tilde Q_L}^2 + 2 m_q \de m_q 
- M_Z^2 c_{2 \beta}\, Q_q \, \de \sw^2 + (T_q^3 - Q_q \sw^2) 
  ( c_{2 \beta}\, \de M_Z^2 + M_Z^2\, \de c_{2\beta})~, \\\label{proc1b}
\de\matr{M}_{\sq_{12}} &= (\Aq^*  - \mu \kappa)\, \de m_q 
+ m_q (\de \Aq^* - \mu\, \de \kappa - \kappa \, \de \mu)~, \\\label{proc1c}
\de\matr{M}_{\sq_{21}} &=\de\matr{M}_{\sq_{12}}^*~, \\\label{proc1d}
\de\matr{M}_{\sq_{22}} &= \de M_{\tilde{q}_R}^2 
+ 2 m_q \de m_q +  M_Z^2 c_{2 \beta}\, Q_q \, \de \sw^2
+ Q_q \sw^2 ( c_{2 \beta}\, \de M_Z^2+ M_Z^2\, \de c_{2 \beta})
\end{align}
with $\kappa$ given in \refeq{kappa}.

Another possibility for the parameter renormalization is to start out
with the physical parameters which corresponds to
the replacement:
\begin{align} \label{proc2}
\matr{U}_{\tilde{q}}\, \matr{M}_{\tilde{q}} \, 
{\matr{U}}_{\tilde{q}}^\dagger &\to\matr{U}_{\tilde{q}}\, \matr{M}_{\tilde{q}} \, 
{\matr{U}}_{\tilde{q}}^\dagger + \matr{U}_{\tilde{q}}\, \de \matr{M}_{\tilde{q}} \, 
{\matr{U}}_{\tilde{q}}^\dagger =
\begin{pmatrix} \msqe^2 & Y_q \\ Y_q^* & \msqz^2 \end{pmatrix} +
\begin{pmatrix}
\de \msqe^2 & \de Y_q \\ \de Y_q^* & \de \msqz^2
\end{pmatrix}~,
\end{align}
where $\de \msqe^2$ and  $\de \msqz^2$ are the counterterms 
 of the squark masses squared. $\de Y_q$ is the
 counter\-term\footnote{The unitary 
     matrix $\matr{U}_{\tilde{q}}$ can be expressed by a mixing angle
     $\theta_{\tilde{q}}$ and
     a corresponding phase $\varphi_{\tilde{q}}$. Then the
     counterterm  $\de Y_q$ can be related to the counterterms of the
     mixing angle and the phase (see \citere{mhcMSSM2L}).}   to the squark
 mixing parameter $Y_q$ (which vanishes
   at tree level, $Y_q = 0$, and corresponds to the 
 off-diagonal entries in $\matr{D}_{\tilde{q}} =\matr{U}_{\tilde{q}}\,
 \matr{M}_{\tilde{q}} \,  
{\matr{U}}_{\tilde{q}}^\dagger$, see~\refeq{transformationkompl}). Using
\refeq{proc2} 
 one can express $\de\matr{M}_{\sq}$ by the counterterms $\de \msqe^2$,
 $\de \msqz^2$ and $\de Y_q$. Especially for $\de\matr{M}_{\sq_{12}}$
 one yields
\begin{align}\label{dMsq12physpar}
\de\matr{M}_{{\sq}_{12}} &=
U^*_{\sq_{11}} U_{\sq_{12}}
(\de \msqe^2 - \de \msqz^2) +
U^*_{\sq_{11}} U_{\sq_{22}} \de Y_q + U_{\sq_{12}}
U^*_{\sq_{21}} \de Y_q^*~.
\end{align}
In the following the relation given by \refeq{proc1b} and
\refeq{dMsq12physpar} will be used to express either $\de Y_q$, $\de
A_q$ or $\de m_q$ by the other counterterms.

For the field renormalization the following procedure is applied,
\begin{align}
\begin{pmatrix} \sqe \\ \sqz \end{pmatrix} &\to 
  \KL \id + \edz \de\matr{Z}_{\sq} \KR 
  \begin{pmatrix} \sqe \\ \sqz \end{pmatrix} 
  \quad {\rm with} \quad
\de\matr{Z}_{\sq} = \begin{pmatrix} 
                   \de Z_{\sq_{11}} & \de Z_{\sq_{12}} \\
                   \de Z_{\sq_{21}} & \de Z_{\sq_{22}} 
                   \end{pmatrix}~.
\end{align}

This yields for the renormalized self-energies
\begin{align}
\hSi_{\sq_{11}}(k^2) &= \Si_{\sq_{11}}(k^2) 
  + \edz (k^2 - \msqe^2) (\dZ{\sq_{11}} + \dZ{\sq_{11}}^*)
  - \de\msqe^2~, \\
\hSi_{\sq_{12}}(k^2) &= \Si_{\sq_{12}}(k^2)
  + \edz (k^2 - \msqe^2) \dZ{\sq_{12}}
  + \edz (k^2 - \msqz^2) \dZ{\sq_{21}}^* 
  - \de Y_q~, \\
\hSi_{\sq_{21}}(k^2) &= \Si_{\sq_{21}}(k^2)
  + \edz (k^2 - \msqe^2) \dZ{\sq_{12}}^*
  + \edz (k^2 - \msqz^2) \dZ{\sq_{21}} 
  - \de Y_q^*~, \\
\hSi_{\sq_{22}}(k^2) &= \Si_{\sq_{22}}(k^2) 
  + \edz (k^2 - \msqz^2) (\dZ{\sq_{22}} + \dZ{\sq_{22}}^*)
  - \de\msqz^2~.
\end{align}
In order to complete the quark/squark sector renormalization also for the
corresponding quark (i.e. its mass, $m_q$, and the quark
field, $q$) renormalization constants have to be introduced:
\begin{align}
m_q &\to m_q + \de m_q~,\\
\omega_{\mp} \,q  &\to (1 + \frac{1}{2}\dZ{q}^{L/R})\, \omega_{\mp} \,q~
\end{align}
with $\de m_q$ being the quark mass counterterm and $\dZ{q}^L$ and
$\dZ{q}^R$ being the $Z$~factors of the left-handed and the right-handed
component of the quark field $q$, respectively. $\omega_{\mp} =
\frac{1}{2}(\id \mp \gamma_5)$
are the left- and right-handed projectors, respectively.
Then the renormalized self energy, $\Hat{\Si}_{q}$, can be decomposed
into left/right-handed and scalar left/right-handed parts, 
${\hSi}_q^{L/R}$ and ${\hSi}_q^{SL/SR}$, respectively,
\begin{align}\label{decomposition}
 \Hat{\Si}_{q} (k) &= \not\! k\, \omega_{-} \Hat{\Si}_q^L (k^2)
                   + \not\! k\, \omega_{+} \Hat{\Si}_q^R (k^2)
                   + \omega_{-} \Hat{\Si}_q^{SL} (k^2) 
                   + \omega_{+} \Hat{\Si}_q^{SR} (k^2)~,
\end{align}
where the components are given by
\begin{align}
\Hat{\Si}_q^{L/R} (k^2) &= {\Si}_q^{L/R} (k^2) 
   + \frac{1}{2} (\dZ{q}^{L/R} + {\dZ{q}^{L/R}}^*)~, \\
\Hat{\Si}_q^{SL} (k^2) &=  {\Si}_q^{SL} (k^2) 
   - \frac{m_q}{2} (\dZ{q}^L + {\dZ{q}^R}^*) - \de m_q~,  \\
\Hat{\Si}_q^{SR} (k^2) &=  {\Si}_q^{SR} (k^2) 
   - \frac{m_q}{2} (\dZ{q}^R + {\dZ{q}^L}^*) - \de m_q~.
\end{align}
Note that $\Hat{\Si}_q^{SR} (k^2) = {\Hat{\Si}_q^{SL} (k^2)}^*$ 
holds due to ${\cal CPT}$ invariance.


\section{Field renormalization of the quark/squark sector}

We first discuss the field renormalization of the top and bottom
quark/squark sector and turn to the parameter renormalization in the next
section \ref{sec:stop}.

The field renormalization, meaning the determination of the $Z$~factors,
is done within an on-shell scheme for squarks and quarks. We impose
  equivalent renormalization conditions for the top as well as for
  the bottom quark/squark sector: 
\begin{itemize}

\item[(a)]
The diagonal $Z$~factors of the squark fields are determined such that
the real part of the residua of propagators is set to unity, 
\begin{align}
\label{residuumStopOS}
\wtre \frac{\dd \hSi_{\sq_{ii}}(k^2)}{\dd k^2}
      \Big|_{k^2 = \msqi^2} &= 0 \qquad (i = 1,2)~.
\end{align}
This condition fixes the real parts of the diagonal $Z$~factors to
\begin{align}
\re\dZ{\sq_{ii}} = - \wtre \frac{\dd \Si_{\sq_{ii}}(k^2)}{\dd k^2}
                  \Big|_{k^2 = \msqi^2} \qquad (i = 1,2)~.
\end{align}
$\wtre$ above denotes the real part with respect to
contributions from the loop integral, but leaves the complex
couplings unaffected.

The imaginary parts of the diagonal $Z$~factors are so far undetermined
and are set to zero, 
\begin{align}
\im \dZ{\sq_{ii}} &= 0 \qquad (i = 1,2)~.
\end{align}
This is possible since they do not contain divergences.

\item[(b)]
For the non-diagonal $Z$~factors of the squark fields we impose the
condition that for 
on-shell squarks no transition from one squark to the other occurs, 
\begin{align}
\wtre\hSi_{\sq_{12}}(\msqe^2) &= 0~, \\
\wtre\hSi_{\sq_{12}}(\msqz^2) &= 0~.
\end{align}
This yields
\begin{align}
\dZ{\sq_{12}} &= + 2 \frac{\wtre\Si_{\sq_{12}}(\msqz^2) - \de Y_q}
     {(\msqe^2 - \msqz^2)}~, \non \\
\dZ{\sq_{21}} &= - 2 \frac{\wtre\Si_{\sq_{21}}(\msqe^2) - \de Y_q^*}
     {(\msqe^2 - \msqz^2)}~.
\label{dZstopoffdiagOS}
\end{align}
The counterterm $\de Y_q$ is determined in the corresponding parameter
renormalization scheme. This means the non-diagonal $Z$~factors of the
squark fields do also depend on the choice of the parameter
renormalization scheme.

\item[(c)] 
The quark fields are also defined via an on-shell condition. We impose
\begin{align}\label{ZquarkOS}
\lim_{k^2\rightarrow m_q^2}\frac{ \not\! k + m_q}{k^2 - m_q^2} \wtre
\Hat{\Si}_{q} (k) u(k) &= 0~,\quad\ \lim_{k^2\rightarrow m_q^2} \bar{u}(k)\wtre
\Hat{\Si}_{q} (k)\frac{ \not\! k + m_q}{k^2 - m_q^2}  = 0~,
\end{align}
where $u(k)$, $\bar{u}(k)$ are the spinors of the external fields. 
This yields
\begin{align}
\re \dZ{q}^{L/R} &= - \wtre \Big\{ {\Si}_q^{L/R} (m_q^2)  \\ 
&\quad  + m_q^2 \KKL {{\Si}_q^{L}}'(m_q^2) + {{\Si}_q^{R}}'(m_q^2) \KKR
              + m_q \KKL {{\Si}_q^{SL}}'(m_q^2) 
              + {{\Si}_q^{SR}}'(m_q^2) \KKR \Big\}~,  \non \\
\mq \KL \im \dZ{q}^L - \im\dZ{q}^R \KR &= 
    i\, \wtre\KKKL {\Si}_q^{SR}(m_q^2) - {\Si}_q^{SL}(m_q^2) \KKKR
    = 2 \im \KKKL \wtre {\Si}_q^{SL}(m_q^2) \KKKR~,
\end{align}
with $\Si'(k^2) \equiv \frac{\partial \Si(k^2)}{\partial k^2}$. 
Choosing also $\im \dZ{q}^L = - \im\dZ{q}^R$, the imaginary parts of the
$Z$~factors can be expressed as
\begin{align}
\im \dZ{q}^{L/R} &= \pm \frac{i}{2\, m_q} 
        \wtre \KKKL {\Si}_q^{SR}(m_q^2) - {\Si}_q^{SL}(m_q^2) \KKKR
      = \pm \frac{1}{m_q} \im \KKKL \wtre {\Si}_q^{SL}(m_q^2) \KKKR~.
\end{align}
Note that the renormalization condition \refeq{ZquarkOS} can only be 
fully satisfied if the corresponding quark mass is defined as on-shell, 
too.

The $Z$~factors of the quark fields are not needed for the calculation
of the considered decay modes of the scalar top quarks (see, however,
\citere{Stop2decay}). 

\end{itemize}


\section{Parameter renormalization of the top and bottom 
quark/squark sector}
\label{sec:stop}

Within the top and bottom quark/squark sector nine real parameters are
defined: The real
soft SUSY-breaking parameters $M_{\tilde Q_L}^2$, $M_{{\tilde{t}}_R}^2$
and $M_{{\tilde{b}}_R}^2$, the complex trilinear couplings $A_t$ and
$\Ab$ and the top and bottom Yukawa couplings $y_t$ and $y_b$ which both
can be chosen to be real. ($\mu$ and $\tan \beta$ as well as the
gauge boson masses and the weak mixing angle are determined within other
sectors, see the beginning of \refse{sec:generic}). Note that the
soft SUSY-breaking parameter $M_{\tilde Q_L}^2$ is the same in the top
as well as in the bottom squark sector due to the 
$SU(2)_L$ invariance of the 
left-handed fields.
As in \citeres{dissHR,mhcMSSM2L}, instead of choosing the five
quantities $M_{\tilde Q_L}^2$, $M_{{\tilde{t}}_R}^2$,
$M_{{\tilde{b}}_R}^2$ and $y_t$, $y_b$ the
squark masses $\mste^2$, $\mstz^2$, $\msbz^2$ as well as the top and
bottom quark masses $\mt$, $\mb$ were taken as independent parameters. 

If a regularization scheme is applied which does not break the symmetries
of the model, it is sufficient to use counterterms which respects the
underlying symmetries. Such counterterms are
 generated by multiplicative
renormalization of
parameters and fields of the MSSM. The parameter counterterms can be fixed by
as many renormalization conditions as independent parameters exist
\cite{MSSMrenormierung}.
Concerning the top and bottom quark/squark sector we have to set
nine renormalization conditions to define all indepedent parameters.

 For the renormalization of the top quark/squark sector we follow
\citeres{dissHR,mhcMSSM2L} but we also include electroweak contributions.%

We impose five renormalization conditions, (A)--(E), to fix the
parameters of the top quark/squark sector:
\begin{itemize}
\item[]
\begin{itemize}
\item[(A)] The top-quark mass is determined via an on-shell condition,
  yielding the one-loop counterterm $\de \mt$:
\begin{align}\label{dmt}
\de \mt &= \frac{1}{2} \wtre \KKKL 
    \mt \KKL\Si_t^L (\mt^2) + \Si_t^R (\mt^2) \KKR  
  + \KKL \Si_t^{SL} (\mt^2) + \Si_t^{SR} (\mt^2) \KKR \KKKR~.
\end{align}
\item[(B), (C)]
The two top squark masses are also defined on-shell, yielding the real
counterterms 
\begin{align}
\label{dmst}
\de\msti^2 &= \wtre\Si_{\Stop_{ii}}(\msti^2) \qquad (i = 1,2)~.
\end{align}

\item[(D), (E)]
Finally, the non-diagonal entry in the matrix of \refeq{proc2} is fixed
as  
\begin{align}
\de Y_t &= \frac{1}{2} \wtre 
    \KKKL \Si_{\Stop_{12}}(\mste^2) + \Si_{\Stop_{12}}(\mstz^2) \KKKR~,
\end{align}
which corresponds to two seperate conditions as $\de Y_t$ is complex.
\end{itemize}
\end{itemize}

The counterterm of the trilinear coupling $\de A_t$ is then given via the
relation of \refeqs{proc1b} and \eqref{dMsq12physpar} as:
\begin{align}
\de A_t &= \frac{1}{m_t}\bigl[U_{\Stop_{11}} U_{\Stop_{12}}^*
           (\de \mste^2 - \de \mstz^2) 
        +  U_{\Stop_{11}} U_{\Stop_{22}}^* \de Y_t^*
        + U_{\Stop_{12}}^* U_{\Stop_{21}} \de Y_t  
        - (\At - \mu^* \cot\beta)\, \de\mt  \bigr] \non \\
&\quad +  (\de\mu^* \cot\be - \mu^* \cot^2\be\, \dtanb)~.
\end{align}
The definition of $\dtanb$ and $\de\mu$ is indicated in
\refse{sec:generic}. 

\bigskip
For the bottom quark/squark sector we are left with four independent
parameters which are not defined yet. We choose the following
four renormalization conditions, (i)--(iv):
\begin{itemize}
\item[]
\begin{itemize}
\item[(i)] The $\Sbotz$~mass is defined on-shell:
\begin{align}\label{sbotzOS}
\de\msbz^2 &= \wtre\Si_{\Sbot_{22}}(\msbz^2)~.
\end{align}
\item[(ii)--(iv)] These three renormalization conditions are chosen
  according to the different renormalization conditions listed in
  \refta{tab:RS} and to the corresponding subsections 
  \ref{sec:OS}--\ref{sec:AbOS_ReYbOS}. 
  They yield the counterterms $\de \mb$, $\de \Ab$
  and $\de Y_b$ where only three of these five real counterterms are
  independent (counting each of the complex counterterms, $\de \Ab$
  and $\de Y_b$, as two real counterterms). The two dependent
  counterterms can be expressed as a combination of the other ones.
\end{itemize}
\end{itemize}
Applying these renormalization conditions fixes the counterterms
generated by multiplicative renormalization which fulfill the
symmetry relations \cite{MSSMrenormierung}.

\begin{table}[ht!]
\renewcommand{\arraystretch}{1.5}
\BC
\begin{tabular}{|c||c|c|c|c||c|c|}
\hline
scheme & $m_{\tilde b_{1,2}}$ & $\mb$ & $\Ab$ & $Y_b$ & Sect. & name 
                                                           \\ \hline\hline
{\small analogous to the $t/\Stop$ sector:} 
 ``OS'' & OS & OS & & OS  
& \ref{sec:OS} & RS1 \\ \hline
``$\mb,\,\Ab$~\DRbar'' & OS & \DRbar\ & \DRbar\ & 
& \ref{sec:mbDRbar_AbDRbar} & RS2 \\ \hline
``$\mb,\, Y_b$~\DRbar'' & OS & \DRbar\ & & \DRbar\ 
& \ref{sec:mbDRbar_YbDRbar} & RS3 \\ \hline
``$\mb$~\DRbar, $Y_b$~OS'' & OS & \DRbar\ & & OS
& \ref{sec:mbDRbar_YbOS} & RS4 \\ \hline
``$\Ab$~\DRbar, $\re Y_b$~OS'' & OS & & \DRbar & $\re Y_b$:\, OS
& \ref{sec:AbDRbar_ReYbOS} & RS5 \\ \hline
``$\Ab$~vertex, $\re Y_b$~OS'' & OS & & vertex & $\re Y_b$:\, OS
& \ref{sec:AbOS_ReYbOS} & RS6 \\ \hline
\end{tabular}
\caption{Summary of the six renormalization schemes for the
  $b/\Sbot$~sector investigated below. Blank entries indicate dependent
  quantities. $\re Y_b$ denotes that only the real part of 
  $Y_b$ is renormalized on-shell, while the imaginary part is a
  dependent parameter. The rightmost columns indicates the section that
  contains the detailed description of the respective renormalization
  and the abbreviated notation used in our analysis.}
\label{tab:RS}
\renewcommand{\arraystretch}{1.0}
\EC
\end{table}

While the $\Sbotz$~mass is defined on-shell, the
$\Sbote$~mass receives a shift due to the radiative corrections: 
\begin{align}
\msbeOS^2 &= \msbe^2 + \KL
 \de\msbe^2 - \wtre\Si_{\Sbot_{11}}(\msbe^2)\KR~.
\end{align}
The term in parentheses is the shift from $\msbe^2$ to the on-shell mass
squared. The value of 
$\msbe^2$ is derived from the diagonalization of the 
sbottom mass matrix, see \refeq{transformationkompl}, and $\de\msbe^2$
is defined as a dependent
quantity~\cite{hr,mhiggsFDalbals}. $\msbeOS^2$ is the on-shell
$\Sbote$~mass squared. In ~\citere{hr} the size of the shift was
analyzed while in~\citere{mhiggsFDalbals} bottom squarks appeared only as
``internal'' particles, i.e.\ as particles inside the loop
diagrams. Concerning the scalar top quark decay, \refeqs{stsbH} and
\eqref{stsbW}, we are now dealing with 
scalar bottom quarks as ``external'' particles, which are defined as
incoming or outgoing particles. These ``external'' particles should
fulfill on-shell properties. At this point there are two options to 
proceed:
\begin{itemize}
\item[($\cO 1$)] 
  The first option is to use different mass values, $\msbe$ and
  $\msbeOS$, for the ``internal'' and the ``external'' particles,
  respectively, which can cause problems for charged particles as,
  for instance, scalar bottom quarks (see below).
\item[($\cO 2$)] 
  The second option is to impose a further renormalization
  condition which ensures that the $\Sbote$~mass is on-shell:
\begin{align}\label{sboteOS}
 \de\msbe^2 &=\wtre\Si_{\Sbot_{11}}(\msbe^2)~.
\end{align}
In this case the input has to be chosen such that the symmetry
relations are fulfilled at the one-loop level.
\end{itemize}

As mentioned above, the option~($\cO 1$) leads to a problem. 
The IR-divergences originating from the loop diagrams involve the
``inner'' (i.e.\ tree-level) 
mass $\msbe$. These have to cancel with the real
Bremsstrahlung IR-divergences, which are evaluated with the help of the
``external'' (i.e.\ one-loop on-shell) mass $\msbeOS$, which is inserted into 
the tree-level diagram (the result can, as usual, be expressed with the
help of the Soft Bremsstrahlung (SB) factor $\de_{\rm SB}$:
$\cM_{\rm tree} \times \de_{\rm SB}$, see \citere{denner}). 
Due to the two different sets of masses the IR-divergences do not
cancel. 
One way out would be the use of tree-level masses in all diagrams
contributing to the part $2 \re \{\cM_{\rm tree} \cM_{\rm loop}\}$,
i.e.\ in all loop diagrams and in the hard and soft Bremsstrahlung diagrams. 
However, this would lead to inconsistencies in the evaluation of the complete
loop corrected amplitude squared
$\propto (|\cM_{\rm tree}|^2 + 2 \re \{\cM_{\rm tree} \cM_{\rm loop}\} )$ 
due to the different masses entering the phase space evaluation.
A consistent phase space integration requires the use of the same
``external'' masses for all outgoing particles in all parts of 
the calculation.

To circumvent the problem of the non-cancellation of IR-divergences we
choose the option~($\cO 2$) and impose the further renormalization 
condition \refeq{sboteOS}. This requires to choose an input that restores 
the symmetries. Relating $(\matr{M}_{\tilde q})_{11}$ of 
\refeq{Sfermionmassenmatrix} and 
$(\matr{U_{\tilde{q}}}^{\dagger} \matr{D}_{\tilde{q}} \matr{U_{\tilde{q}}})_{11}$
with $\matr{D}_{\tilde{q}}$ of \refeq{transformationkompl} yields an
expression for the soft SUSY-breaking parameter $M_{\tilde{Q}_L}^2$
(depending on the squark flavor),
\begin{align}
M_{\tilde{Q}_L}^2(\tilde{q}) = |U_{\tilde{q}_{11}}|^2 \msqe^2
   + |U_{\tilde{q}_{12}}|^2 \msqz^2 
   - \MZ^2 c_{2\be} (T_q^3 - Q_q \sw^2) - \mq^2
\end{align}
with $\tilde{q} = \{\tilde{t}, \tilde{b}\}$. Requiring the $SU(2)_L$ relation
to be valid at the one-loop level induces the following shift in 
$M^2_{\tilde{Q}_L}$ (see also
\citeres{squark_q_V_als,stopsbot_phi_als,dr2lA}): 
\begin{align}
M_{\tilde{Q}_L}^2(\Sbot) = M_{\tilde{Q}_L}^2(\tilde{t}) 
   + \de M_{\tilde{Q}_L}^2(\tilde{t}) - \de M_{\tilde{Q}_L}^2(\tilde{b})
\label{MSbotshift}
\end{align}
with
\begin{align}
\de M_{\tilde{Q}_L}^2(\tilde{q}) &= |U_{\tilde{q}_{11}}|^2 \de\msqe^2
   + |U_{\tilde{q}_{12}}|^2 \de\msqz^2
   - U_{\tilde{q}_{22}} U_{\tilde{q}_{12}}^* \de Y_q
   - U_{\tilde{q}_{12}} U_{\tilde{q}_{22}}^* \de Y_q^* - 2 \mq \de\mq \non \\
&\quad  + \MZ^2\, c_{2\be}\, Q_q\, \de \sw^2 
        - (T_q^3 - Q_q \sw^2) (c_{2\be}\, \de \MZ^2 + \MZ^2\, \de c_{2\be})~.
\label{MSbotshift-detail}
\end{align}
In other words, everywhere in the calculation the masses and mixing
matrix elements coming 
from the diagonalization of the bottom squark mass matrix, see
\refeq{transformationkompl}, are used with $M_{\tilde{Q}_L}^2(\tilde{b})$
including the above shift as in \refeq{MSbotshift}. 
In this way the problems concerning UV- and IR-finiteness are
avoided. (An exception is the field renormalization of the $W$-boson
field: In the corresponding selfenergies the $SU(2)_L$ relation is needed
at tree-level to ensure UV-finiteness. In this case, tree-level bottom
squark masses are used.)

The various renormalization schemes, following the general
choice~($\cO 2$), are summarized in \refta{tab:RS} and
outlined in detail in the following subsections.

Comparing with the literature, several of the renormalization
schemes (or variants of them) have been used to calculate higher-order
corrections to squark or Higgs decays.
The older calculations of the loop corrections have all been
performed in the rMSSM. 

\begin{itemize}

\item
A renormalization scheme employing an ``OS'' renormalization for
$\mb$ and $Y_b$ was used in 
\citeres{stopsbot_phi_als,sbot_stop_Hpm_altb} for the calculation of
stop and sbottom decays. (The calculation of
\citere{stopsbot_phi_als} is also implemented in
\citere{sdecay}.) In order to check our
implementation given in \refse{sec:OS} we calculated the decay 
$\Sbot_{1,2} \to \Stop_1 H^{-}$ (see \refse{sec:calc} for our set-up) and
found good agreement with \citere{stopsbot_phi_als}.

\item
A renormalization scheme similar to the real version of RS2,
i.e.\ ``$\mb,\,\Ab$~\DRbar''
has been employed in \citere{H_sferm_sferm_full} for the
calculation of Higgs decays to scalar fermions. In the scalar top and
the Higgs sector they apply an on-shell scheme 
(partially 
based on \citeres{sbot_top_cha_alt,sfermprod_alf}), 
which differs in some points from our renormalization scheme.

\item
An on-shell scheme was also used in \citere{sferm_f_V_full} (based on
\citeres{sbot_top_cha_alt,squark_q_chi_full}) to evaluate the decay
$\Sferm \to \Sfermp V$ ($V = W^\pm, Z$). 

\item
In \citere{stop_stop_H_alt}, as a starting point, an on-shell renormalization
scheme was used for the calculation of the electroweak corrections to
$\Ga(\Stopz \to \Stope \phi)$, ($\phi = h, H, A$).
To improve the calculation, the parameters $\mb$, $\mt$, $\At$ and $\Ab$ have
also been used as running parameters.

\item
Other ``early'' papers considered QCD corrections to various scalar
quark decays~\cite{squark_q_chi_als,squark_q_gl_als,stop_top_gl_als}. 
They mostly employed an on-shell 
scheme for the quark/squark masses and the squark mixing angle 
$\tsq$, where the counterterm to the mixing angle is
$\de \tsq \propto \de Y_q$.

\item
The renormalization scheme ``$\Ab$~vertex, $\re Y_b$~OS'' is the
complex version of the renormalization used in
\citeres{sbotrenold,mhiggsFDalbals} for the 
\order{\alb\als} corrections to the neutral Higgs boson self-energies
and thus to the lightest MSSM Higgs boson mass, $\Mh$.

\end{itemize}

\bigskip
In the following subsections we define in detail the various
renormalization schemes. As explained before and indicated in
\refta{tab:RS} the two bottom squark
masses are renormalized on-shell in all the schemes, as in 
\refeqs{sbotzOS} and \eqref{sboteOS}, and taking into account
the shift of $M_{\tilde Q_L}^2(\tilde{b})$ in \refeq{MSbotshift}. 
Within the subsections only the remaining conditions and
renormalization constants are defined explicitly
(where $\de \mu$ and $\dtanb$ are
defined within the chargino/neutralino sector and the Higgs sector,
respectively, in all the different renormalization schemes and are not
discussed any further).


\subsection{On-shell (RS1)}
\label{sec:OS}

This renormalization scheme is analogous to the OS scheme employed for
the top quark/squark sector.

\begin{itemize}
\item[]
\begin{itemize}
\item[(ii)]
The bottom-quark mass is defined OS, yielding the one-loop 
counterterm $\de \mb$:
\begin{align} 
\label{dmb_OS}
\de\mb = \frac{1}{2} \wtre \KKKL
  \mb \KKL \Si_b^L (\mb^2) + \Si_b^R (\mb^2) \KKR
    + \KKL \Si_b^{SL} (\mb^2) + \Si_b^{SL} (\mb^2) \KKR \KKKR~.
\end{align}

\item[(iii), (iv)]
We choose an OS renormalization condition for the non-diagonal
entry in the matrix of \refeq{proc2}, analogous to the one
applied in the top quark/squark sector, setting
\begin{align}
\de Y_b =  \frac{1}{2} \wtre \KKKL
    \Si_{\Sbot_{12}}(\msbe^2) + \Si_{\Sbot_{12}}(\msbz^2) \KKKR~.
\label{dYb_OS}
\end{align}
\end{itemize}
\end{itemize}
The conditions (i)--(iv) fix all independent parameters and their
respective counterterms. 
Analogous to the calculation of the counterterm of the trilinear
coupling $A_t$, relating \refeq{proc1b} and \refeq{dMsq12physpar} yields
the following condition for $\de \Ab$, 
\begin{align}\label{Ab_OS}
\de\Ab &= \frac{1}{\mb} \bigl[U_{\Sbot_{11}} U_{\Sbot_{12}}^*
          (\de\msbe^2 - \de\msbz^2) 
        + U_{\Sbot_{11}} U_{\Sbot_{22}}^* \de Y_b^*
        + U_{\Sbot_{12}}^* U_{\Sbot_{21}} \de Y_b 
        - (\Ab - \mu^*\tb)\, \de\mb \bigr] \non \\
&\quad  + (\de\mu^* \tb + \mu^*\, \dtanb)
\end{align} 
with $\de\msbe^2$ and $\de\msbz^2$ given  in \refeqs{sboteOS} and 
\eqref{sbotzOS}, respectively.


\subsection{\boldmath{$\mb$ \DRbar\ } and \boldmath{$\Ab$ \DRbar\ } (RS2)}
\label{sec:mbDRbar_AbDRbar}

\begin{itemize}
\item[]
\begin{itemize}
\item[(ii)]
The bottom-quark mass is defined \DRbar, yielding the one-loop
  counterterm $\de \mb$:
\begin{align} 
\de\mb = \frac{1}{2} \wtre \KKKL
  \mb \KKL \Si_b^L (\mb^2) + \Si_b^R (\mb^2) \KKR_{\rm div}
+ \KKL \Si_b^{SL} (\mb^2) + \Si_b^{SR} (\mb^2) \KKR_{\rm div} \KKKR~.
\end{align}
The $|_{\rm div}$ terms are the ones proportional to
$\De = 2/\eps - \ga_{\rm E} + \log(4 \pi)$, when using dimensional
regularization/reduction in $D = 4 - \eps$ dimensions; $\ga_{\rm E}$ is
the Euler constant.

\item[(iii), (iv)]
The complex parameter $\Ab$ is renormalized \DRbar,
\begin{align}
\de\Ab &= \frac{1}{\mb} \Bigl[ U_{\Sbot_{11}} U_{\Sbot_{12}}^*
         \KL  \wtre\Si_{\Sbot_{11}}(\msbe^2)\ddiv 
             -\wtre\Si_{\Sbot_{22}}(\msbz^2)\ddiv \KR \non \\
&\quad + \frac{1}{2}\, U_{\Sbot_{12}}^* U_{\Sbot_{21}} 
          \KL \wtre\Si_{\Sbot_{12}}(\msbe^2)\ddiv
             +\wtre\Si_{\Sbot_{12}}(\msbz^2)\ddiv \KR \non \\
&\quad + \frac{1}{2}\, U_{\Sbot_{11}} U_{\Sbot_{22}}^* 
          \KL \wtre\Si_{\Sbot_{12}}(\msbe^2)\ddiv
             +\wtre\Si_{\Sbot_{12}}(\msbz^2)\ddiv \KR^* \non \\
&\quad - \frac{1}{2}(\Ab - \mu^* \tb)\, 
\wtre \bigl\{
  \mb \KKL \Si_b^L (\mb^2) + \Si_b^R (\mb^2) \KKR_{\rm div} \non \\
&\qquad + \KKL \Si_b^{SL} (\mb^2) + \Si_b^{SR} (\mb^2) \KKR_{\rm
    div} \bigr\} 
 \Bigr]  
       + \de\mu^*\ddiv \tb + \mu^*\, \dtanb~. 
\end{align} 
\end{itemize}
\end{itemize}

All independent parameters are defined by the conditions (i)--(iv) and
the corresponding counterterms are determined.
Solving \refeqs{proc1b} and
(\ref{dMsq12physpar}) for $\de Y_b$ yields
\begin{align}
\de Y_b &= \frac{1}{|U_{\Sbot_{11}}|^2 - |U_{\Sbot_{12}}|^2} \Big[ 
           U_{\Sbot_{11}} U_{\Sbot_{21}}^* 
          (\de\msbe^2 - \de\msbz^2) \non \\
&\quad  + \mb \Big( U_{\Sbot_{11}} U_{\Sbot_{22}}^* 
          \KL \de\Ab^* - \mu\, \dtanb - \tb\, \de\mu \KR 
                  - U_{\Sbot_{12}} U_{\Sbot_{21}}^* 
          \KL \de\Ab - \mu^* \dtanb - \tb\, \de\mu^* \KR \Big) \non \\
&\quad  + \KL U_{\Sbot_{11}} U_{\Sbot_{22}}^* (\Ab^* - \mu \tb)
            - U_{\Sbot_{12}} U_{\Sbot_{21}}^* (\Ab - \mu^* \tb) \KR\, \de\mb
\Big]~,
\label{dYb_mbDRbar_AbDRbar}
\end{align}
where $\de\msbe^2$ and $\de\msbz^2$ are given in  \refeqs{sboteOS} and  
\eqref{sbotzOS}, respectively.


\subsection{\boldmath{$\mb$ \DRbar\ } and \boldmath{$Y_b$ \DRbar\ } (RS3)}
\label{sec:mbDRbar_YbDRbar}

\begin{itemize}
\item[]
\begin{itemize}
\item[(ii)]
The bottom-quark mass is defined \DRbar, yielding the one-loop
  counterterm $\de \mb$:
\begin{align} 
\label{dmb_mbDRbar_YbDRbar}
\de\mb = \frac{1}{2} \wtre \KKKL
  \mb \KKL \Si_b^L (\mb^2) + \Si_b^R (\mb^2) \KKR_{\rm div}
+ \KKL \Si_b^{SL} (\mb^2) + \Si_b^{SR} (\mb^2) \KKR_{\rm div} \KKKR~.
\end{align}

\item[(iii), (iv)]
The complex counterterm  $\de Y_b$ is determined via a \DRbar\
renormalization condition, setting
\begin{align}
\de Y_b =  \frac{1}{2} \wtre \KKKL
 \Si_{\Sbot_{12}}(\msbe^2)\ddiv + \Si_{\Sbot_{12}}(\msbz^2)\ddiv \KKKR~.
\label{dYb_mbDRbar_YbDRbar}
\end{align}
\end{itemize}
\end{itemize}

As in Sect.~\ref{sec:OS}, the renormalization conditions 
(ii), (iii) and (iv) fix the
counterterms $\de \mb$ and $\de Y_b$, respectively. Together with the
renormalization conditions for $\de\msbe^2$ and
$\de\msbz^2$ (see \refeq{sboteOS} and \refeq{sbotzOS}, respectively),
$\de \Ab$ is given by the linear 
combination of these counterterms as 
\begin{align}
\de\Ab &= \frac{1}{\mb} \bigl[ U_{\Sbot_{11}} U_{\Sbot_{12}}^*
          (\de\msbe^2 - \de\msbz^2) 
          + U_{\Sbot_{11}} U_{\Sbot_{22}}^* \de Y_b^*
          + U_{\Sbot_{12}}^* U_{\Sbot_{21}} \de Y_b 
          - (\Ab - \mu^*\tb)\, \de\mb \bigr] \non \\
&\quad + (\de\mu^* \tb + \mu^*\, \dtanb)~,
\label{dAb_mbDRbar_YbDRbar}
\end{align}
which, of course, shows the same analytical dependence of the independent
counterterms as $\de\Ab$ in \refeq{Ab_OS} in \refse{sec:OS}.


\subsection{\boldmath{$\mb$ \DRbar\ } and \boldmath{$Y_b$} on-shell (RS4)}
\label{sec:mbDRbar_YbOS}

\begin{itemize}
\item[]
\begin{itemize}
\item[(ii)]
The bottom-quark mass is defined \DRbar, yielding the one-loop
  counterterm $\de \mb$:
\begin{align} 
\label{dmb_mbDRbar_YbOS}
\de\mb = \frac{1}{2} \wtre \KKKL
  \mb \KKL \Si_b^L (\mb^2) + \Si_b^R (\mb^2) \KKR_{\rm div}
+ \KKL \Si_b^{SL} (\mb^2) + \Si_b^{SR} (\mb^2) \KKR_{\rm div} \KKKR~.
\end{align}

\item[(iii), (iv)]
The complex counterterm $\de Y_b$ is fixed by an on-shell 
renormalization condition, as in Sect.~\ref{sec:OS},
\begin{align}\label{dYb_mbDRbar_YbOS}
\de Y_b =  \frac{1}{2} \wtre \KKKL
  \Si_{\Sbot_{12}}(\msbe^2) + \Si_{\Sbot_{12}}(\msbz^2) \KKKR~.
\end{align}
\end{itemize}
\end{itemize}
As in Sect.~\ref{sec:OS} and in \refse{sec:mbDRbar_YbDRbar}, the
  renormalization conditions (i)--(iv) fix the 
counterterms $\de\msbz^2$, $\de \mb$ and $\de Y_b$. The further
renormalization condition \refeq{sboteOS} determines the counterterm
$\de\msbe^2$. Analogous to Sect.~\ref{sec:OS} and to
\refse{sec:mbDRbar_YbDRbar}, 
$\de \Ab$  can be expressed in terms of these counterterms, 
\begin{align}
\de\Ab &= \frac{1}{\mb} \bigl[ U_{\Sbot_{11}} U_{\Sbot_{12}}^*
          (\de\msbe^2 - \de\msbz^2) 
          + U_{\Sbot_{11}} U_{\Sbot_{22}}^* \de Y_b^*
          + U_{\Sbot_{12}}^* U_{\Sbot_{21}} \de Y_b 
          - (\Ab - \mu^*\tb)\, \de\mb \bigr] \non \\
&\quad + (\de\mu^* \tb + \mu^*\, \dtanb)~,
\end{align}
which, of course, has the same form as in \refeqs{Ab_OS} and
\eqref{dAb_mbDRbar_YbDRbar}.


\subsection{\boldmath{$\Ab$ \DRbar\ } and \boldmath{$\re Y_b$} on-shell (RS5)}
\label{sec:AbDRbar_ReYbOS}

\begin{itemize}
\item[]
\begin{itemize}
\item[(ii)]
In the subsections \ref{sec:OS}--\ref{sec:mbDRbar_YbOS} the second
renormalization condition defines the bottom quark mass. In this scheme,
we choose an on-shell renormalization condition for the real part of the
counterterm $\de Y_b$ which determines $\re \de Y_b$ as following
\begin{align}
\re \de Y_b = \frac{1}{2} \re \KKKL
              \wtre{\Si}_{\Sbot_{12}}(\msbe^2) +
              \wtre{\Si}_{\Sbot_{12}}(\msbz^2) \KKKR~.
\label{dReYb_AbDRbar_ReYbOS}
\end{align}

\item[(iii), (iv)]
The complex $\Ab$ parameter is defined \DRbar
\begin{align}
\de\Ab &= \frac{1}{\mb} \Bigl[ U_{\Sbot_{11}} U_{\Sbot_{12}}^*
         \KL \wtre\Si_{\Sbot_{11}}(\msbe^2)\ddiv 
            -\wtre\Si_{\Sbot_{22}}(\msbz^2)\ddiv \KR \non \\
&\quad + \frac{1}{2} U_{\Sbot_{12}}^* U_{\Sbot_{21}} 
         \KL \wtre\Si_{\Sbot_{12}}(\msbe^2)\ddiv
            +\wtre\Si_{\Sbot_{12}}(\msbz^2)\ddiv \KR \non \\
&\quad + \frac{1}{2} U_{\Sbot_{11}} U_{\Sbot_{22}}^* 
         \KL \wtre\Si_{\Sbot_{12}}(\msbe^2)\ddiv
            +\wtre\Si_{\Sbot_{12}}(\msbz^2)\ddiv \KR^* \non \\
&\quad - \frac{1}{2}(\Ab - \mu^* \tb)\, 
\wtre \bigl\{
  \mb \KKL \Si_b^L (\mb^2) + \Si_b^R (\mb^2) \KKR_{\rm div} \non\\
&\qquad + \KKL \Si_b^{SL} (\mb^2) + \Si_b^{SR} (\mb^2) \KKR_{\rm
    div} \bigr\} 
       + \de\mu^*\ddiv \tb + \mu^*\, \dtanb~. 
\label{dAb}
\end{align} 
\end{itemize}
\end{itemize}

With the conditions (i)--(iv) the independent counterterms $\delta
\msbz^2$, $\re \de Y_b$ and $\de \Ab$ are determined, 
and $\de\msbe^2$ is given by \refeq{sboteOS}. The missing
counterterms $\de \mb$ and $\im \de Y_b$ can be expressed by the
independent counterterms. Relating \refeq{proc1b}, here explicitly
written as
\begin{align}\label{proc1bexpl}
(\de \matr{M}_{\Sbot})_{12} &= (\Ab^* - \mu\tb)\, \de\mb 
                              + \mb \KL \de\Ab^*
                              - \mu\, \dtanb - \de\mu \tb \KR~,
\end{align}
and \refeq{dMsq12physpar}, here with $\de Y_b$ explicitly split into a
real and an imaginary part
\begin{align}\label{dMsq12physparsplit}
(\de \matr{M}_{\Sbot})_{12} &= 
  U_{\Sbot_{11}}^* U_{\Sbot_{12}} (\de\msbe^2 - \de\msbz^2) \non \\
&\quad
 + U_{\Sbot_{11}}^* U_{\Sbot_{22}} (\re \de Y_b + i \im \de Y_b)
 + U_{\Sbot_{12}} U_{\Sbot_{21}}^* (\re \de Y_b - i \im \de Y_b)~,
\end{align}
results in the two equations
\begin{align}
\label{eq:dM12R}
\re \KKKL \Ab^* - \mu \tb \KKKR\, \de\mb
&= - \mb \re \de\Ab - \re \de S 
   - \im \KKKL U_{\Sbot_{11}}^* U_{\Sbot_{22}} 
             - U_{\Sbot_{12}} U_{\Sbot_{21}}^* \KKKR \im \de Y_b~, \\
\label{eq:dM12I}
\im \KKKL \Ab^* - \mu \tb \KKKR\, \de\mb 
&= + \mb \im \de\Ab - \im \de S 
   + \re \KKKL U_{\Sbot_{11}}^* U_{\Sbot_{22}} 
             - U_{\Sbot_{12}} U_{\Sbot_{21}}^* \KKKR \im \de Y_b
\end{align}
with
\begin{align}
\de S &= - \mb\, (\mu\,\dtanb + \de\mu \tb)
         - U_{\Sbot_{11}}^* U_{\Sbot_{12}} (\de \msbe^2 - \de\msbz^2) \non \\
&\quad   - \KL U_{\Sbot_{11}}^* U_{\Sbot_{22}} 
         + U_{\Sbot_{12}} U_{\Sbot_{21}}^* \KR \re \de Y_b~,
\end{align}
where $\de \msbe^2$ and $\de\msbz^2$ are given by \refeq{sboteOS} and
\refeq{sbotzOS}. 

\medskip
The above two equations, 
(\ref{eq:dM12R}) and (\ref{eq:dM12I}), 
can be solved for $\im\de Y_b$ and $\de\mb$, yielding\\[1em]
\begin{align}\label{dmb_AbDRbar_ReYbOS}
\de \mb &= \frac{b_r c_i - b_i c_r}{a_r b_i - a_i b_r}~,\\[2mm]
\label{dImYb_AbDRbar_ReYbOS}
\im\de Y_b &= \frac{a_i c_r - a_r c_i}{a_r b_i - a_i b_r}
\end{align}
with
\begin{align}
a_r &= \re \KKKL \Ab^* - \mu \tb \KKKR~, \\
a_i &= \im \KKKL \Ab^* - \mu \tb \KKKR~, \\
b_r &= + \im \KKKL U_{\Sbot_{11}}^* U_{\Sbot_{22}} 
                 - U_{\Sbot_{12}} U_{\Sbot_{21}}^* \KKKR~, \\
b_i &= - \re \KKKL U_{\Sbot_{11}}^* U_{\Sbot_{22}} 
                 - U_{\Sbot_{12}} U_{\Sbot_{21}}^* \KKKR~, \\
c_r &= + \mb \re \de\Ab + \re\de S~, \\
c_i &= - \mb \im \de\Ab + \im\de S~.
\end{align}


\subsection{\boldmath{$\Ab$} via vertex and \boldmath{$\re Y_b$ on-shell} (RS6)}
\label{sec:AbOS_ReYbOS}

\begin{itemize}
\item[]
\begin{itemize}
\item[(ii)]
An on-shell renormalization condition is imposed for the real part of
the counterterm $\de Y_b$ which determines  $\re \de Y_b$ as
\begin{align}
\re \de Y_b = \frac{1}{2} \re \KKKL
 \wtre{\Si}_{\Sbot_{12}}(\msbe^2) +
 \wtre{\Si}_{\Sbot_{12}}(\msbz^2) \KKKR~.
\label{dReYb_AbOS_ReYbOS}
\end{align}

\item[(iii), (iv)]
The renormalization conditions introduced here are analogous to 
the prescriptions used in \citeres{dissHR,sbotrenold,mhiggsFDalbals}, 
but extended to the complex MSSM.
The complex parameter $\Ab$ is renormalized via the vertex
$A\, \Sbote^\dagger \Sbotz$, denoting the renormalized vertex as 
$\hat\La(p_{A}^2, p_{\Sbote}^2, p_{\Sbotz}^2)$, see \reffi{fig:vertex}.
\vspace{-2ex}
\begin{figure}[htb!]
\BC
\setlength{\unitlength}{1pt}
\begin{picture}(350, 180)
\DashArrowLine(160,105)(195,125){5}
\DashArrowLine(195,055)(160,075){5}
\DashLine(80,90)(140,90){5}
\put(65,85){$A$}
\put(200,50){$\Sbotz$}
\put(200,125){$\Sbote$}
\put(240,85){$\Hat{=} \quad
             i\, \hat{\La}(p_{A}^2, p_{\Sbote}^2, p_{\Sbotz}^2)$}\,
\GCirc(140,90){20}{.6}
\end{picture}
\vspace{-2ex}
\caption{The renormalized vertex 
$\hat\La(p_{A}^2, p_{\Sbote}^2, p_{\Sbotz}^2)$.}
\label{fig:vertex}
\EC
\end{figure}

The tree-level vertex $A\, \Sbote^\dagger \Sbotz$, denoted as $V_{A\,
  \Sbote^\dagger \Sbotz}$, is given  as 
\begin{align}\nonumber
V_{A\, \Sbote^\dagger \Sbotz} = \frac{i e\,\mb}{2\MW\sw \cos \beta} 
\Bigl[&
    U_{\Sbot_{11}} U_{\Sbot_{22}}^* (\mu \cos \ben + \Ab^* \sin \ben) 
\non \\
- & U_{\Sbot_{12}} U_{\Sbot_{21}}^* (\mu^* \cos \ben + \Ab \sin \ben)
\Bigr]~,
\end{align}
where $\ben$ is the mixing angle of the $\cp$-odd Higgs boson fields
with $\ben = \beta$ at tree-level. Note that in our renormalization
prescription we do not renormalize the mixing angles but only $\tan
\beta$ appearing in the Lagrangian before the transformation of the 
$\cp$-odd Higgs boson fields into mass eigenstate fields is performed.
The renormalized vertex reads,
\begin{align}
& \hat\La(p_A^2, p_{\Sbote}^2, p_{\Sbotz}^2) \; = \;
   \La(p_A^2, p_{\Sbote}^2, p_{\Sbotz}^2) \; 
   + \frac{i e\,\mb}{2\MW\sw} \Bigg\{ \non \\
&\qquad\;   \tb \KKL U_{\Sbot_{11}} U_{\Sbot_{22}}^* \de\Ab^*
                   - U_{\Sbot_{12}} U_{\Sbot_{21}}^* \de \Ab \KKR 
              + \KKL U_{\Sbot_{11}} U_{\Sbot_{22}}^* \de\mu
                   - U_{\Sbot_{12}} U_{\Sbot_{21}}^* \de \mu^* \KKR \non \\
&\quad + \KKL  U_{\Sbot_{11}} U_{\Sbot_{22}}^* \KL \mu + \tb\Ab^* \KR
             - U_{\Sbot_{12}} U_{\Sbot_{21}}^* \KL \mu^* + \tb\Ab \KR 
         \KKR \non \\
&\qquad  \times \KKL \frac{\de\mb}{\mb} + \edz (\de \bar Z_{\Sbot_{11}}^* 
                    + \de\bar Z_{\Sbot_{22}} + \dZ{AA}) 
                    + \sinb\cosb \dtanb \KKR \non \\
&\quad + \KKL U_{\Sbot_{11}} U_{\Sbot_{22}}^* (\mu + \tb\Ab^*)
            - U_{\Sbot_{12}} U_{\Sbot_{21}}^* (\mu^* + \tb\Ab)
         \KKR \KL \dZ{e} - \frac{\de\MW^2}{2\,\MW^2}
            - \frac{\de\sw}{\sw} \KR  \non \\ 
&\quad  + i \im \KKKL U_{\Sbot_{11}} U_{\Sbot_{12}}^* 
                           \KL \mu + \tb\Ab^* \KR
                     \KKKR \dZ{\Sbot_{12}}
       + i \im \KKKL U_{\Sbot_{21}} U_{\Sbot_{22}}^* 
                     \KL \mu + \tb\Ab^* \KR
               \KKKR \dZ{\Sbot_{21}}^*  \non \\
&\quad  - \edz \KKL U_{\Sbot_{11}} U_{\Sbot_{22}}^* \KL \Ab^* - \mu \tb \KR
                 - U_{\Sbot_{12}} U_{\Sbot_{21}}^* \KL \Ab - \mu^* \tb \KR
              \KKR \dZ{AG}
\Bigg\}~.
\end{align}
The off-diagonal $Z$~factors are determined according to
\refeq{dZstopoffdiagOS}, 
\begin{align}
\dZ{\Sbot_{12}} &= + 2\, 
\frac{\wtre\Si_{\Sbot_{12}}(\msbz^2) - \re\de Y_b - i \im \de Y_b}
     {(\msbe^2 - \msbz^2)} \; 
=: \dZ{\Sbot_{12}}^{\rm c} - \frac{2 i \im \de Y_b}{\msbe^2 - \msbz^2}~, \non\\
\dZ{\Sbot_{21}} &= - 2\,
\frac{\wtre\Si_{\Sbot_{21}}(\msbe^2) - \re \de Y_b + i \im \de Y_b}
     {(\msbe^2 - \msbz^2)} \;
=: \dZ{\Sbot_{21}}^{\rm c} - \frac{2 i \im \de Y_b}{\msbe^2 - \msbz^2}~.
\end{align}
Introducing appropriate abbreviations we get
\begin{align}
& \hat\La(p_A^2, p_{\Sbote}^2, p_{\Sbotz}^2) \; = \;
  \La(p_A^2, p_{\Sbote}^2, p_{\Sbotz}^2) \; + \\
&\quad\;\, \frac{i e}{2\MW\sw} \Big\{ 
\mb \tb\; (U_{\Sbot_{11}} U_{\Sbot_{22}}^* \de\Ab^*  
         - U_{\Sbot_{12}} U_{\Sbot_{21}}^* \de\Ab) + \de M 
         + i\, U_Y\, \im \de Y_b \non \\
&\quad + \KKL  U_{\Sbot_{11}} U_{\Sbot_{22}}^* \KL \mu + \tb\Ab^* \KR
             - U_{\Sbot_{12}} U_{\Sbot_{21}}^* \KL \mu^* + \tb\Ab \KR 
         \KKR 
         (\de\mb + \de Z_{\rm d}) \Big\} + \de Z_{\rm o} \non
\end{align}

with

\begin{align}
\de M &= \mb \KKL U_{\Sbot_{11}} U_{\Sbot_{22}}^* \de\mu 
                - U_{\Sbot_{12}} U_{\Sbot_{21}}^* \de\mu^* 
             \KKR~, \\[2mm]
U_Y &= \frac{4\, i\, \mb}{\msbe^2 - \msbz^2}
      \im \Big\{ U_{\Sbot_{11}}^* U_{\Sbot_{12}} \KL \mu^* + \tb\Ab \KR
          \Big\}~, \\[2mm]
\dZ{\rm d} &= \mb \KKL \edz (\de \bar Z_{\Sbot_{11}}^* 
                             + \de \bar Z_{\Sbot_{22}} + \dZ{AA} )
                    + \sinb\cosb \dtanb \KKR~,\\[2em]
\dZ{\rm o} &= \frac{i e\, \mb}{2\MW\sw} \Bigg\{ \non \\
&\quad \KKL U_{\Sbot_{11}} U_{\Sbot_{22}}^* (\mu + \tb\Ab^*)
          - U_{\Sbot_{12}} U_{\Sbot_{21}}^* (\mu^* + \tb\Ab)
       \KKR  \KL \dZ{e} - \frac{\de\MW^2}{2\,\MW^2}
                        - \frac{\de\sw}{\sw} \KR  \non \\
&\quad + i \im \KKKL U_{\Sbot_{11}} U_{\Sbot_{12}}^* 
                          \KL \mu + \tb\Ab^* \KR
                    \KKKR \dZ{\Sbot_{12}}^{\rm c}
       + i \im \KKKL U_{\Sbot_{21}} U_{\Sbot_{22}}^* 
                     \KL \mu + \tb\Ab^* \KR
               \KKKR \dZ{\Sbot_{21}}^{{\rm c}\,*} \non \\
&\quad - \edz \KKL U_{\Sbot_{11}} U_{\Sbot_{22}}^* \KL \Ab^* - \mu \tb \KR
                 - U_{\Sbot_{12}} U_{\Sbot_{21}}^* \KL \Ab - \mu^* \tb \KR
              \KKR \dZ{AG}
\Bigg\}~. 
\end{align}
The renormalization condition reads~\cite{dissHR,sbotrenold}
\begin{align}
\wtre\hat\La(0, \msbe^2, \msbe^2) + \wtre\hat\La(0, \msbz^2, \msbz^2) = 0~,
\end{align}
which corresponds to the two conditions
\begin{align}
\label{eq:ReLam}
\re \KKKL \wtre\hat\La(0, \msbe^2, \msbe^2) 
        + \wtre\hat\La(0, \msbz^2, \msbz^2) \KKKR = 0~, \\
\label{eq:ImLam}
\im \KKKL \wtre\hat\La(0, \msbe^2, \msbe^2) 
        + \wtre\hat\La(0, \msbz^2, \msbz^2) \KKKR = 0~.
\end{align}

\end{itemize}
\end{itemize}

The conditions (i)--(iv), are sufficient to fix all independent parameters
and their respective counterterms. As in Sect.~\ref{sec:AbDRbar_ReYbOS},
relating \refeqs{proc1bexpl} and \eqref{dMsq12physparsplit}, one
derives \refeqs {eq:dM12R} and  \eqref{eq:dM12I} which can also be
written in the form
\begin{align}
\label{eq:dM12R_AbOS_ReYbOS}
\re \KKKL \Ab^* - \mu \tb \KKKR\, \de\mb 
&= - \mb \re \de\Ab - \re \de S + \im U_+ \im \de Y_b~, \\
\label{eq:dM12I_AbOS_ReYbOS}
\im \KKKL \Ab^* - \mu \tb \KKKR\, \de\mb 
&= + \mb \im \de\Ab - \im \de S + \re U_- \im \de Y_b
\end{align}
with
\begin{align}
\de S &= -\mb (\mu\,\dtanb + \de\mu \tb)
         - U_{\Sbot_{11}}^* U_{\Sbot_{12}} (\de\msbe^2 - \de\msbz^2)
         - \KL \re U_+ - i \im U_- \KR \re \de Y_b~, \\
U_{\pm} &= U_{\Sbot_{11}} U_{\Sbot_{22}}^* \pm U_{\Sbot_{12}} U_{\Sbot_{21}}^*~.
\end{align}
$\de\msbe^2$ and $\de\msbz^2$
are fixed by \refeqs{sboteOS} and \eqref{sbotzOS}. 

\medskip
The above four equations (\ref{eq:ReLam}), (\ref{eq:ImLam}),
(\ref{eq:dM12R_AbOS_ReYbOS}) and 
(\ref{eq:dM12I_AbOS_ReYbOS}), 
can be solved for $\re\de \Ab$, $\im\de \Ab$, $\im\de Y_b$ and
$\de\mb$. Though, we still consider $\re\de \Ab$ and $\im\de \Ab$ as
independent counterterms we first calculate  $\im\de Y_b$ and
$\de\mb$ in dependence of  $\re\de \Ab$ and $\im\de \Ab$ for
economically solving the systems of equations. The solution for  $\im\de
Y_b$ and $\de\mb$ is\\[1em]
\begin{align}
\label{dmb_AbOS_ReYbOS}
\de \mb &= \frac{d_i f_r - d_r f_i}{e_r f_i - e_i f_r}~,\\[2mm]
\im\de Y_b &= \frac{d_r e_i - d_i e_r}{e_r f_i - e_i f_r}
\end{align}
with
\begin{align}
d_r &= 2 \tb \KL \im U_+ \im \de S - \re U_- \re \de S \KR \\
&\quad + 2 \re \KKL \frac{\MW \sw}{i\,e} 
               \KL 2 \dZ{\rm o} + \wtre\La(0, \msbe^2, \msbe^2) 
                                + \wtre\La(0, \msbz^2, \msbz^2) \KR
                + \de M + \dZ{\rm d} U_m \KKR~, \non \\
d_i &= -2 \tb \KL \re U_+ \im \de S + \im U_- \re \de S \KR \\
&\quad + 2 \im \KKL \frac{\MW \sw}{i\,e} 
               \KL 2 \dZ{\rm o} + \wtre\La(0, \msbe^2, \msbe^2) 
                                + \wtre\La(0, \msbz^2, \msbz^2) \KR
                + \de M + \dZ{\rm d} U_m \KKR~, \non \\
e_r &= +2 \tb \KKL \im U_+ \im \KKKL \Ab^* - \mu\tb \KKKR 
                  -\re U_- \re \KKKL \Ab^* - \mu\tb \KKKR 
              \KKR + 2 \re U_m~,\\
e_i &= -2 \tb \KKL \re U_+ \im \KKKL \Ab^* - \mu\tb \KKKR 
                  +\im U_- \re \KKKL \Ab^* - \mu\tb \KKKR 
              \KKR + 2 \im U_m~, \\
f_r &= - 2 \im U_Y~, \\
f_i &= 2 \tb \KL |U_{\Sbot_{11}}|^2 - |U_{\Sbot_{12}}|^2 \KR
\end{align}
and
\begin{align}
\label{def:Um}
U_m &=   U_{\Sbot_{11}} U_{\Sbot_{22}}^* (\Ab^* \tb + \mu)
       - U_{\Sbot_{12}} U_{\Sbot_{21}}^* (\Ab \tb + \mu^*)~.
\end{align}
From the \refeqs{eq:dM12R_AbOS_ReYbOS} and 
\eqref{eq:dM12I_AbOS_ReYbOS} we immediately obtain $\de \Ab$ as
\begin{align}
\re\de\Ab &= \ed{\mb} \KKL + \im\de Y_b \im U_+ - \re \de S
             - \de\mb \re \KKKL \Ab^* - \mu \tb \KKKR \KKR~, \\
\im\de\Ab &= \ed{\mb} \KKL - \im\de Y_b \re U_- + \im \de S
             + \de\mb \im \KKKL \Ab^* - \mu \tb \KKKR \KKR~.
\end{align}

Finally the $\bar{Z}$~factors in $\hat{\Lambda}$ have to be
determined. The following 
condition is used 
\begin{align}
\wtre \hSi_{\Sbot_{ii}}(\msbe^2) - 
\wtre \hSi_{\Sbot_{ii}}(\msbz^2) = 0 \qquad (i = 1,2)~.
\end{align}    
This condition results in the following $\bar{Z}$~factors
\begin{align}
\delta\bar{Z}_{\Sbot_{ii}} = 
-\frac{\wtre \Si_{\Sbot_{ii}}(\msbe^2) - 
       \wtre \Si_{\Sbot_{ii}}(\msbz^2)}{\msbe^2 - \msbz^2} 
       \qquad (i = 1,2)~,
\end{align}
which guarantees the IR finiteness of the renormalized vertex
$\hat{\Lambda}$~\cite{sbotrenold}.\\

\smallskip
Another subtlety has to be explained here:
due to the fact that we have infrared divergent $C$-functions
at $p_1 = 0$ in $\Lambda(p_1^2=0,p^2,p^2)$, we must deal with vanishing 
Gram-determinants. 
Therefore we follow \citere{cfunc} (and references therein) and 
replace the corresponding $C$-functions by well behaving linear 
combinations of $B$-functions. 
Details can be found in the appendix.


\subsection{Parameter definition}

The input parameters in the $b/\Sbot$ sector have to correspond to the
chosen renormalization scheme. We start by defining the bottom quark
mass, where the 
experimental input is the SM \MSbar\ mass \cite{pdg},
\begin{align}
\label{def:mbMB}
\mb^{\MSbar}(\mb) & = 4.2 \gev~.
\end{align}
The value of $\mb^{\MSbar}(\mu_R)$ (at the renormalization scale 
$\mu_R$) is calculated from $\mb^{\MSbar}(\mb)$ at the three loop
level following the prescription given in~\citere{RunDec}.

An ``on-shell'' mass is derived from the \MSbar\ mass via
\begin{align}
\mb^{\OS} &= \mb^{\MSbar}(\mu_R) \; 
   \KKL 1 + \frac{\als^{\MSbar}(\mu_R)}{\pi} 
        \KL \frac{4}{3} + 2\, \ln \frac{\mu_R}{\mb^{\MSbar}(\mu_R)} \KR 
   \KKR~.
\end{align}
The $\DRbar$ bottom quark mass is calculated iteratively from%
\footnote{
In case of complex $\db$ the replacement $(1 + \db) \to |1 + \db|$
should be performed~\cite{komplexDb}.}
\begin{align}
\label{eq:mbDR}
\mb^{\DRbar} &= \frac{\mb^{\OS} (1 + \db) + \de\mb^{\OS} - \de\mb^{\DRbar}}
            {1 + \db}
\end{align}
with an accuracy of $|1 - (\mb^{\DRbar})^{(n)}/(\mb^{\DRbar})^{(n-1)}| < 10^{-5}$
reached in the $n$th step of the iteration.
The bottom quark mass of a special renormalization scheme is then obtained
from 
\begin{align}\label{eq:mbcorr}
\mb &= \mb^{\DRbar} + \de\mb^{\DRbar} - \de\mb~.
\end{align}
Here we have used
\begin{align}
\de\mb^{\OS} &= \frac{1}{2} \wtre \KKKL
  \mb \KKL \Si_b^L (\mb^2) + \Si_b^R (\mb^2) \KKR
    + \KKL \Si_b^{SL} (\mb^2) + \Si_b^{SL} (\mb^2) \KKR \KKKR~, \non \\[2mm]
\de\mb^{\DRbar} &= \frac{1}{2} \wtre \KKKL
  \mb \KKL \Si_b^L (\mb^2) + \Si_b^R (\mb^2) \KKR_{\rm div}
+ \KKL \Si_b^{SL} (\mb^2) + \Si_b^{SR} (\mb^2) \KKR_{\rm div} \KKKR~, 
\end{align}
and $\de\mb$ as given in \refses{sec:OS}--\ref{sec:AbOS_ReYbOS}.
The quantity $\db$~\cite{deltab1,deltab2} resums the \order{(\als\tb)^n}
and \order{(\alt\tb)^n} terms and is given by 
\begin{align}
\db &= \frac{2\als(\mt)}{3\pi} \, \tb \, M_3^* \, \mu^* \,
                          I(\msbe^2, \msbz^2, \mgl^2) \;
     + \frac{\alt(\mt)}{4\pi} \, \tb \, \At^* \, \mu^* \, 
                          I(\mste^2, \mstz^2, |\mu|^2)
\end{align}
with
\begin{align}
I(a, b, c) &= - \frac{a b\, \ln(b/a) + a c\, \ln(a/c) + b c\, \ln(c/b)}
                     {(a - c) (c - b) (b - a)}~.
\end{align}
Here $\alt$ is defined in terms of the top Yukawa coupling 
$y_t(\mt) = \sqrt{2} \mt(\mt)/v$ as
$\alt(\mt) = y_t^2(\mt)/(4\pi)$ with 
$v = 1/\sqrt{\sqrt{2}\, G_F} = 246.218 \gev$
and 
$\mt(\mt)\approx \mt/(1-\frac{1}{2\,\pi} \alt(\mt) +\frac{4}{3\,\pi}\als(\mt))$.
$M_3$ is the soft SUSY-breaking parameter
for the gluinos, with the gluino mass given as $\mgl := |M_3|$.

\newpage


\section{Renormalization scheme analysis}
\label{sec:RSana}

\subsection{Calculation of loop diagrams}
\label{sec:calc}

\begin{figure}[t!]
\begin{center}
\includegraphics[width=0.90\textwidth]{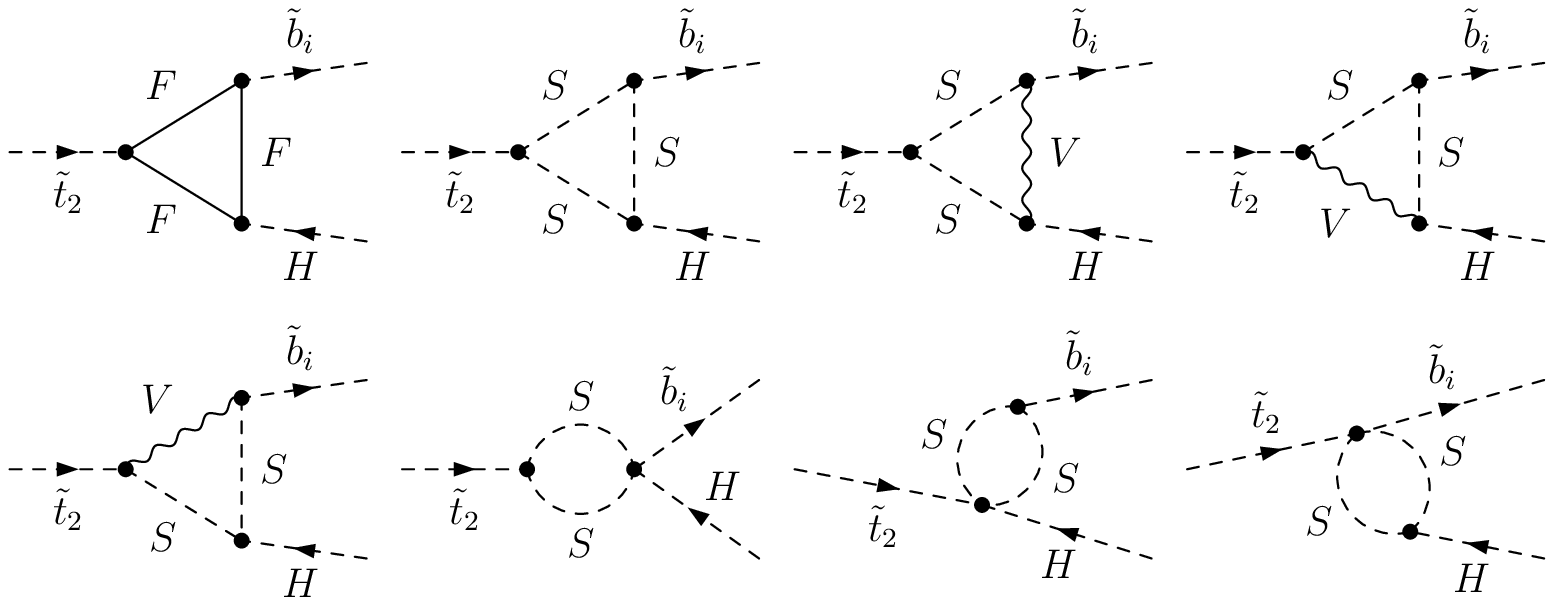}
\caption{
Generic Feynman diagrams for the decay 
$\Stopz \to \Sboti H^+$ ($i = 1,2$).
$F$ can be a SM fermion, a chargino or neutralino or a gluino, $S$
can be a 
sfermion or a Higgs boson, $V$ can be a $\ga$, $Z$, $W^\pm$ or $g$. 
Not shown are the diagrams with a $W^+$--$H^+$ or $G^+$--$H^+$ transition
contribution on the external Higgs boson leg. 
}
\label{fig:fdsbotHpm}
\vspace{2cm}
\includegraphics[width=0.90\textwidth]{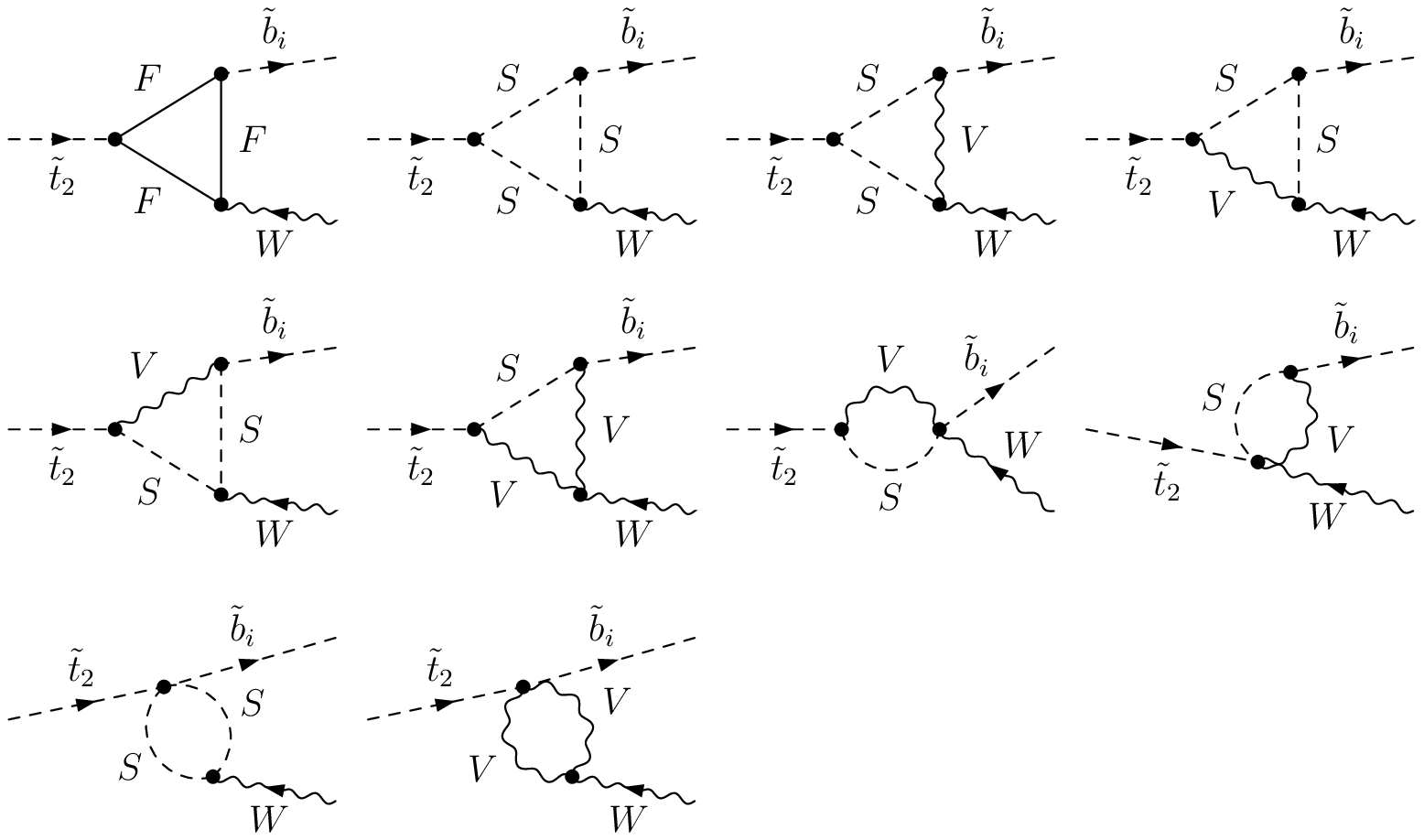}
\caption{
Generic Feynman diagrams for the decay 
$\Stopz \to \Sboti W^+$ ($i = 1,2$).
$F$ can be a SM fermion, a chargino or neutralino or a gluino, $S$ can be a
sfermion or a Higgs boson, $V$ can be a $\ga$, $Z$, $W^\pm$ or $g$. 
}
\label{fig:fdsbotW}
\end{center}
\end{figure}

In this section we give the relevant details about the calculation of the
higher-order corrections to the decay channels (\ref{stsbH},\ref{stsbW}). 
Sample diagrams are shown in \reffis{fig:fdsbotHpm}, \ref{fig:fdsbotW}. 
Not shown are the diagrams for real (hard or soft) photon and gluon
radiation (which, however, can become numerically very important). 
They are obtained from the corresponding tree-level diagrams
by attaching a photon (gluon) to the electrically (color) charged
particles. The internal, in a generical way depicted particles in
\reffis{fig:fdsbotHpm}, \ref{fig:fdsbotW} are labeled as follows:
$F$ can be a SM fermion, a chargino or neutralino or a gluino, $S$
can be a sfermion 
or a Higgs boson, $V$ can be a photon $\ga$, a $Z$ or $W^\pm$ boson or a
gluon $g$. 
Not shown are the diagrams with a gauge boson (Goldstone $G^\pm$)--Higgs
selfenergy 
contribution on the external Higgs boson leg that can appear in 
the decay $\Stopz \to \Sboti H^+$.
On the other hand, in our calculation, the wave function corrections for
$\Stopz \to \Sboti W^+$ vanish as all the external particle fields are
renormalized on-shell.

The diagrams and corresponding amplitudes have been obtained with the
program \fa~\cite{feynarts}. 
The further evaluation has been performed with 
\fc~\cite{formcalc}. As regularization scheme for the UV-divergences we
have used constrained differential renormalization~\cite{cdr}, 
which has been shown to be equivalent to 
dimensional reduction~\cite{dred} at the \onel\ level~\cite{formcalc}. 
Thus the employed regularization preserves SUSY~\cite{dredDS,dredDS2}. 
It was checked that all UV-divergences cancel in the final result.

The IR-divergences from diagrams with an internal photon or gluon have
to cancel with the ones from the corresponding real soft radiation.
In the case of QED we have included the soft photon contribution
following the description given in \citere{denner}. 
In the case of QCD we have modified this prescription by replacing the
product of electric charges by the appropriate combination of color
charges (linear combination of $C_A$ and $C_F$ times $\als$).
More details will be given in \citere{Stop2decay}.
Using the sbottom masses at the one-loop level, see \refse{sec:stop},
we found cancellation beyond one-loop order
of the related IR and UV divergences for
the decay  $\Stopz \to \Sboti H^+$, and a cancellation, as required, at the
one-loop level for the decay $\Stopz \to \Sboti W^+$.%
\footnote{Using tree-level masses yields a cancellation of IR divergences
beyond one-loop order also for $\Stopz \to \Sboti W^+$.}

For completness we show here also the formulas that have been
used to calculate the tree-level decay widths:
\begin{align}
\Gamma^{\rm tree}(\decaySbiH) &= \frac{|C(\Stopz, \Sboti, H^+)|^2\,
                                \lambda^{1/2}(\mstz^2,\msbi^2,\MHp^2)}
                                {16\, \pi\, \mstz^3}\qquad (i = 1,2)~, \\
\Gamma^{\rm tree}(\decaySbiW) &= \frac{|C(\Stopz, \Sboti, W)|^2\,
                                \lambda^{3/2}(\mstz^2,\msbi^2,\MW^2)}
                                {16\, \pi\, \MW^2\, \mstz^3}\qquad (i = 1,2)~,
\end{align}
where $\lambda(x,y,z) = (x - y - z)^2 - 4yz$ and the couplings 
$C(a, b, c)$ can be found in the \fa~model files~\cite{feynarts-mf}.
The bottom-Yukawa couplings generically are enhanced with $\tb$.

\newpage
\pagebreak
\clearpage


\subsection{Numerical examples for the six renormalization schemes}
\label{sec:numpar}

We start our analysis by showing some representative numerical
examples. We evaluate the tree-level results and the one-loop correction
for $\Ga(\decayH)$ including wave function corrections. 
The parameters are chosen according to the two scenarios, \SE\ and \SZ, 
shown in \refta{tab:para}.\footnote{It should be noted that we do not
  include any further 
shifts in the parameters than the one given in \refeq{MSbotshift}. 
Correspondingly, the values for the parameters $\Ab$ and $M_{\tilde{b}_R}$ in
\refta{tab:para} do not reflect the actual values for the input
parameters with respect to the 
chosen renormalization scheme. For example, the
$\Sbotz$~mass  --- though considered as an input in the
renormalization scheme and defined as on-shell mass --- receives a shift
going from tree- to one-loop level when starting out with the values in
\refta{tab:para} and including only the shift  \refeq{MSbotshift}. 
To circumvent this shift of the $\Sbotz$~mass, additional shifts to the
tree-level values of 
$\Ab$ and $M_{\tilde b_R}$ would be required (depending on the
renormalization scheme).
\vspace{1mm}}

\begin{table}[t!]
\renewcommand{\arraystretch}{1.5}
\BC
\begin{tabular}{|c||r|r|r|r|r|r|r|r|}
\hline
Scen.\ & $\MHp$ & $\mstz$ & $\mu$ & $\At$ & $\Ab$ & $M_1$ & $M_2$ & $M_3$ 
\\ \hline\hline
\SE & 150 & 600 & 200 &  900 &  400 & 200 & 300 & 800 
\\ \hline
\SZ & 180 & 900 & 300 & 1800 & 1600 & 150 & 200 & 400  
\\ \hline
\end{tabular}
\caption{MSSM parameters for the initial numerical investigation; all
parameters are in GeV. 
We always set $\mb^{\MSbar}(\mb) = 4.2 \gev$.
In our analysis we use 
$M_{\tilde Q_L}(\Stop) = M_{\Stop_R} = M_{\Sbot_R} =: \msusy$, where $\msusy$
  is chosen such that the above value of $\mstz$ is realized.
For the $\Sbot$~sector the shift in $M_{\tilde Q_L}(\Sbot)$ as defined in
\refeq{MSbotshift} is taken into account.
The parameters entering the scalar lepton sector and/or the first two
generations do not play a relevant role in our analysis.
The values for $\At$ and $\Ab$ are chosen such that charge- or
color-breaking minima are avoided~\cite{ccb}.
}
\label{tab:para}
\EC
\renewcommand{\arraystretch}{1.0}
\end{table}

So far we concentrate on the rMSSM: if a scheme shows deficiencies in
the rMSSM, the same problems occur in the cMSSM. The final numerical
examples in \refse{sec:numex} will also show complex parameters as well
as results for $\Ga(\Stopz \to \Sbot_{1,2} W^+)$.
It should be noted that $\tb \lsim 9.6\, (4.6)$
is excluded for \SE\ (\SZ) 
due to the MSSM Higgs boson searches at LEP~\cite{LEPHiggsSM,LEPHiggsMSSM}. 
However, we are interested in the general behavior of the renormalization 
schemes. If certain features appear in the two numerical scenarios 
(\SE\ and \SZ) only for experimentally excluded $\tb$ values, 
other parameter choices may exhibit these features also in unexcluded 
parts of the MSSM parameter space. 
Consequently, in order to investigate the various renormalization
schemes on 
general grounds, in the following we show the results for $\tb > 1$. 
A similar reasoning applies to the limits on the MSSM parameter space 
due to SUSY searches. Nevertheless, to avoid completely unrealistic spectra, 
the following exclusion limits \cite{pdg} hold in our two
scenarios:
\begin{align}
\mste &> 95 \gev, \;
\msbe > 89 \gev, \;
m_{\tilde{q}} > 379 \gev, \;
m_{\tilde{e}_1} > 73 \gev, \non \\
\mneu{1} &> 46 \gev, \;
\mcha{1} > 94 \gev, \;
\mgl > 308 \gev .
\end{align}

A few examples of the scalar top and bottom quark masses 
at the one-loop level%
\footnote{For the scalar top quark masses the
  tree-level and the one-loop values are the same (according to
    our renormalization conditions).}%
~(using 
$M_{\tilde{Q}_L}^2(\tilde{b})$ in \refeq{MSbotshift} for the
one-loop result) 
in the scenarios \SE\ and \SZ\ are 
shown in \refta{tab:squark}. The values of $\mstz$ allow copious
production of the heavier scalar top quark at the LHC. For other
choices of the 
gluino mass, $\mgl > \mstz$, which would leave no visible effect for
most of the decay modes of the $\Stopz$, the heavier
scalar top quark could also be
produced from gluino decays at the LHC. 
Furthermore, in \SE\ (even for the nominal value of $\mstz$ as given in
\refta{tab:para}) the production of $\Stopz$ at the ILC(1000), i.e.\ with 
$\sqrt{s} = 1000 \gev$, via $e^+e^- \to \Stopz\Stope$ will be possible,
with the subsequent decay modes (\ref{stsbH}) and (\ref{stsbW})
being open. The clean environment of the ILC would permit a detailed
study of the scalar top quark decays.
Depending on the combination of allowed decay
channels a determination of the branching ratios at the few per-cent
level might be achievable in the high-luminosity running of the ILC(1000).
More details will be discussed elsewhere~\cite{Stop2decay}.

\begin{table}[t!]
\renewcommand{\arraystretch}{1.5}
\BC
\begin{tabular}{|c|c||r|r|r|r|}
\hline
Scen. & $\tb$ & $\mste$~~ & $\mstz$~~ & $\msbe$~~ & $\msbz$~~  
\\ \hline\hline
    &  2 & 293.391 & 600.000 & 441.987 & 447.168
\\ \cline{2-6}
\SE & 20 & 235.073 & 600.000 & 418.824 & 439.226
\\ \cline{2-6}
    & 50 & 230.662 & 600.000 & 400.815 & 449.638
\\ \hline\hline
    &  2 & 495.014 & 900.000 & 702.522 & 707.598
\\ \cline{2-6}
\SZ & 20 & 445.885 & 900.000 & 678.531 & 695.180
\\ \cline{2-6}
    & 50 & 442.416 & 900.000 & 628.615 & 697.202
\\ \hline
\end{tabular}
\caption{The top and bottom squark masses 
  at the one-loop level (see text) in the
  scenarios S1 and S2 and 
  at different $\tb$ for the numerical investigation; 
  all masses are in GeV and rounded to one MeV.
}
\label{tab:squark}
\EC
\renewcommand{\arraystretch}{1.0}
\end{table}

Later we will also analyze numerical results for complex input parameters.
Here it should be noted that the results for physical observables are
affected only 
by certain combinations of the complex phases of the 
parameters $\mu$, the trilinear couplings $A_f$, 
$f = \{u,c,t,d,s,b,e,\mu,\tau\}$, the 
gaugino mass parameters $M_1$, $M_2$,
$M_3$ and the Higgs soft SUSY breaking parameter
$m_{12}^2$~\cite{MSSMcomplphasen,SUSYphases}. 
It is possible, for instance, to eliminate the phase $\varphi_{M_2}$ and
the phase $\varphi_{m_{12}^2}$.
Experimental constraints on the (combinations of) complex phases 
arise in particular from their contributions to electric dipole moments of
heavy quarks~\cite{EDMDoink}, of the electron and 
the neutron (see \citeres{EDMrev2,EDMPilaftsis} and references therein), 
and of the deuteron~\cite{EDMRitz}. While SM contributions enter 
only at the three-loop level, due to its
complex phases the MSSM can contribute already at one-loop
order.
Large phases in the first two generations of sfermions can only be 
accommodated if these generations are assumed to be very
heavy~\cite{EDMheavy} or large cancellations occur~\cite{EDMmiracle},
see however the discussion in \citere{EDMrev1,plehnix}.
A recent review can be found in \citere{EDMrev3}.
Accordingly, using the convention that $\varphi_{M_2} =0$ and
$\varphi_{m_{12}^2} =0$, as done in this paper, in particular 
the phase $\varphi_\mu$ is tightly constrained~\cite{plehnix}, 
while the bounds on the phases of the third generation
trilinear couplings are much weaker.
The phase of $\mu$ enters in the combinations 
$(\varphi_{A_{t,b}} + \varphi_{\mu} - \varphi_{m_{12}^2})$. Setting
$\varphi_\mu = 0$ (and $\varphi_{M_2} =\varphi_{m_{12}^2} =0$, see above) 
leaves us with $\At$ and $\Ab$ as complex valued
parameters. Since we are interested in the renormalization of the
$b/\Sbot$~sector, in our numerical analysis we will focus on a
complex~$\Ab$ and keep $\At$ real (see, however, \citere{Stop2decay}).

\begin{table}[t!] 
\renewcommand{\arraystretch}{1.5}
\BC
\begin{tabular}{|cc||c||r|r|r||r|r|r|}
\hline
\multicolumn{3}{|c||}{$\Ga(\decayH)$ for \SE} & 
\multicolumn{3}{|c||}{$\tb = 2$} &
\multicolumn{3}{|c|}{$\tb = 50$} \\
\hline
& renorm.\ scheme & $\mu_R$ & tree & loop & 
$\mb$ & tree & loop & $\mb$ 
\\
\hline \hline
RS1: &``OS'' & $\mstz$ & 
0.0017 & -0.0011 & 3.29 & 2.5930 & -53.3469 & 3.84 \\ \hline
RS2: & ``$\mb,\,\Ab$~\DRbar'' & $\mstz$ & 
0.0009 &  0.0002 & 2.38 & 0.9653 &  -0.0311 & 2.16 \\ \hline
RS3: & ``$\mb,\, Y_b$~\DRbar'' & $\mstz$ & 
0.0009 &  0.0004 & 2.38 & 0.9484 &  -1.5404 & 2.16 \\ \hline
RS4: & ``$\mb$~\DRbar, $Y_b$~OS'' & $\mstz$ & 
0.0009 &  0.0000 & 2.38 & 0.9593 &  -0.3411 & 2.16 \\ \hline
RS5: & ``$\Ab$~\DRbar, $\re Y_b$~OS'' & $\mstz$ & 
------ & ------ & ------ & 0.9399 & -0.0481 & 2.13 \\ \hline
RS6: & ``$\Ab$~vertex, $\re Y_b$~OS'' & $\mstz$ & 
0.0007 &  0.0001 & 2.19 & 0.9390 &  -0.0347 & 2.13 \\ \hline
\end{tabular}\\[3ex]
\begin{tabular}{|cc||c||r|r|r||r|r|r|}
\hline
\multicolumn{3}{|c||}{$\Ga(\decayH)$ for \SZ} & 
\multicolumn{3}{|c||}{$\tb = 2$} &
\multicolumn{3}{|c|}{$\tb = 50$} \\
\hline
& renorm.\ scheme & $\mu_R$ & tree & loop & 
$\mb$ & tree & loop & $\mb$ 
\\
\hline \hline
RS1: & ``OS'' & $\mstz$ & 
2.0928 & -0.0776 & 3.23 & 8.5163 & -106.9700 & 3.70 \\ \hline
RS2: & ``$\mb,\,\Ab$~\DRbar'' & $\mstz$ & 
2.2171 & -0.1449 & 2.33 & 1.8173 &   -0.5125 & 2.11 \\ \hline
RS3: & ``$\mb,\, Y_b$~\DRbar'' & $\mstz$ & 
0.0077 &  0.0582 & 2.33 & 3.1409 &  -11.6833 & 2.11 \\ \hline
RS4: & ``$\mb$~\DRbar, $Y_b$~OS'' & $\mstz$ & 
2.2564 & -0.1031 & 2.33 & 2.9230 &   -4.5506 & 2.11 \\ \hline
RS5: & ``$\Ab$~\DRbar, $\re Y_b$~OS'' & $\mstz$ & 
2.2332 & -0.1004 & 2.45 & 2.3018 &  0.2924 & 1.84 \\ \hline
RS6: & ``$\Ab$~vertex, $\re Y_b$~OS'' & $\mstz$ & 
2.2925 & -0.1067 & 2.14 & 2.3558 & -0.0710 & 1.86 \\ \hline
\end{tabular}
\caption{Examples for tree-level and full one-loop contributions
  (see text) to $\Ga(\decayH)$ for \SE\ (upper table) and \SZ\ (lower table); 
  all values are in GeV (no comparison of the renormalization
    schemes, see text). 
  In S1 using RS5 a divergence is reached for $\tb = |\Ab|/|\mu| = 2$ and no
  value can be computed (see text below). The different
    renormalization schemes are listed in \refta{tab:RS}.
}
\label{tab:numex}
\EC
\renewcommand{\arraystretch}{1.0}
\end{table}

\begin{figure}[t!]
\begin{center}
\begin{tabular}{c}
\includegraphics[width=0.49\textwidth,height=7.5cm]{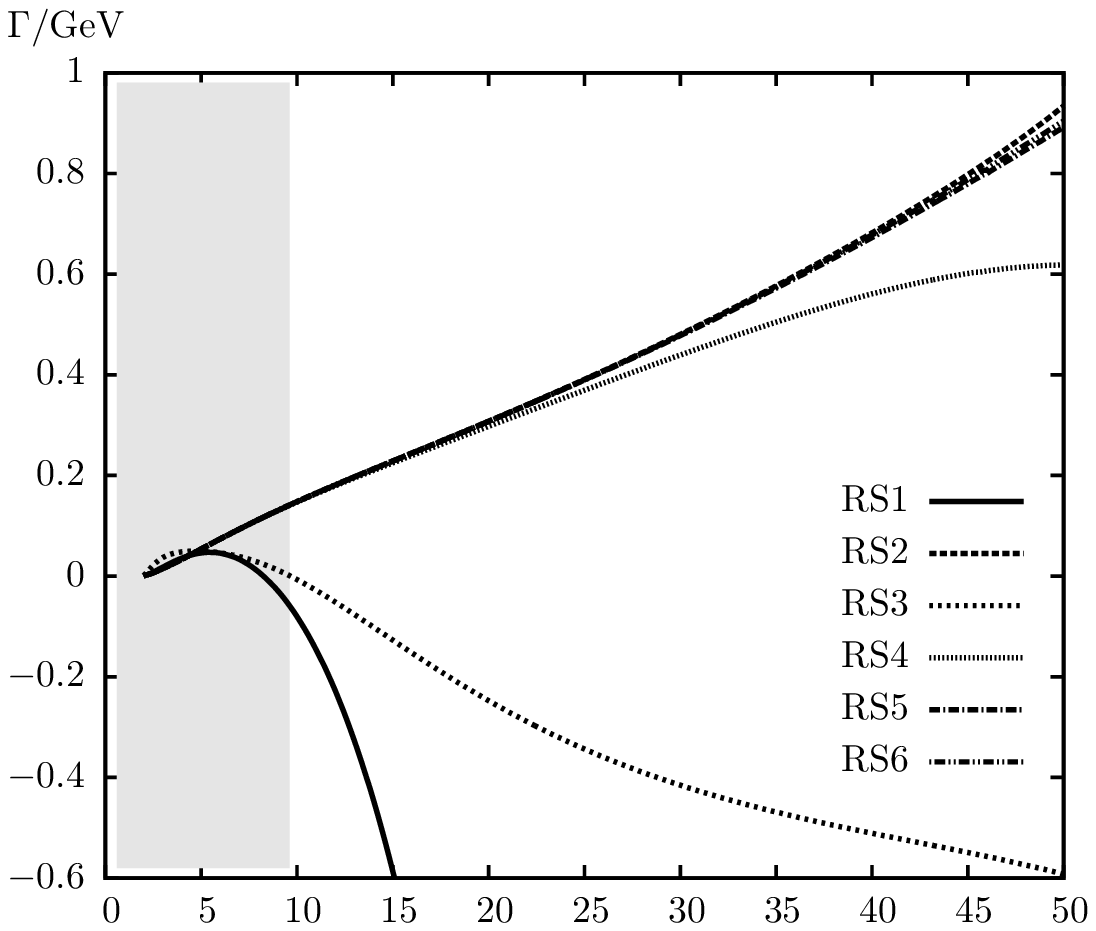}
\hspace{-4mm}
\includegraphics[width=0.49\textwidth,height=7.5cm]{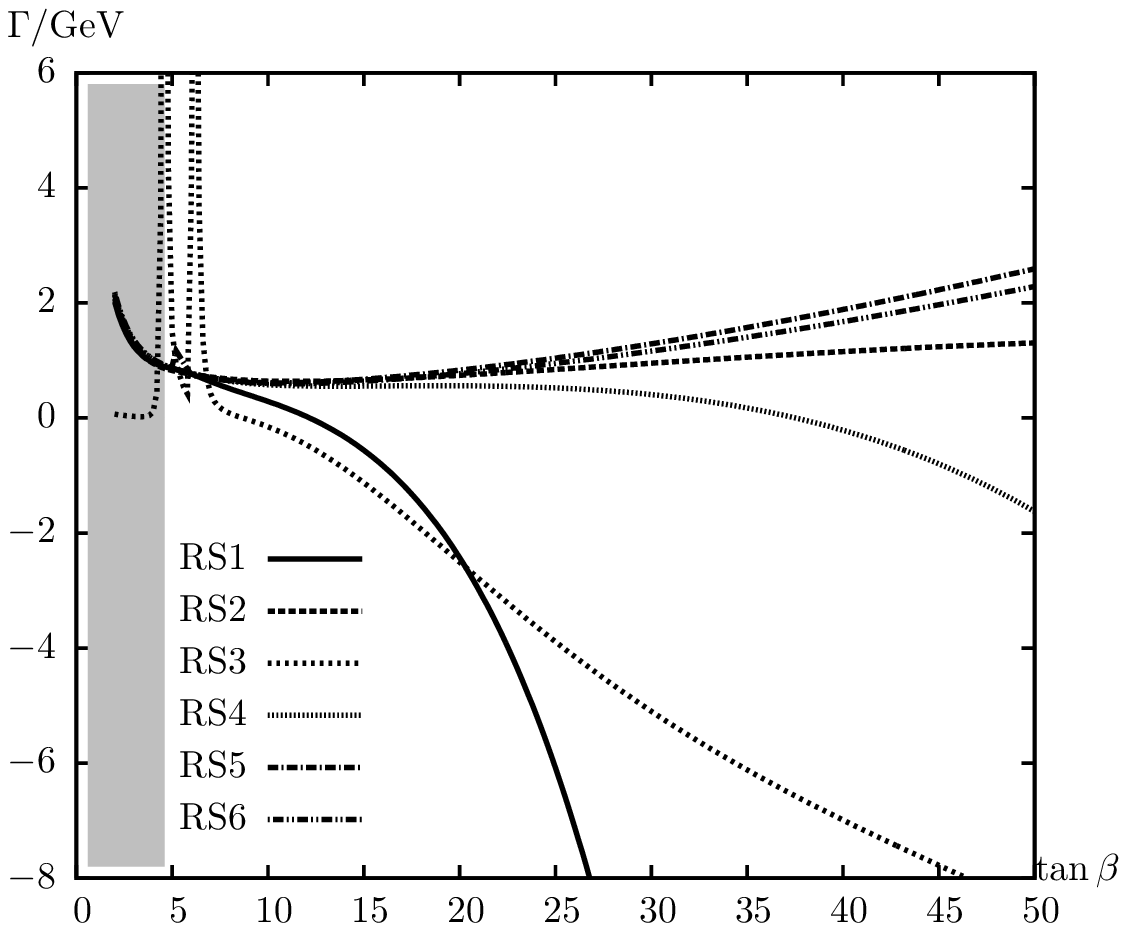} 
\end{tabular}
\caption{$\Ga(\decayH)$. 
  Full one-loop corrected partial decay widths for the different
  renormalization schemes (no comparison, see text). 
  The parameters are chosen according to \SE\ 
  in the
  left plot and \SZ\ in the right plot. For \SE\ the grey region 
  and for \SZ\ the dark grey region is excluded by LEP Higgs searches
  (see text).}
\label{fig:st2sb1H.RS}
\end{center}
\end{figure}

\bigskip
We start our numerical examples with the evaluation of $\Ga(\decayH)$ in
\SE\ and \SZ\ for $\tb = 2$ and $\tb = 50$ as shown in
\refta{tab:numex}. The corresponding results as a continuous function of
$\tb$ can be seen in \reffi{fig:st2sb1H.RS}.
It must be emphasized here that the table and the plots do not
constitute a {\em comparison} of the various schemes, but ``only''
individual numerical examples that are used to exhibit certain problems
of the various schemes. A numerical comparison of the schemes requires
that the input parameters are converted from one scheme into another,
see, for instance, \citere{mhiggsFDalbals}, which is not performed
within this analysis. In our numerical examples 
the renormalization scale, $\mu_R$, has been set to the mass of the
decaying particle, i.e.\ $\mu_R = \mstz$. 
In \refta{tab:numex} the two main columns, labeled ``$\tb = 2$'' and 
``$\tb = 50$'', are divided into three columns where ``tree'' contains the 
tree-level results and ``loop'' the one-loop 
contribution. $\mb$ denotes the corrected bottom quark value
corresponding to the respective renormalization, see \refeq{eq:mbcorr}.

The two values of $\tb$ were chosen as an example of a very low 
and a very high value. It should be kept in mind that the low value is
possibly already in conflict with MSSM Higgs boson 
searches~\cite{LEPHiggsSM,LEPHiggsMSSM}, but kept to show an
``extreme'' example as explained above.
It can be seen that RS1, RS3, RS4 and RS5 yield 
relatively large absolute values of loop contributions  with respect to
the tree-level 
result, either for $\tb = 2$ {\em or} for $\tb = 50$, at least in one of
the two numerical scenarios.
This simple example shows that (by choosing a specific scenario) 
already all except two renormalization schemes fail in part
of the parameter space.

More problems of the renormalization schemes RS1, RS3, RS4 and RS5
become visible in \reffi{fig:st2sb1H.RS}.
In the left (right) plot of  \reffi{fig:st2sb1H.RS} we show the results of
\SE (\SZ) as a function of $\tb$. For \SE\ the grey region 
and for \SZ\ the dark grey region at low values of $\tb$ are excluded
by LEP Higgs searches~\cite{LEPHiggsMSSM}. 
It can be seen in \reffi{fig:st2sb1H.RS} that RS1 and RS3 deviate
strongly from the (see the end of \refse{sec:calc})
expected behavior of increasing
$\Ga(\decayH)$ with growing $\tb$ that the other schemes exhibit. The
same is observed for RS4 in \SZ\ for $\tb \gsim 35$.
Problems in RS2 are discussed in \refse{sec:mbAb}, problems in RS6
have been found for complex parameters, see \refse{sec:problem_non_mb}.
The various spikes and dips can be understood as follows:
\begin{itemize}

\item
For RS3 in \SZ\ a ``peak'' appears at $\tb \approx 4.6$ and at 
$\tb \approx 6.2$. This is discussed in \refse{sec:problem_non_Ab} below.

\item
For RS5 in \SE\ a ``peak'' appears (not visible) at 
$\tb = |\Ab|/|\mu| = 2$.
This is caused by large corrections to the bottom quark mass as discussed 
further in \refse{sec:problem_non_mb}. This is also the reason why 
there is no entry in \refta{tab:numex} for RS5, S1 at $\tb = 2$.

\item
For RS5 in \SZ\ a ``peak'' appears at $\tb = |\Ab|/|\mu| = 5.33$.
This is caused by large corrections to the bottom quark mass as 
discussed further in \refse{sec:problem_non_mb}.

\end{itemize}


\subsection{Generic considerations for the \boldmath{$b/\Sbot$} sector
  renormalization (I)}
\label{sec:genericI}

As discussed in \refse{sec:stop}, a bottom quark/squark sector
renormalization scheme always contains dependent counterterms which can
be expressed by the independent ones.
According to our six definitions, this can be
$\de\mb$, $\de\Ab$ or $\de Y_b$. 
A problem can occur when the MSSM parameters are chosen such that the
independent counterterms (nearly) drop out of the relation determining
the dependent counterterms.
As will be shown below, even
restricting to the two numerical examples, \SE\ and \SZ, it is 
possible to find a set of MSSM parameters which show this behaviour for
each of the chosen
renormalization schemes.
Consequently, it appears to be difficult {\em by construction} to
define a 
renormalization scheme for the bottom quark/squark sector (once
the top quark/squark
sector has been defined) that behaves well for the full MSSM parameter
space. One possible exception could be a pure \DRbar\ scheme, which,
however, is not well suited for processes with external top
squarks and/or bottom squarks. 

Assuming that SUSY, and more specifically the MSSM, will be discovered
at the LHC and its parameters will be measured, the problem will have
disappeared. For a specific set of MSSM parameters, renormalization
schemes can (easily) be found that behave well. 
However, due to our ignorance about the actual values of the SUSY
parameters, scans over large parts of the MSSM parameter space are
performed, see also \refse{sec:numex}. For this kind of analysis a
careful choice of the renormalization scheme has to be made.

In the following subsections we will analyze in more detail, analytically
and numerically, the deficiencies of the various schemes.


\subsection{Problems of the ``OS'' renormalization}
\label{sec:problem_OS}

The ``OS'' renormalization as described in \refse{sec:OS} does not yield
reasonable results in perturbative 
calculations as shown 
already, e.g., in \citere{mhiggsFDalbals,sbotrenold}. 
For the sake of completeness we briefly repeat the results. 
The ``OS'' scheme of \refse{sec:OS} is the renormalization scheme
analogous to the one used 
in the $t/\Stop$ sector and thus would be the ``naive'' choice. It
includes an on-shell renormalization 
condition on the sbottom mixing parameter $Y_b$ that contains the
combination $(\Ab - \mu^* \tb)$.  
In parameter regions where $(\mu\tb)$ is much larger than $\Ab$, the
counterterm $\de\Ab$ receives a very large finite shift when calculated
from the counterterm $\de Y_b$. More specifically, $\de\Ab$ as given in
\refeq{Ab_OS} contains the contribution
\begin{align}
\label{dAb_OSproblem}
\de\Ab = \frac{1}{\mb} \KKL -(\Ab - \mu^*\tb)\, \de\mb + \ldots \KKR
\end{align}
that can give rise to very large corrections to $\Ab$.
This is also visible in \reffi{fig:dAb} below, where we show the 
numerical values of $\de\Ab$ as a function of $\tb$ for various 
renormalization schemes. 
In \citere{mhiggsFDalbals} it was shown that, because of
\refeq{dAb_OSproblem}, the ``OS'' renormalization yields
huge corrections to the lightest MSSM Higgs mass. 
Also the numerical results shown in \refta{tab:numex} and 
\reffi{fig:st2sb1H.RS} show extremely large one-loop corrections 
for $\tb = 50$. 

This problem is (more or less) avoided in the other renormalization
schemes introduced in \refta{tab:RS}, where the renormalization
condition is applied directly to $\Ab$, rather than deriving $\de\Ab$
from a renormalization condition fixing $\de Y_b$.
Also the renormalization schemes 
RS3 (``$\mb,\, Y_b$~\DRbar'') and RS4 (``$\mb$~\DRbar, $Y_b$~OS'')
avoid this severe problem by renormalizing the bottom quark mass \DRbar.

\begin{figure}[t!]
\begin{center}
\begin{tabular}{c}
\includegraphics[width=0.49\textwidth,height=7.5cm]{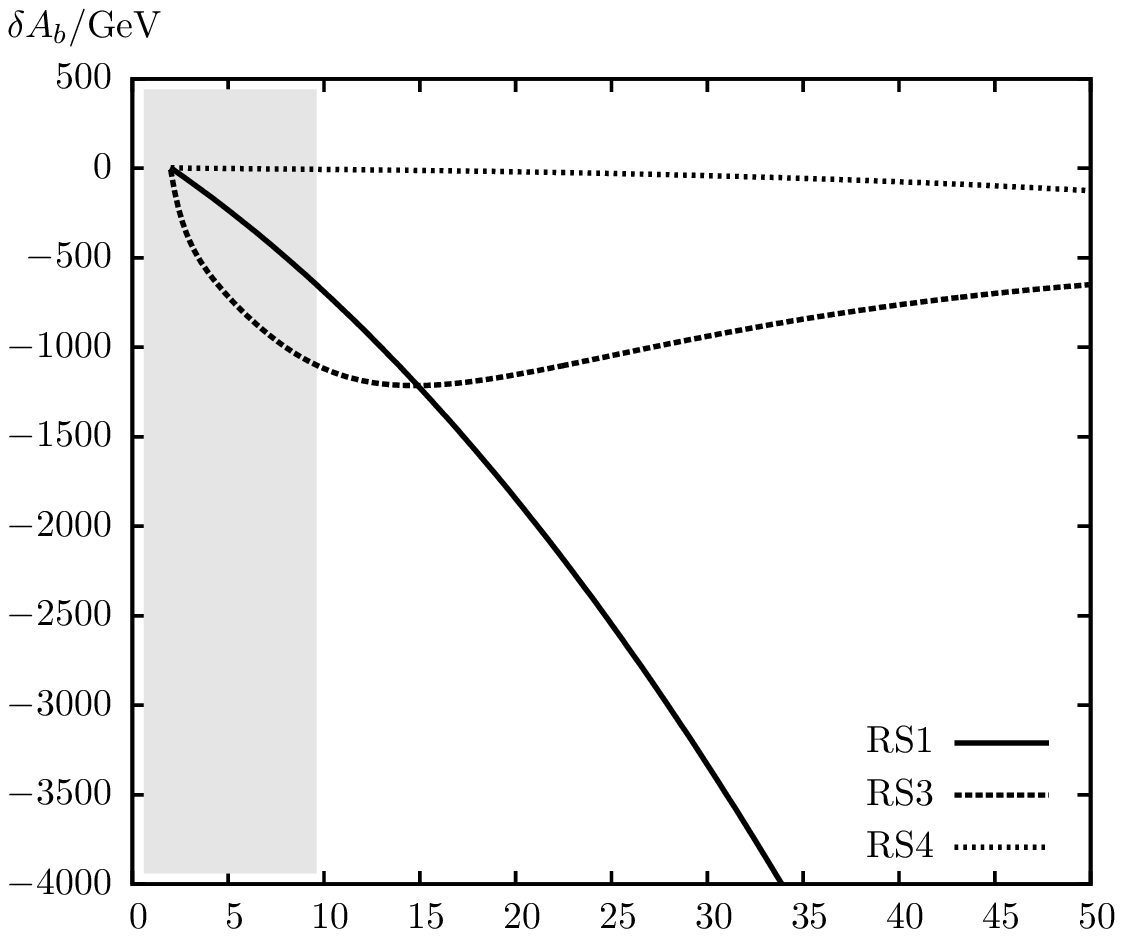}
\hspace{-4mm}
\includegraphics[width=0.49\textwidth,height=7.5cm]{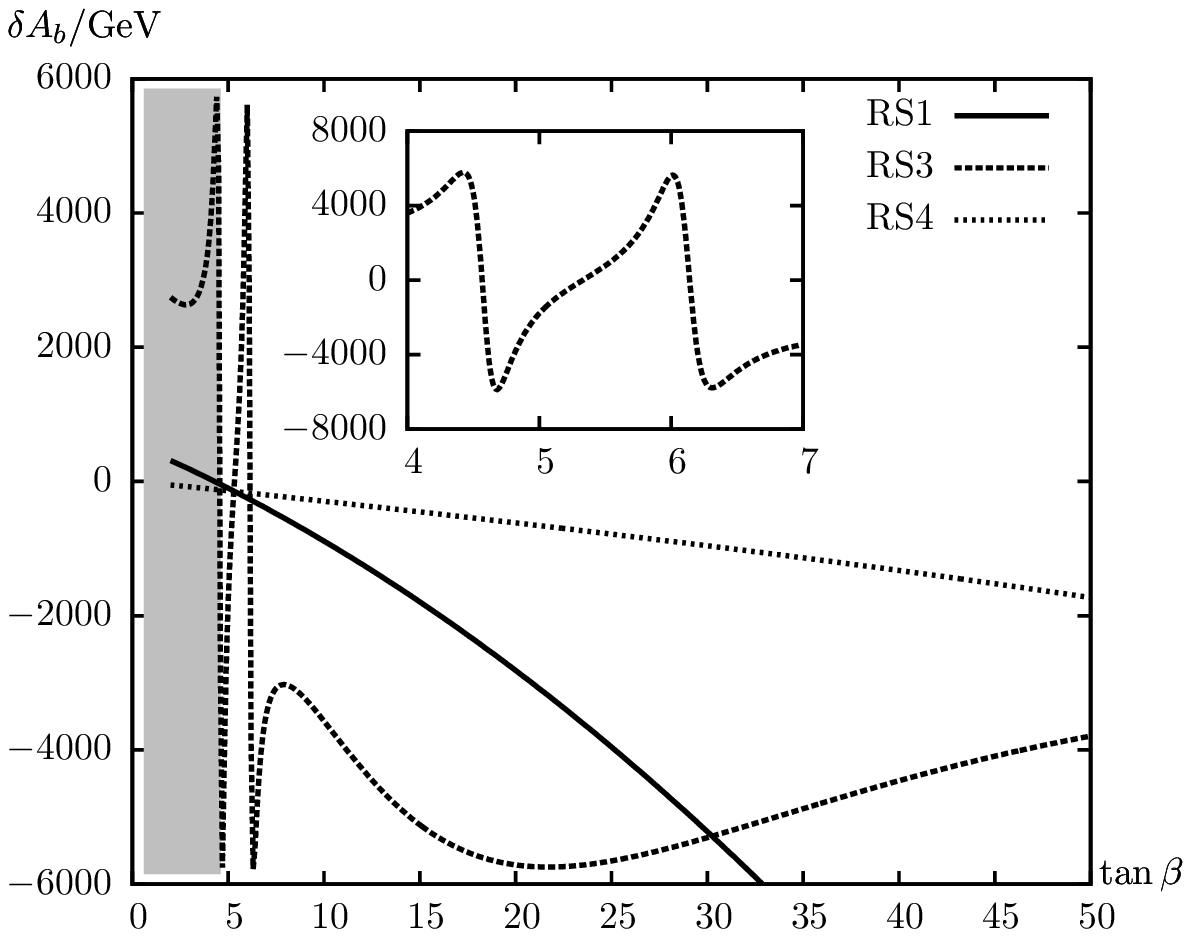} 
\end{tabular}
\caption{
  Finite parts of $\de\Ab$ in various renormalization schemes. The
  parameters are chosen according to \SE\ 
  left plot and \SZ\ right plot. 
  For \SE\ the grey region is excluded 
  and for \SZ\ the dark grey region is excluded.}
\label{fig:dAb}
\end{center}
\end{figure}


\subsection{Problems of non-\boldmath{$\Ab$} renormalization schemes}
\label{sec:problem_non_Ab}

Two of our schemes, besides the ``OS'' scheme (RS1), do not employ
a renormalization of $\Ab$: 
RS3 (``$\mb,\, Y_b$~\DRbar'') and RS4 (``$\mb$~\DRbar, $Y_b$~OS''). 
As argued in \refse{sec:problem_OS} a huge contribution to $\de\Ab$ as
evaluated in that section is avoided by the \DRbar\ renormalization of
$\mb$. However, following \refeq{Ab_OS} with $\de\msbe^2$, $\de\msbz^2$,
$\de Y_b$ and $\de \mb$  chosen according to the renormalization
schemes RS3 and RS4, respectively, one finds for the finite parts of $\de\Ab$:
\begin{align}
\mbox{RS3} &:~ \de\Ab|_{\text{fin}} = \ed{\mb} \KKL U_{\Sbot_{11}} U_{\Sbot_{12}}^*
          \KL \de\msbe^2 - \de\msbz^2 \KR \KKR_{\text{fin}} 
          + \ldots~, \\
\label{dAb_problem}
\mbox{RS4} &:~ \de\Ab|_{\text{fin}} = \ed{\mb} \KKL U_{\Sbot_{11}} U_{\Sbot_{12}}^*
          \KL \de\msbe^2 - \de\msbz^2 \KR
          + U_{\Sbot_{11}} U_{\Sbot_{22}}^* \de Y_b^*
          + U_{\Sbot_{12}}^* U_{\Sbot_{21}} \de Y_b \KKR_{\text{fin}} + \ldots~,
\end{align}
where the ellipses denote contributions from $\de\mu$ which, however, are
not relevant for our argument.
It can be seen that still $\de\Ab$ depends on parameters (diagonal and
off-diagonal sbottom self-energies) that are independent of $\Ab$. As
an example, Higgs boson loops in the sbottom self-energy contain
contributions $\sim \mu \tb$, which can become very large, independently
of the value of $\Ab$. This can be seen in the right plot of
\reffi{fig:dAb}, where 
we show $\de\Ab$ as a function of $\tb$ in \SZ. In both renormalization
schemes, RS3 and RS4, $\de\Ab$ becomes very large and negative for
large $\tb$. This yields the very large and negative loop corrections to
$\Ga(\decayH)$ shown in the right plot of \reffi{fig:st2sb1H.RS}. 
In \SE\ this problem is less pronounced, as can be seen in the left 
plot of \reffi{fig:dAb} ($\de\Ab$) and \reffi{fig:st2sb1H.RS} 
($\Ga(\decayH)$). 

But also for lower $\tb$ values, $\tb \lsim 10$, problems can occur. 
The (finite) ``multiple spike structure'' in RS3 for \SZ\ around 
$\tb \approx 5.33$ (for details see the small insert within the 
right plot of \reffi{fig:dAb}) 
is due to an interplay of top/chargino contributions to the two 
diagonal sbottom self-energies, invalidating this scenario also for 
this part of the parameter space.


\subsection{Problems of an \boldmath{$\mb$}--\boldmath{$\Ab$} 
            renormalization}
\label{sec:mbAb}

If $\mb$ and $\Ab$ are renormalized, the sbottom mixing parameter $Y_b$
is necessarily a dependent 
parameter, see \refta{tab:RS}. This situation is realized in the scheme
RS2 (``$\mb,\,\Ab$~\DRbar''), see \refse{sec:mbDRbar_AbDRbar}.
$\de Y_b$ enters prominently into $\dZ{\Sbot_{21}}$. For real parameters
we have,
\begin{align}
\dZ{\Sbot_{21}} &=- 2\, \frac{\re\Si_{\Sbot_{21}}(\msbz^2) - \de Y_b}
                             {\msbe^2 - \msbz^2}~.
\label{dZSbot21}
\end{align}
In this way $\de Y_b$ (or the interplay between $\de Y_b$ and
$\re\Si_{\Sbot_{21}}(\msbz^2)$) can induce large loop corrections to the
scalar top quark decay width.
$\de Y_b$ can be decomposed according to \refeq{dYb_mbDRbar_AbDRbar}
(concentrating again on the case of real parameters),
\begin{align}
\de Y_b &= \frac{U_{\Sbot_{11}} U_{\Sbot_{21}}}
                {|U_{\Sbot_{11}}|^2 - |U_{\Sbot_{12}}|^2}
           \KL \de\msbe^2 - \de\msbz^2 \KR
           + \ldots~,
\label{dYb_problem1}
\end{align}
where the ellipses denote terms with only divergent
contributions (due to the chosen renormalization scheme RS2) as well as
finite contributions from $\de\mu$, 
which, however, do not play a role for our argument.
For ``maximal sbottom mixing'', 
$|U_{\Sbot_{11}}| \approx |U_{\Sbot_{12}}|$,
$\de Y_b$ diverges, and the loop calculation does not yield a reliable
result. In our two parameter scenarios, \SE\ and \SZ, this is not the
case. Such a 
large sbottom mixing is often associated with large values of $|\Ab|$
that may be in conflict with charge- or color-breaking minima~\cite{ccb}.

\begin{figure}[ht!]
\begin{center}
\begin{tabular}{c}
\includegraphics[width=0.49\textwidth,height=7.5cm]{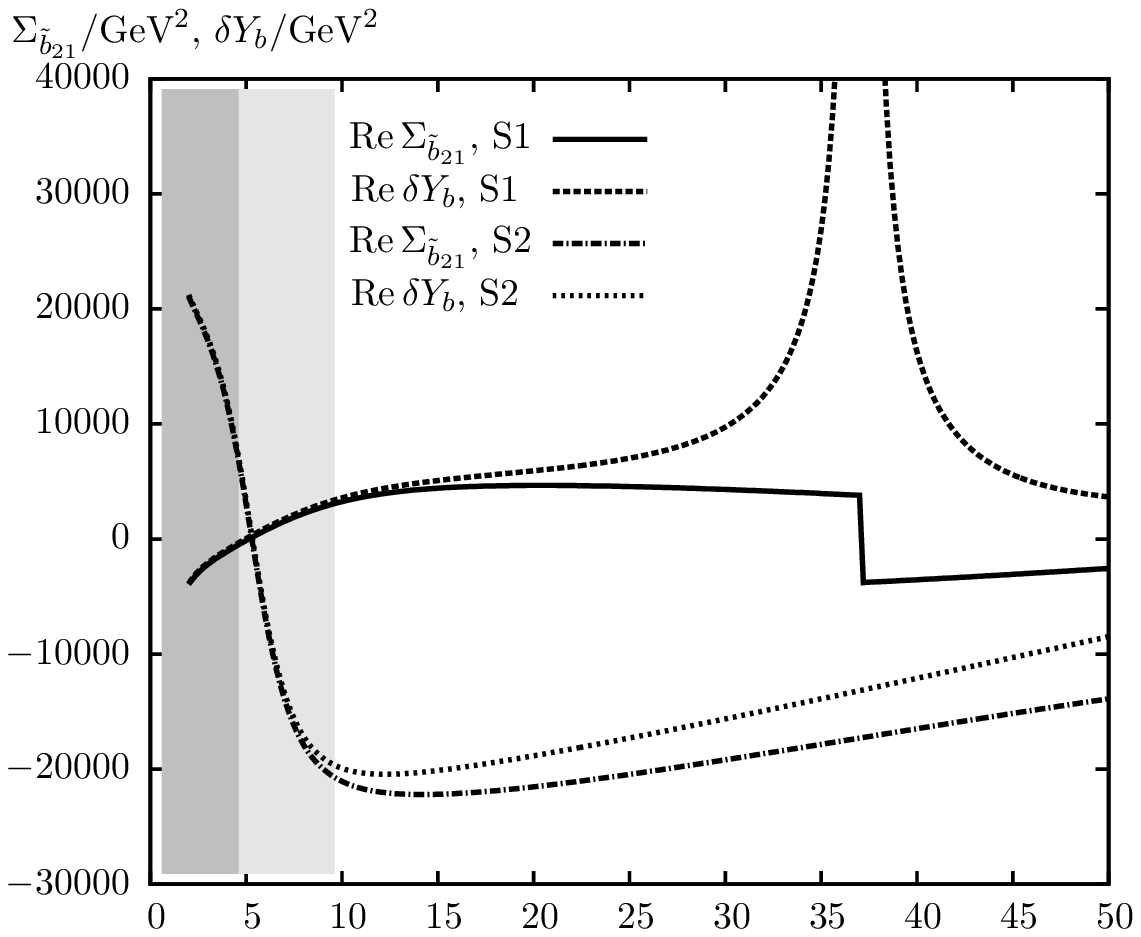}
\hspace{-4mm}
\includegraphics[width=0.49\textwidth,height=7.5cm]{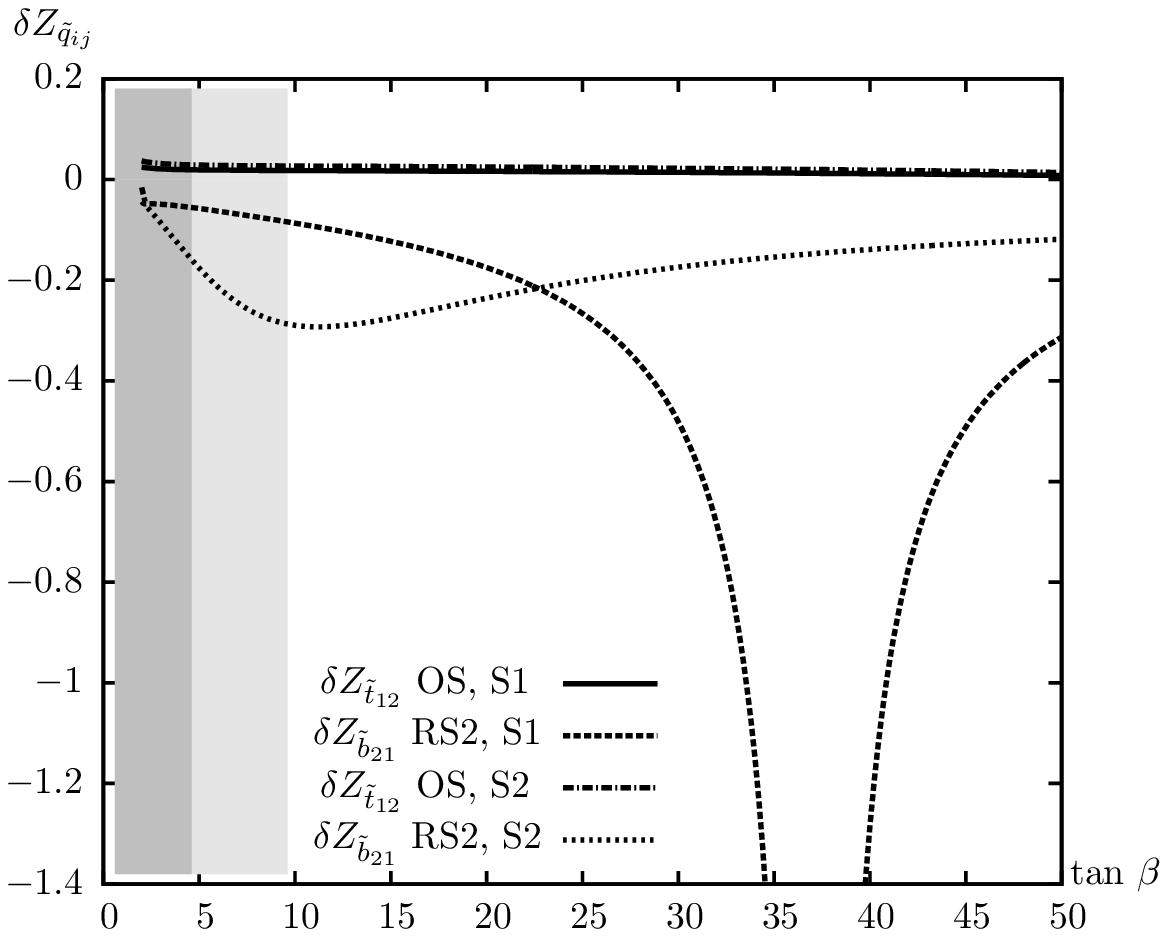} 
\end{tabular}
\caption{
  $\Ga(\decayH)$.
  Left plot: size of $\de Y_b$ and $\re\Si_{\Sbot_{21}}(\msbe^2)$, the
  two contributions to $\dZ{\Sbot_{21}}$, in RS2.
  Right plot: comparison of the size of $\dZ{\Sbot_{21}}$ in the scheme RS2
  (``$\mb,\,\Ab$~\DRbar''). For both plots the parameters are chosen
  according to \SE (but here with $\Ab = 1000$ GeV), \SZ\ in \refta{tab:para}.
  For \SE\ the grey region is excluded and for \SZ\ the
  dark grey region is excluded via LEP Higgs searches (see text).}
\label{fig:mbAb}
\end{center}
\end{figure}

However, in order to show an example with a divergence in $\de
Y_b$ we use a modified version of \SE\ with $\Ab = 1000 \gev$ 
(a value still allowed following \citere{ccb}).
In this scenario at $\tb \approx 37$ we indeed find the case of
``maximal mixing'' in the scalar bottom sector.
As expected this leads to a divergence in $\de Y_b$,
as can be seen in the left plot of
\reffi{fig:mbAb}. This divergence propagates into $\dZ{\Sbot_{21}}$ as
shown in the right plot of \reffi{fig:mbAb}.%
\footnote{
The scalar bottom masses could receive large corrections via 
$M_{\tilde{Q}_L}^2(\tilde{b})$ in \refeq{MSbotshift}, with $\de Y_b$ 
entering via \refeq{MSbotshift-detail}.
}%
(Also $\Si_{\Sbot_{21}}$ exhibits a discontinuity due to a sign change
in $U_{\Sbot}$ for this extreme set of MSSM parameters.)
The $\tb$ value for which this ``divergence'' occurs depends on the
choice of the other MSSM parameters. For (numerical) comparison we
also show $\dZ{\Stop_{21}}$ for the two scenarios.

For the different choice of MSSM parameters in \SZ\ (without a
higher $\Ab$ value) this divergences
does not occur. However, for $\tb \lsim 7$ one finds 
$\de Y_b \gsim \re\Si_{\Sbot_{21}}(\msbe^2)$ 
(with $\de Y_b = \re\Si_{\Sbot_{21}}(\msbe^2)$ for $\tb \approx 7.5$).
In this part of the parameter space we also find $\msbe \approx \msbz$,
yielding a relatively large value of $\dZ{\Sbot_{21}}$ according to
\refeq{dZSbot21}, as can be seen in the right plot of
\reffi{fig:mbAb}. This relatively large (negative) value of
$\dZ{\Sbot_{21}}$ in 
turn induces relatively large corrections to $\Ga(\decayH)$. 
However, the loop corrections do not exceed the tree-level value
of $\Ga(\decayH)$ (for our choice of MSSM parameters).
In summary: while for \SE\ a divergence in $\de Y_b$ and thus in
$\dZ{\Sbot_{12}}$ can appear for very large values of $|\Ab|$
(possibly in conflict with charge- or color-breaking minima),
invalidating the renormalization scheme~RS2 in this part of the 
parameter space, these kind of problems are not encountered in S2. Here
only moderate loop corrections to the respective tree-level values are
found, and RS2 can be applied safely.


\subsection{Problems of non-\boldmath{$\mb$} renormalization schemes}
\label{sec:problem_non_mb}

Two of our schemes do not employ a renormalization condition for $\mb$:
RS5 (``$\Ab$~\DRbar, $\re Y_b$~OS'') and 
RS6 (``$\Ab$~vertex, $\re Y_b$~OS''). 
Since $\Ab$ and $Y_b$ are complex, we chose to renormalize $\Ab$ and the
real part of $Y_b$. 

We start with the discussion of the (simpler) ``$\Ab$~\DRbar, $\re Y_b$~OS''
scheme. We will focus on the real case as a subclass of
the more general complex case. In this renormalization scheme the bottom
quark mass counterterm has the following form for real parameters
(compare to \refeq{dmb_AbDRbar_ReYbOS}),
\begin{align}
\label{dmb_problem}
\de\mb &= - \frac{\mb\, \de\Ab + \de S}{(\Ab - \mu\tb)}~.
\end{align}
For vanishing sbottom mixing one finds $(\Ab - \mu\tb) \to 0$.
In the ``$\Ab$~\DRbar, $\re Y_b$~OS'' scheme this yields a finite (and
negative) numerator in \refeq{dmb_problem}, but a vanishing denominator.

\begin{figure}[t!]
\begin{center}
\begin{tabular}{c}
\includegraphics[width=0.49\textwidth,height=7.5cm]{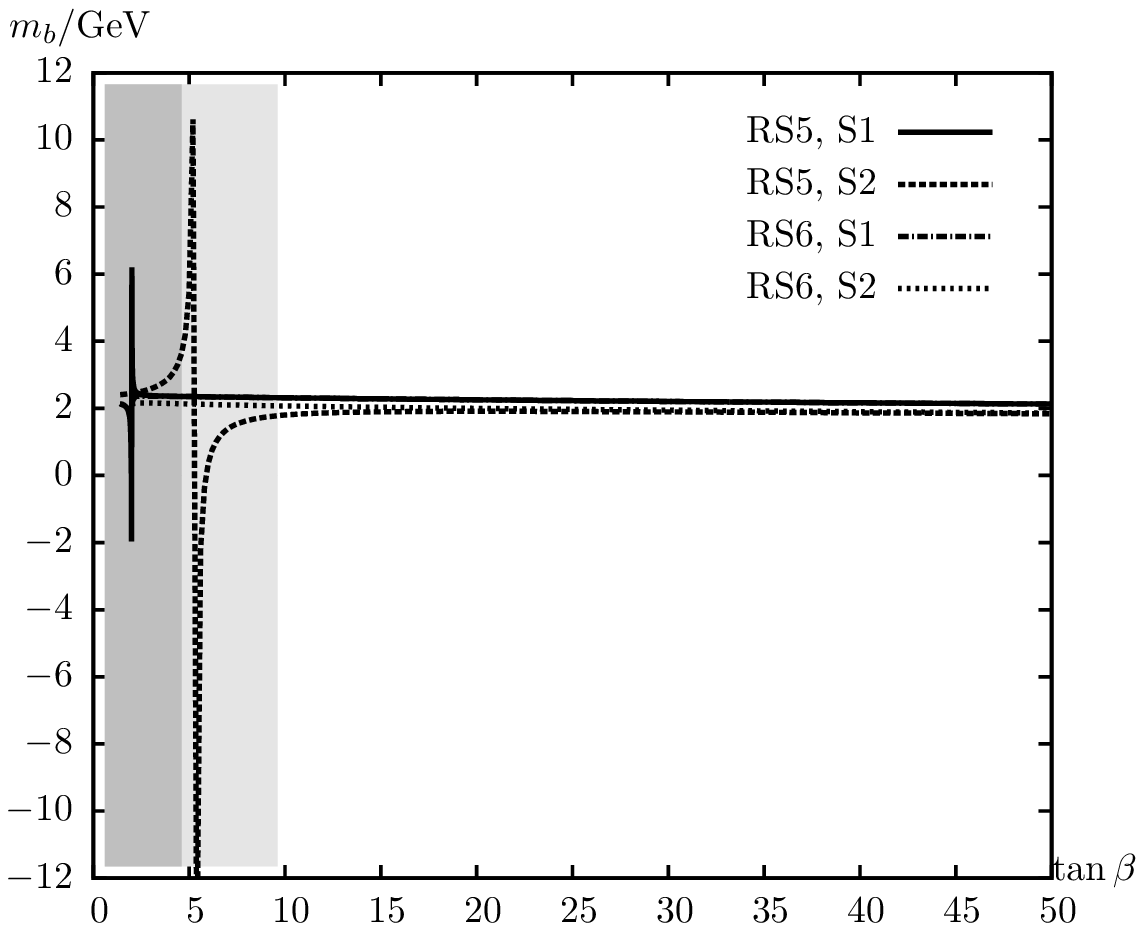}
\hspace{-2mm}
\includegraphics[width=0.49\textwidth,height=7.5cm]{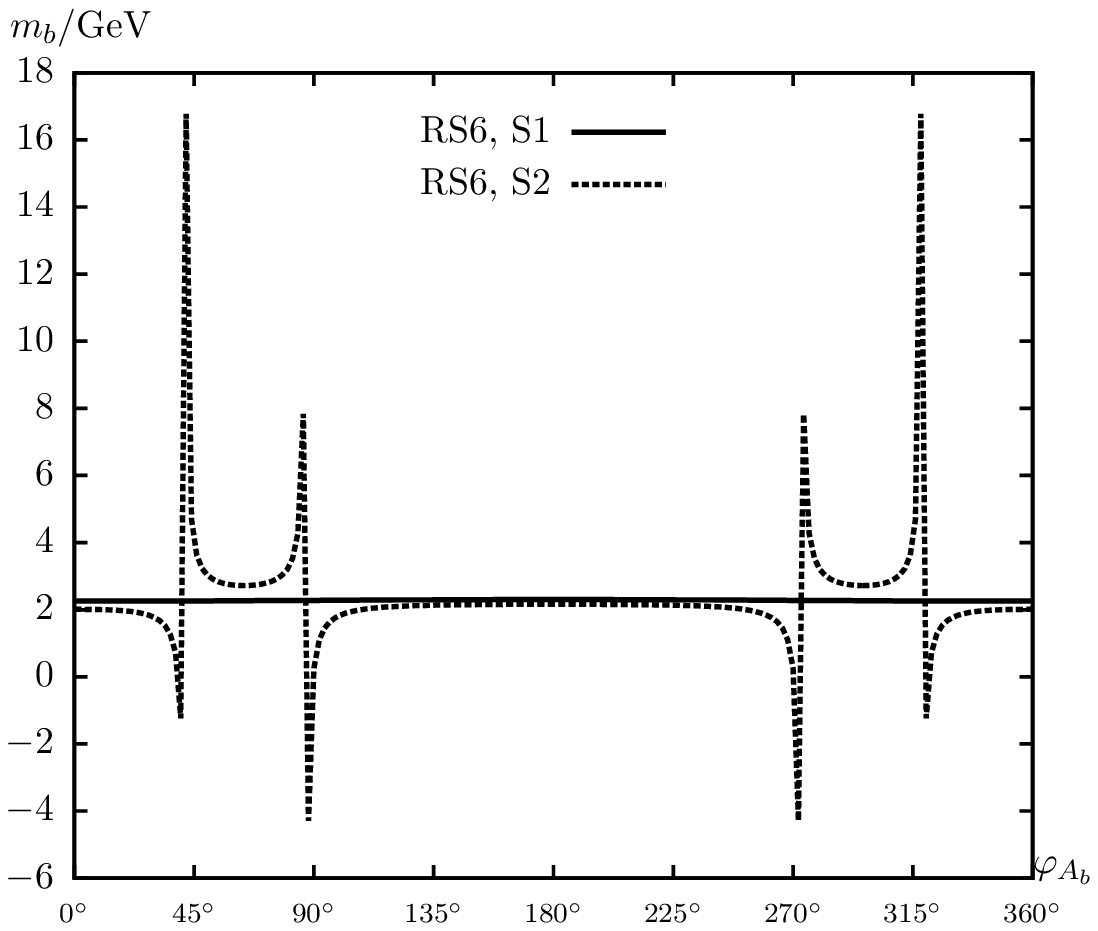} 
\end{tabular}
\caption{
  Left plot: $\mb$ in RS5 and RS6 for \SE, \SZ. 
  For \SE\ the grey region is excluded and for \SZ\ the dark grey region 
  is excluded.
  Right plot: $\mb$ in RS6 for \SE, \SZ\ but both with $\tb = 20$ and 
  $\varphi_{\Ab}$ varied. In \SZ\ we used also $|\mu| = 120 \gev$.}
\label{fig:mb.explanation}
\end{center}
\end{figure}

In a numerical evaluation, starting out with a value for the bottom
quark mass defined as \DRbar~parameter, the actual value of the bottom
quark mass  receives a shift with respect to the \DRbar~bottom quark
mass according to \refeq{eq:mbcorr}. This shift corresponds to the
finite part of $\de \mb$ in \refeq{dmb_problem}.
Consequently, large positive or negative contributions to the bottom quark 
mass can occur, yielding possibly 
negative values for the bottom quark mass and thus invalidating the
renormalization scheme for  
these parts of the parameter space.
This can be seen in the left plot of \reffi{fig:mb.explanation}, 
where we show $\mb$ in RS5 (and RS6) for the two numerical scenarios 
given in \refta{tab:para} as a function of $\tb$. 
$\mb$ exhibits a strong upward/downward shift around the pole reached
for $\tb = \Ab/\mu$ and consequently yields unreliable results in this
part of the parameter space.

\bigskip
We now turn to the RS6 (``$\Ab$~vertex, $\re Y_b$~OS'') scheme. 
Following the same analysis as for the
``$\Ab$~\DRbar, $\re Y_b$~OS'' scheme an additional term in the
denominator of the bottom quark mass counterterm
$\sim U_m/U_-$ appears,
\begin{align}
\de\mb &= - \frac{\de S + F}{(\Ab - \mu\tb) - U_m/(\tb\,U_-)}~,
\end{align}
where $F$ denotes other (relatively small) additional contributions. 
With the help of \refeq{def:Um} one finds for real parameters
\begin{align}
\frac{U_m}{\tb\,U_-} &= \frac{U_{\Sbot_{11}} U_{\Sbot_{22}} (\Ab \tb + \mu)
                             -U_{\Sbot_{12}} U_{\Sbot_{21}} (\Ab \tb + \mu)}
                             {\tb (U_{\Sbot_{11}} U_{\Sbot_{22}} - 
                                   U_{\Sbot_{12}} U_{\Sbot_{21}})}
 = (\Ab + \mu/\tb)~,
\end{align}
and therefore
\begin{align}
\label{dmb_problem2}
\de\mb &= \frac{\de S + F}{\mu\, (\tb + 1/\tb)}~.
\end{align}
The denominator of \refeq{dmb_problem2} can go to zero only for 
$\mu \to 0$, which is experimentally already excluded. 
Consequently, the problem of (too) large contributions to $\mb$ is
avoided in this scheme. This can be seen in the left plot of
\reffi{fig:mb.explanation}, where RS6, contrary to RS5, does not
exhibit any pole-like structure in $\mb$.

In the complex case the above argument is no longer valid, and larger
contributions to $\de\mb$ can arise. In the limit of $\tb \gg 1$ and
$\mu$~real the denominator of $\de\mb$ in \refeq{dmb_AbOS_ReYbOS}
reads
\begin{align}
\ed{\de\mb} &\sim 4\,\mu\, \tan^3\be\, \Big[
         \re U_- \KL |U_{\Sbot_{11}}|^2 - |U_{\Sbot_{12}}|^2 \KR 
         + \im U_- \frac{4\, \mb}{\msbe^2 - \msbz^2} 
         \im \KL U_{\Sbot_{11}}^* U_{\Sbot_{12}} \Ab \KR \Big]~. 
\label{dmb_RS6_pole}
\end{align}
Depending on $\phiab$ this denominator can go to zero and thus yield
unphysically large corrections to $\mb$ in RS6.
In the right plot of \reffi{fig:mb.explanation} we show $\mb$ as 
function of $\phiab$. At 
$\phiab \approx 41.5^\circ$, $87.5^\circ$, $272.5^\circ$, $318.2^\circ$ 
the denominator in \refeq{dmb_RS6_pole} goes to zero and changes its 
sign which explains the corresponding structures.
This divergence in $\de\mb$ enters via \refeq{eq:mbcorr} already
into the tree-level prediction.
To summarize: while in \SE\ the scheme RS6 is well-behaved and can
be safely applied (also for complex $\Ab$), in \SZ\ 
(with $|\mu| = 120$ GeV) severe problems
(divergences in the counterterms) arise once complex parameters are
taken into account. Consequently, for \SZ\ the scheme RS6 cannot be applied.

It should be noted that 
the ``$\Ab$~vertex, $\re Y_b$~OS'' (RS6) scheme is the complex version of the 
renormalization scheme used in \citeres{sbotrenold,mhiggsFDalbals} for the
\order{\alb\als} corrections to the neutral Higgs boson self-energies
and thus to the mass of the lightest MSSM Higgs boson, $\Mh$. For
real parameters, no problems occured. 
 Therefore,
employing this renormalization scheme in \citeres{sbotrenold,mhiggsFDalbals} 
yields numerically stable results.


\subsection{Generic considerations for the \boldmath{$b/\Sbot$} sector
  renormalization (II)} 
\label{sec:genericII}

In the previous subsections we have analyzed analytically 
(and numerically) the
deficiences of the various renormalization schemes. We have shown that
despite of the variety of schemes, even concentrating on the two sets of
parameters, \SE\ and \SZ, severe problems can be encountered
in all schemes.

For the further numerical evaluation of the partial stop quark decay
widths we choose 
RS2 as our ``preferred scheme''. According to our analyses in the
previous subsections, RS2 shows the ``relatively most stable'' behavior,
problems only occur for maximal sbottom mixing, 
$|U_{\Sbot_{11}}| = |U_{\Sbot_{12}}|$, where a divergence in $\de Y_b$
appears. 
Having $\de Y_b$ as a dependent counterterm induces 
effects in the field renormalization constants $\dZ{\Sbot_{12}}$ and
$\dZ{\Sbot_{21}}$ and in
$\de M_{\tilde{Q}_L}^2(\tilde{b})$ entering the scalar bottom quark masses.
In a process with only internal scalar bottom quarks, no problems occur
due to the field renormalization, but 
counterterms to propagators, which
induce a transition from a $\Sbote$ squark to a $\Sbotz$ squark contain
also the term $\de Y_b$. However, $\de Y_b$ appearing in counterterms of
{\em internal} scalar bottom quarks does not exhibit a problem, since in
this case these ``dangerous'' contributions cancel (which we have checked
analytically).
On the other hand,
other schemes with $\de\mb$ or $\de\Ab$ as dependent counterterms 
may exhibit problems in larger parts of the parameter
space and may induce large effects, since $\mb$ (or the bottom Yukawa 
coupling) and $\Ab$ enter prominently into the various couplings of the
Higgs bosons to other particles.

We are not aware of any paper dealing with scalar quark decays (or
decays into scalar quarks) that has employed exactly RS2 
(or its real version), see our discussion in the beginning of \refse{sec:stop}.
Very recently a calculation of the scalar top decay width in the rMSSM
using a pure \DRbar\ scheme for all parameters was
reported~\cite{HelmutLL2010}.


\section{Numerical examples for our favorite scheme}
\label{sec:numex}

Following the discussion in \refse{sec:RSana} we pick the renormalization 
scheme that shows the ``most stable'' behavior over the MSSM parameter space.
We choose the ``$\mb,\,\Ab$~\DRbar''(RS2) scheme.
Tree-level values of the partial decay widths shown in this section have
been obtained including a 
shift in $\mb$ according to \refeq{eq:mbcorr}. 
We will concentrate on the calculation of the  partial $\Stopz$ decay widths
including one scalar bottom quark in the final state. A calculation of
the respective branching ratios 
requires the evaluation of {\em all} partial scalar top quark decay
widths, which in turn 
requires the renormalization of the full cMSSM. This is beyond the scope of
our paper and will be presented elsewhere~\cite{Stop2decay}.
 

\subsection{Full one-loop results}
\label{sec:full1L}

\begin{figure}[htb!]
\begin{center}
\begin{tabular}{c}
\includegraphics[width=0.49\textwidth,height=7.5cm]{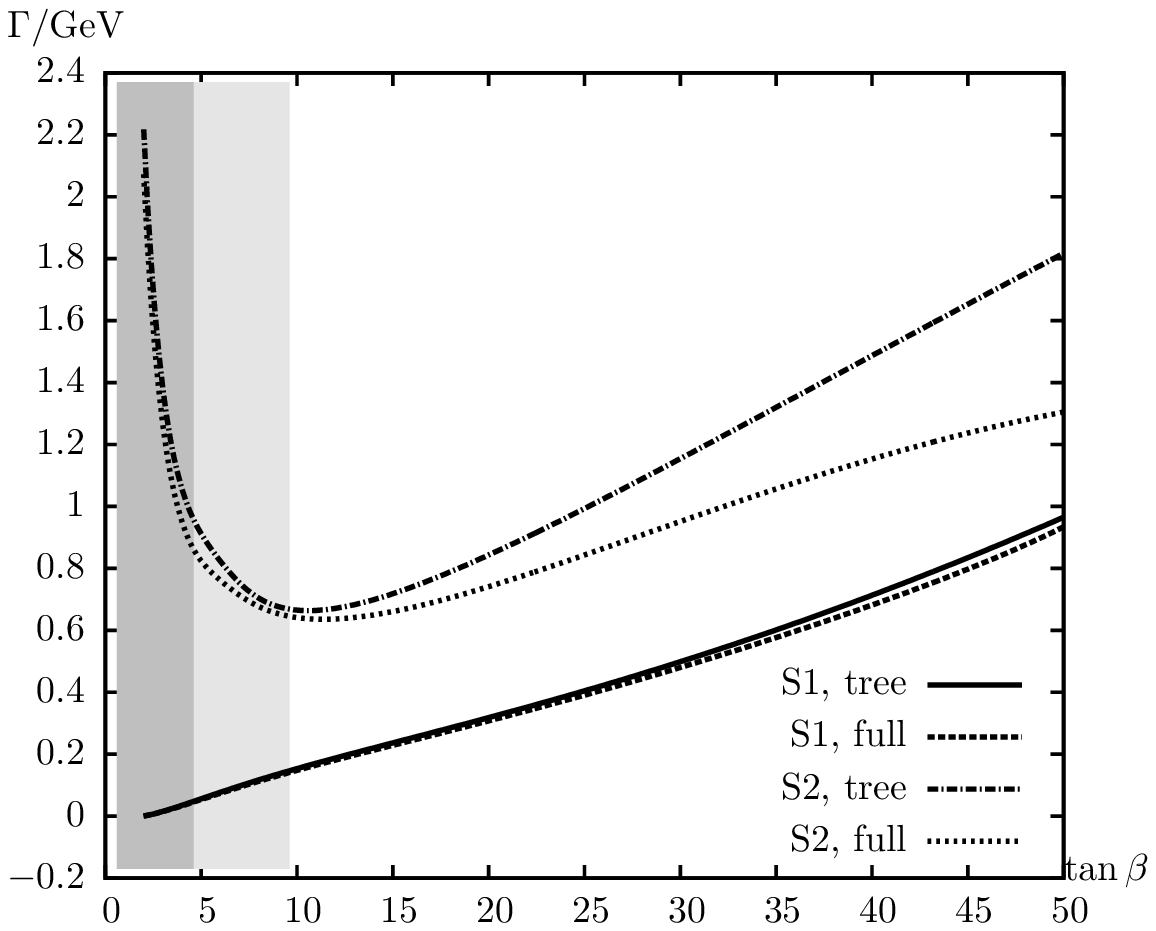}
\hspace{-2mm}
\includegraphics[width=0.49\textwidth,height=7.5cm]{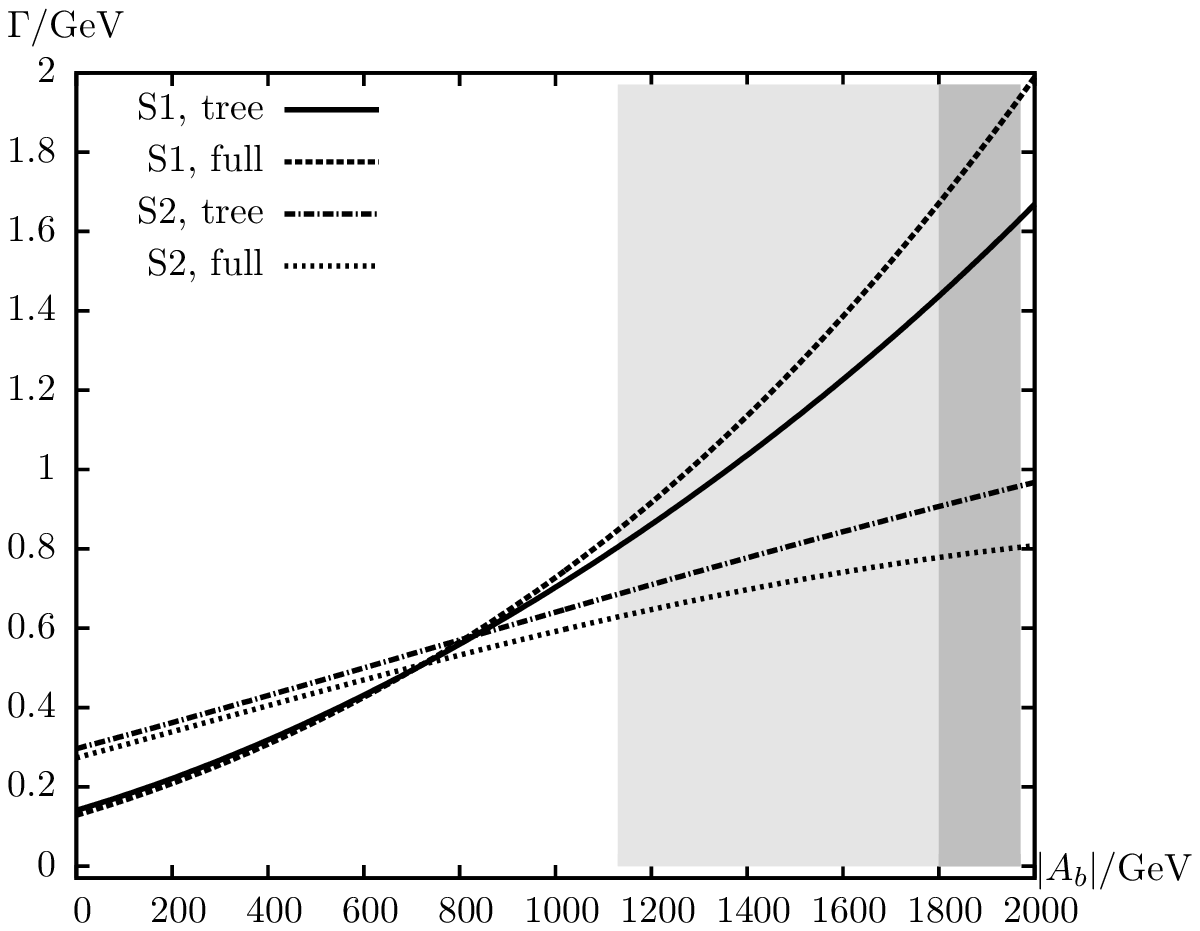} 
\\[4em]
\includegraphics[width=0.49\textwidth,height=7.5cm]{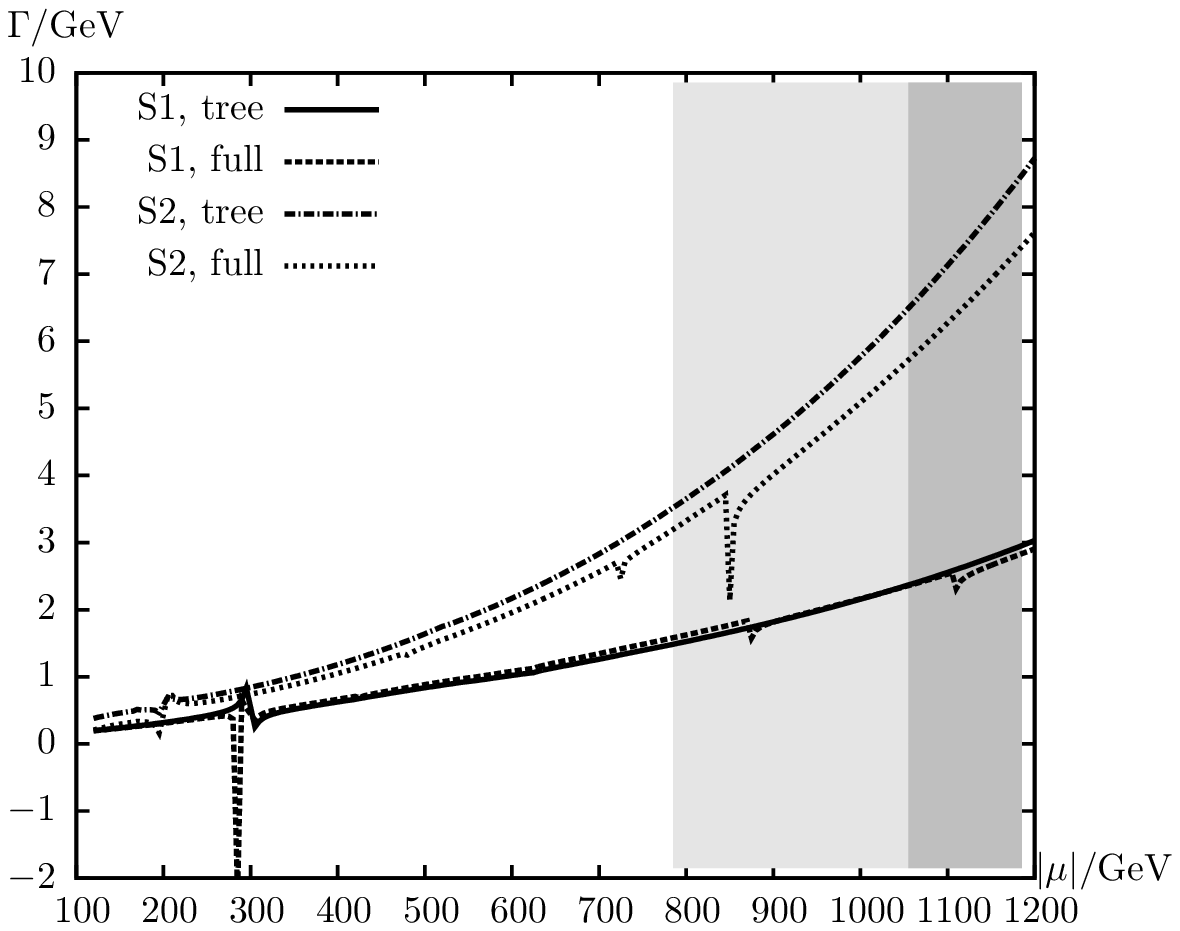}
\hspace{-2mm}
\includegraphics[width=0.49\textwidth,height=7.5cm]{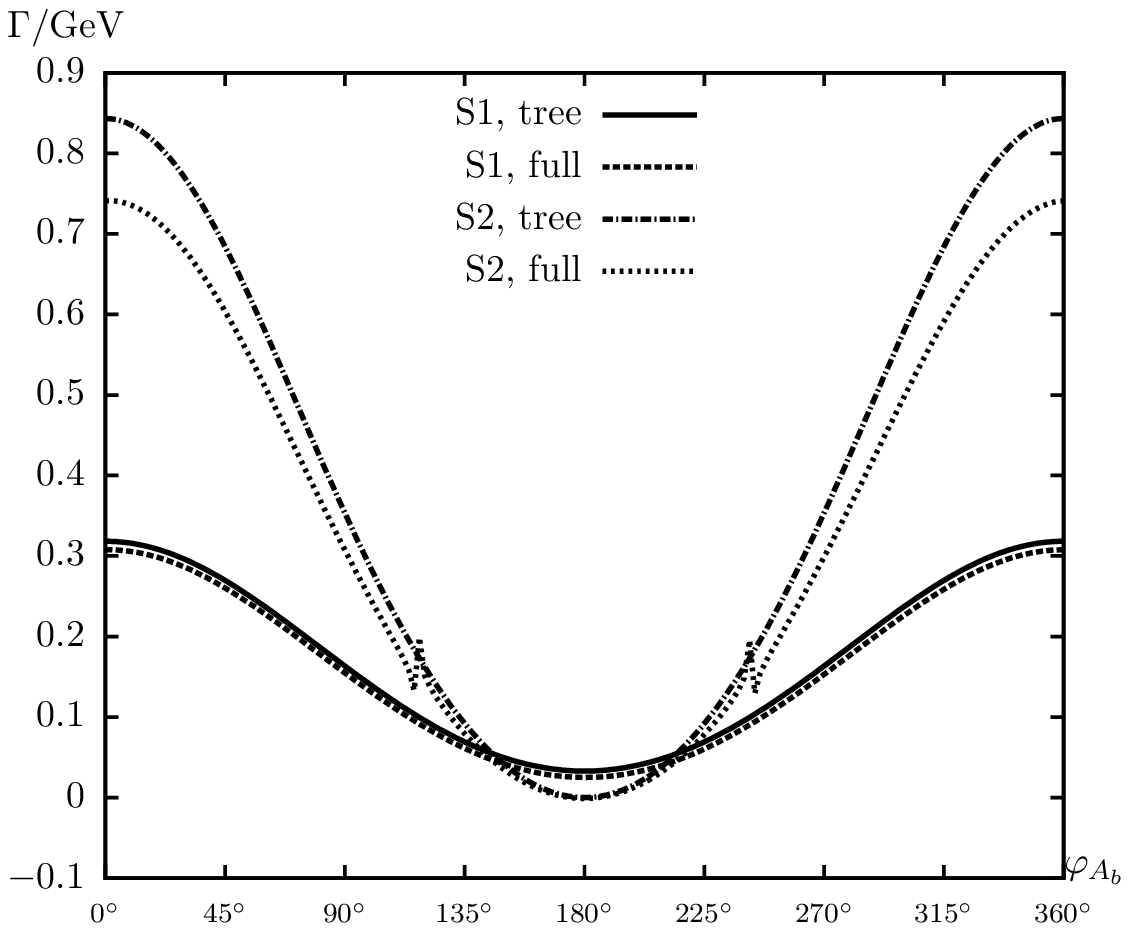}
\end{tabular}
\vspace{2em}
\caption{
$\Ga(\decayH)$. Tree-level and full one-loop corrected partial decay widths 
  for the renormalization scheme RS2. The parameters are chosen according to
  the scenarios \SE\ and \SZ\ (see \refta{tab:para}). 
  For \SE\ the grey region is excluded and for \SZ\ the dark grey region 
  is excluded.
  Upper left plot: $\tb$ varied.
  Upper right plot: $\tb = 20$ and $|\Ab|$ varied.
  Lower left plot: $\tb = 20$ and $|\mu|$ varied.
  Lower right plot: $\tb = 20$ and $\varphi_{\Ab}$ varied.
}
\label{fig:st2sb1H}
\end{center}
\end{figure}

We start our numerical analysis with the upper left plot of
\reffi{fig:st2sb1H}, where we show the partial decay width 
$\Ga(\decayH)$ as a function of $\tb$. ``tree'' denotes the tree-level value
and ``full'' is the decay width including {\em all} one-loop 
corrections as described in \refse{sec:calc}.
As one can see, the full one-loop corrections are negative and rather small 
over the full range of $\tb$, the largest size of the loop corrections
is found to be $\sim 28\%$ of the tree-level value for $\tb = 50$ in \SZ.%
\footnote{
It is interesting to note that at $\tb = |\Ab|/|\mu| = 2\, (5.33)$ in
\SE\ (\SZ) we get $U_{\Sbot_{11,22}} = 1$ and $U_{\Sbot_{12,21}} = 0$,
and consequently $\tilde{b}_{L,R} = \tilde{b}_{1,2}$, respectively.
}%
~In \SE\ the grey region and in \SZ\ the dark grey region is excluded due 
to too small values of the mass of the lightest MSSM Higgs boson,
$\Mh$.

In the upper right plot of \reffi{fig:st2sb1H} we show the partial
decay width varying $|\Ab|$ for $\tb = 20$. 
In \SE\ and \SZ\ the full one-loop corrections grow with 
$\Ab$, but never exceed $\sim 25\%$ of the tree-level result. 
Note, that for S1 $|\Ab| > 1130 \gev$
(grey region) and S2 $|\Ab| > 1800 \gev$ (dark grey region) 
is excluded due to the charge- or color-breaking minima.
Over the full parameter space the loop corrections are smooth and
small with respect to the tree-level results.

In the lower left plot of \reffi{fig:st2sb1H} we analyze 
the partial decay width varying $|\mu|$ for $\tb = 20$. 
Values for $|\mu| \lsim 120 \gev$ are excluded due to 
$\mcha1 < 94 \gev$~\cite{pdg}.
The loop corrected predictions for the partial decay width show several dips and
spikes. In \SE\ 
the first dip at $|\mu| \approx 285 \gev$ is due to 
$|U_{\Sbot_{11}}| \approx |U_{\Sbot_{12}}|$, see the discussion in
\refse{sec:mbAb}.
The second peak/dip (already present in the tree-level prediction) at
$|\mu| = 300 \gev$ is due to the 
renormalization of $\mu$~\cite{dissTF} and will be discussed in more
detail in \citere{Stop2decay}.%
\footnote{
The chosen renormalization exhibits a divergence for $\mu = M_2$. 
$\de \mu$ enters via $\de Y_b$ into $\de M_{\tilde{Q}_L}^2(\tilde{b})$ 
and thus into the values of $\msbi$. Consequently, the dip is already
present in  
the tree-level result.}
~The third dip at $|\mu| \approx 424 \gev$, which is hardly visible, is
due to the production threshold  $\mt + \mneu{3} = \mstz$.
The fourth dip at $|\mu| \approx 873 \gev$ is the threshold
$\mste + \MHp = \msbe$ of the self energy $\Si_{\Sbot_{11}}(\msbe^2)$ 
in the renormalization constants $\dZ{\Sbot_{11}}$ and $\de\msbe^2$.
The fifth dip at $|\mu| \approx 1107 \gev$ is the production threshold 
$\msbz + \MW = \mstz$.
For $|\mu| > 790 \gev$ the value of $\Mh$ drops strongly, and the 
scenario S1 is excluded by LEP Higgs searches as indicated by the 
gray shading.
Apart from the dips analyzed above the loop corrections are very
small and do not exceed $\sim 7\%$ of the tree-level result, 
the prediction for $\Ga(\decayH)$ is well under control.
We now turn to the scenario \SZ. Here, for growing $|\mu|$, the squark mass
splitting in the $\Stop/\Sbot$ sector becomes very large, leading to
large contributions to the electroweak precision observables. The dark
gray region for $|\mu| > 1060 \gev$ yields $W$~boson masses outside the
experimentally favored region at the $2\,\si$ level, 
$\MW \gsim 80.445 \gev$~\cite{lepewwgNEW}. 
Such large $|\mu|$ values are consequently disfavored.
The dip/peak at $|\mu| = 200 \gev$ in the tree and the loop contribution
is due to $\de \mu$, where $\mu = M_2$ is reached, see above. 
The second dip at $|\mu| = 477 \gev$, which is hardly visible, 
is the threshold $\mt + \mcha{2} = \msbe$ of the self energy 
$\Si_{\Sbot_{11}}(\msbe^2)$ in the renormalization constants 
$\dZ{\Sbot_{11}}$ and $\de\msbe^2$.
The third dip at $|\mu| = 725 \gev$ is the production threshold 
$\mt + \mneu{3} = \mstz$.
The fourth dip at $|\mu| = 850 \gev$ is again the threshold 
$\mste + \MHp = \msbe$. 
In \SZ\ the one-loop corrections are negative and growing with $|\mu|$.
Apart from the dips described above, also in this numerical evaluation
the loop corrections stay mostly relatively small 
with respect to the tree-level result, 
reaching the largest relative contribution at the smallest 
$|\mu|$ values, and are thus well under control.

We now turn to the case of complex parameters.  
As discussed in \refse{sec:numpar} we consider only $\Ab$ as a complex
parameter.
In the lower right plot of \reffi{fig:st2sb1H} we show 
the partial decay width depending on $\phiab$ for $\tb = 20$.
In \SE, the tree-level values and the loop corrections are well-behaved. 
The latter ones stay relatively small for the whole parameter space, 
not exceeding $\sim 18\%$ of the tree-level result.
In \SZ, the largest corrections occur for real positive values of 
$\Ab$ and reach $\sim 12\%$ of the tree-level values. 
For negative $\Ab$, the tree-level result becomes very small 
($< 0.01\gev$) and here the size of the loop corrections can be as 
large as the tree-level values. 
A small (and barely visible) asymmetry in the one-loop corrections 
appears in the lower right plot of \reffi{fig:st2sb1H}, 
due to terms $\sim U_{\Sbot_{ij}} \times C_{0,1,2}$-function.
The peak/dip at $\phiab \approx 117^\circ, 243^\circ$ are again due to 
$|U_{\Sbot_{11}}| \approx |U_{\Sbot_{12}}|$, see \refse{sec:mbAb}.
It can be seen that the peaks due to this divergence are relatively
sharp, i.e.\ the region of parameter space that is invalidated remains
relatively small.

\begin{figure}[htb!]
\begin{center}
\begin{tabular}{c}
\includegraphics[width=0.49\textwidth,height=7.5cm]{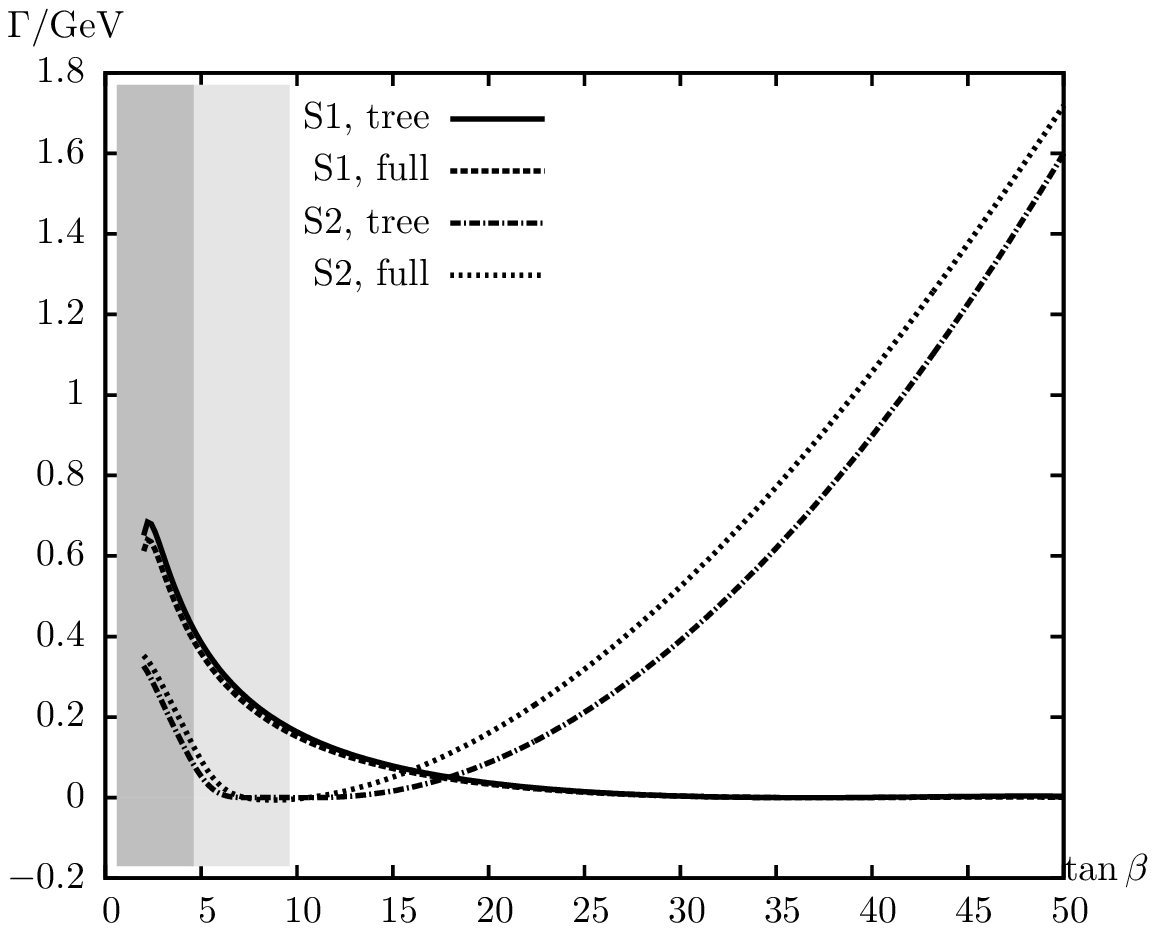}
\hspace{-2mm}
\includegraphics[width=0.49\textwidth,height=7.5cm]{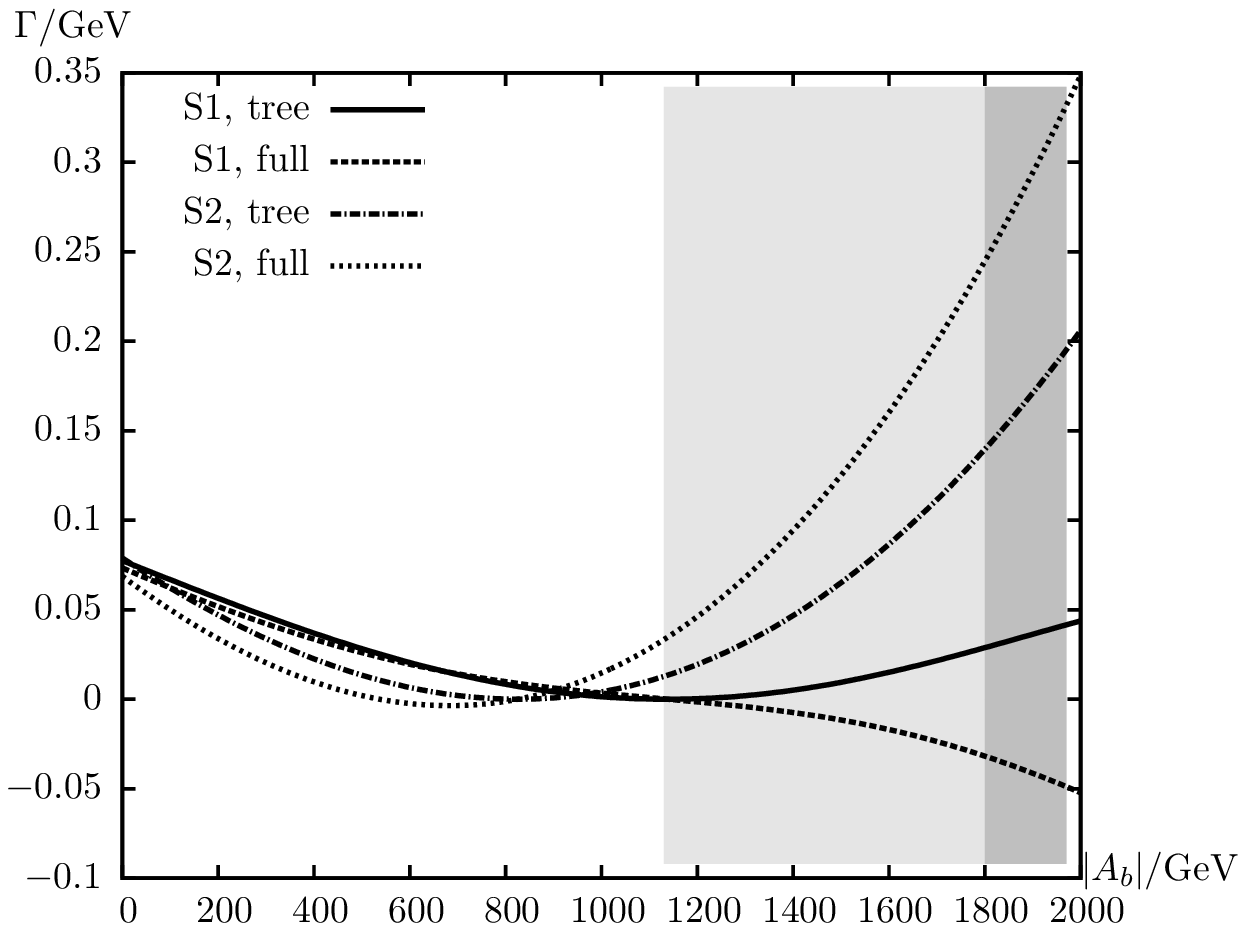} 
\\[4em]
\includegraphics[width=0.49\textwidth,height=7.5cm]{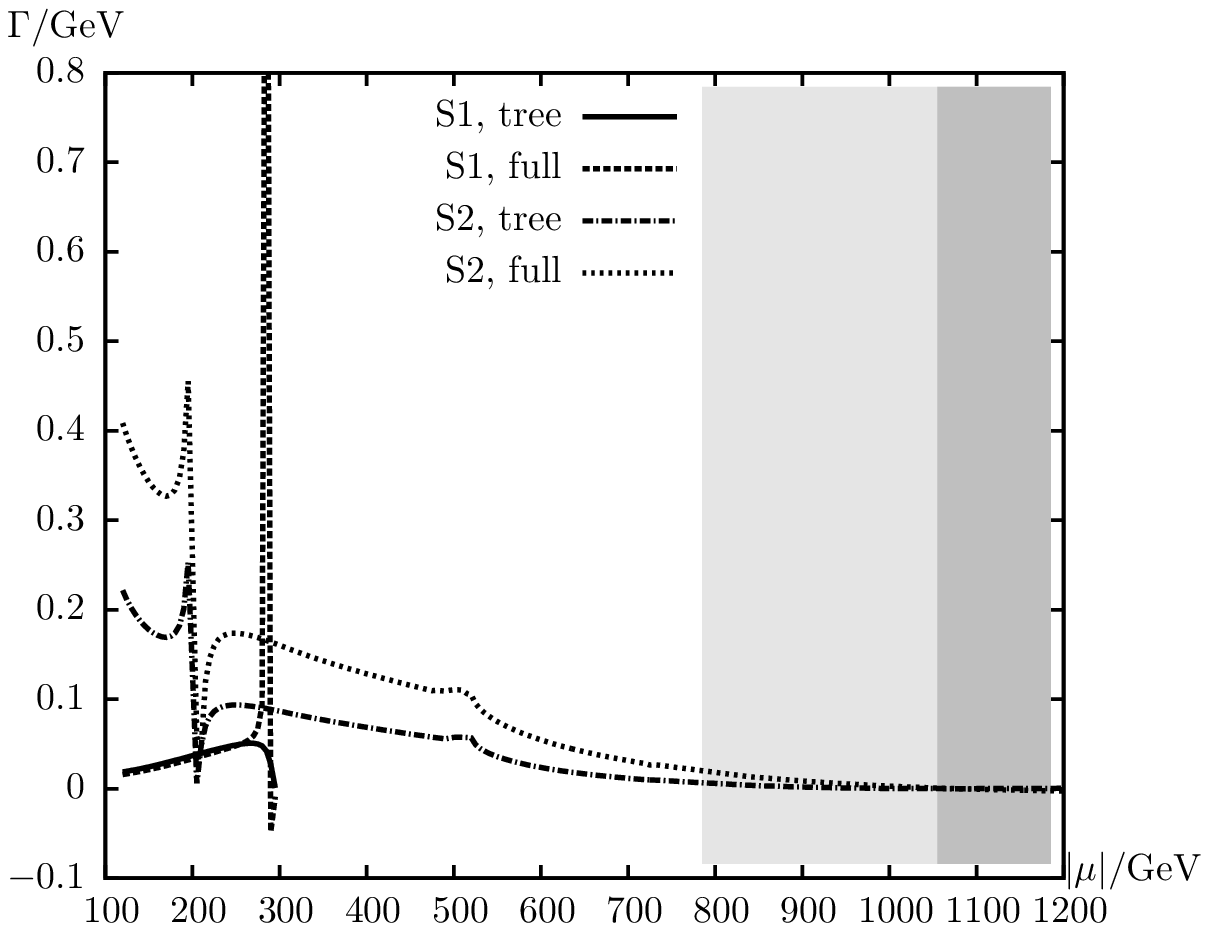}
\hspace{-2mm}
\includegraphics[width=0.49\textwidth,height=7.5cm]{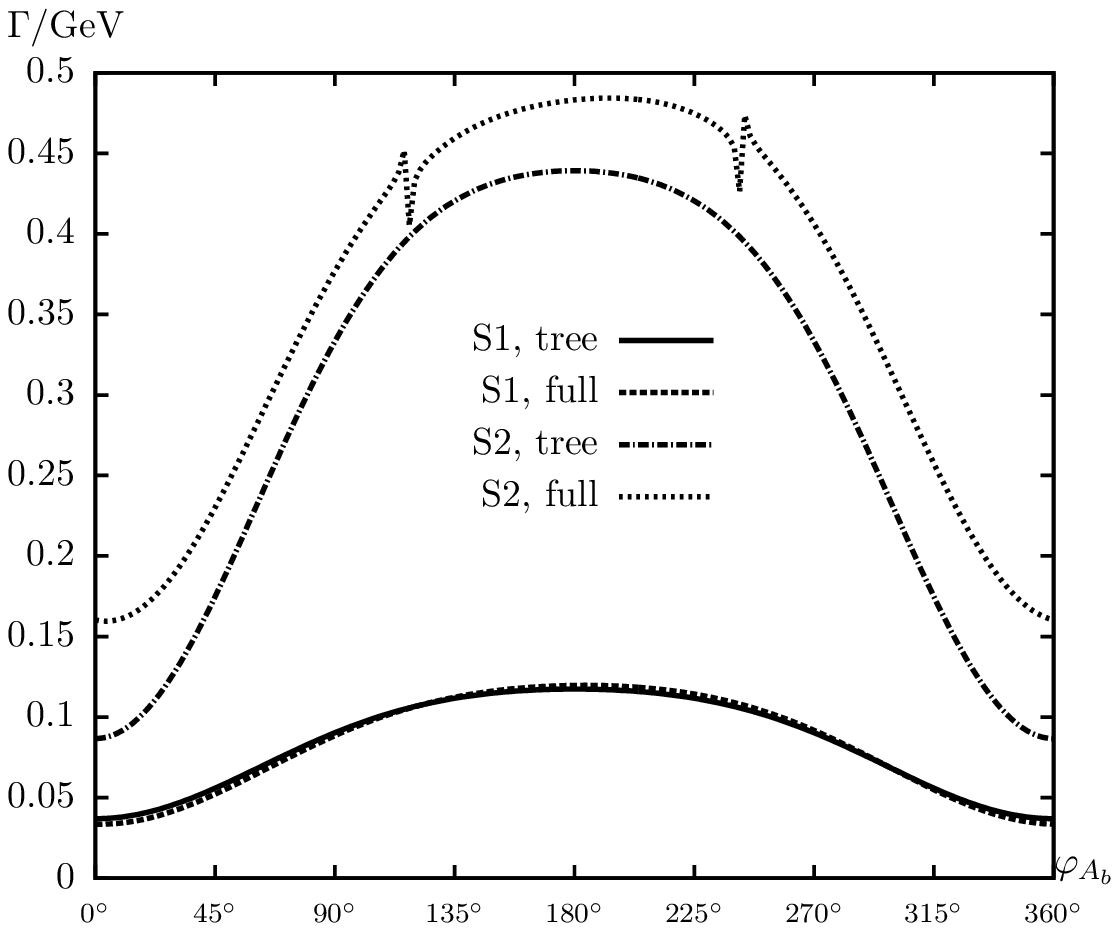}
\end{tabular}
\vspace{2em}
\caption{
$\Ga(\decaySbzH)$. Tree-level and full one-loop corrected partial decay widths 
  for the renormalization scheme RS2. The parameters are chosen
  according to the scenarios \SE\ and \SZ\ 
  (see \refta{tab:para}). 
  For \SE\ the grey region is excluded and for \SZ\ the dark grey region 
  is excluded.
  Upper left plot: $\tb$ varied.
  Upper right plot: $\tb = 20$ and $|\Ab|$ varied.
  Lower left plot: $\tb = 20$ and $|\mu|$ varied.
  Lower right plot: $\tb = 20$ and $\varphi_{\Ab}$ varied.
}
\label{fig:st2sb2H}
\end{center}
\end{figure}

\smallskip
In \reffi{fig:st2sb2H} we show the results for $\Ga(\decaySbzH)$ for 
the same set and variation of parameters as above. 
Consequently, the same peak and dip structures are visible in 
the lower plots of \reffi{fig:st2sb2H}. 
In the lower left plot of \reffi{fig:st2sb2H} in \SE\ both lines 
end because the phase space closes, 
$\msbz + \MHp > \mstz$ for $|\mu| > 300 \gev$.
Overall the partial decay width is much smaller than for
$\Ga(\decayH)$, which can 
partially be attributed to the smaller phase space, see for instance
the results within \SZ\ in
the upper left plot of \reffi{fig:st2sb2H}, and partially to the 
smallness of the tree-level coupling.
Only in \SZ\ for $\tb \gsim 35$ we find $\Ga(\decaySbzH) \gsim 1 \gev$.
The relative corrections become very large for $|\Ab| \gsim 1200 \gev$
as shown in the upper right plot of \reffi{fig:st2sb2H}, 
however these values are disfavored by the constraints from charge and 
color breaking minima as discussed above.
The smallness of $\Ga(\decaySbzH)$ at the tree-level can lead sometimes to 
a ``negative value at the loop level''. 
In this case of (accidental) smallness of the
tree-level partial decay width also $|\cM_{\rm loop}|^2$ would have to
be taken into 
account, yielding a positive value for $\Ga(\decaySbzH)$.
Overall, because of the smallness of the tree-level result due to the tree-level
coupling the {\em relative} size of the loop corrections are a bit larger than
for $\Ga(\decayH)$. Nevertheless, apart from the peaks visible in
the lower plots of \reffis{fig:st2sb2H}, the loop corrections are
well under control also for $\Ga(\decaySbzH)$ using the renormalization
scheme RS2. 
Again a small asymmetry in the one-loop corrections in
the lower right plot of \reffi{fig:st2sb2H} can be observed, 
which is due to terms $\sim U_{\Sbot_{ij}} \times C_{0,1,2}$-function.

\begin{figure}[htb!]
\begin{center}
\begin{tabular}{c}
\includegraphics[width=0.49\textwidth,height=7.5cm]{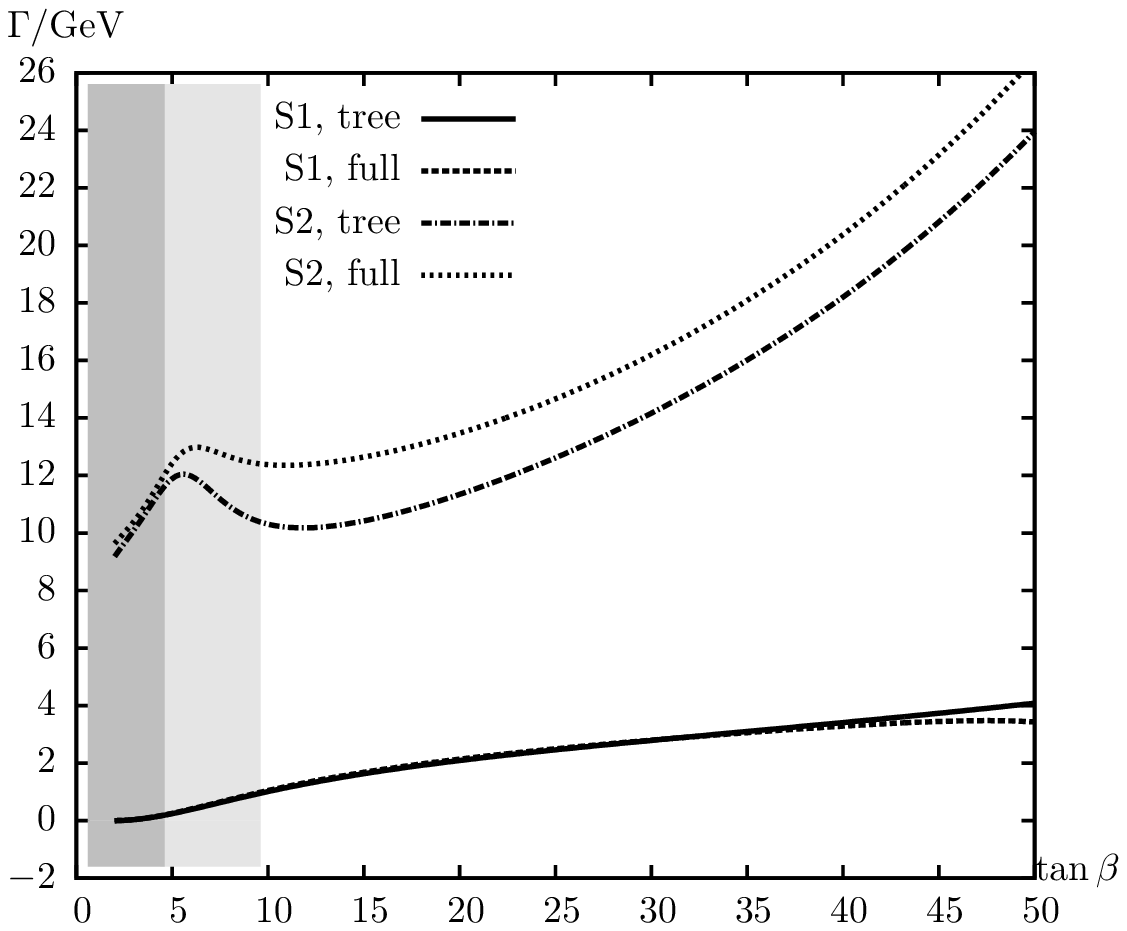}
\hspace{-2mm}
\includegraphics[width=0.49\textwidth,height=7.5cm]{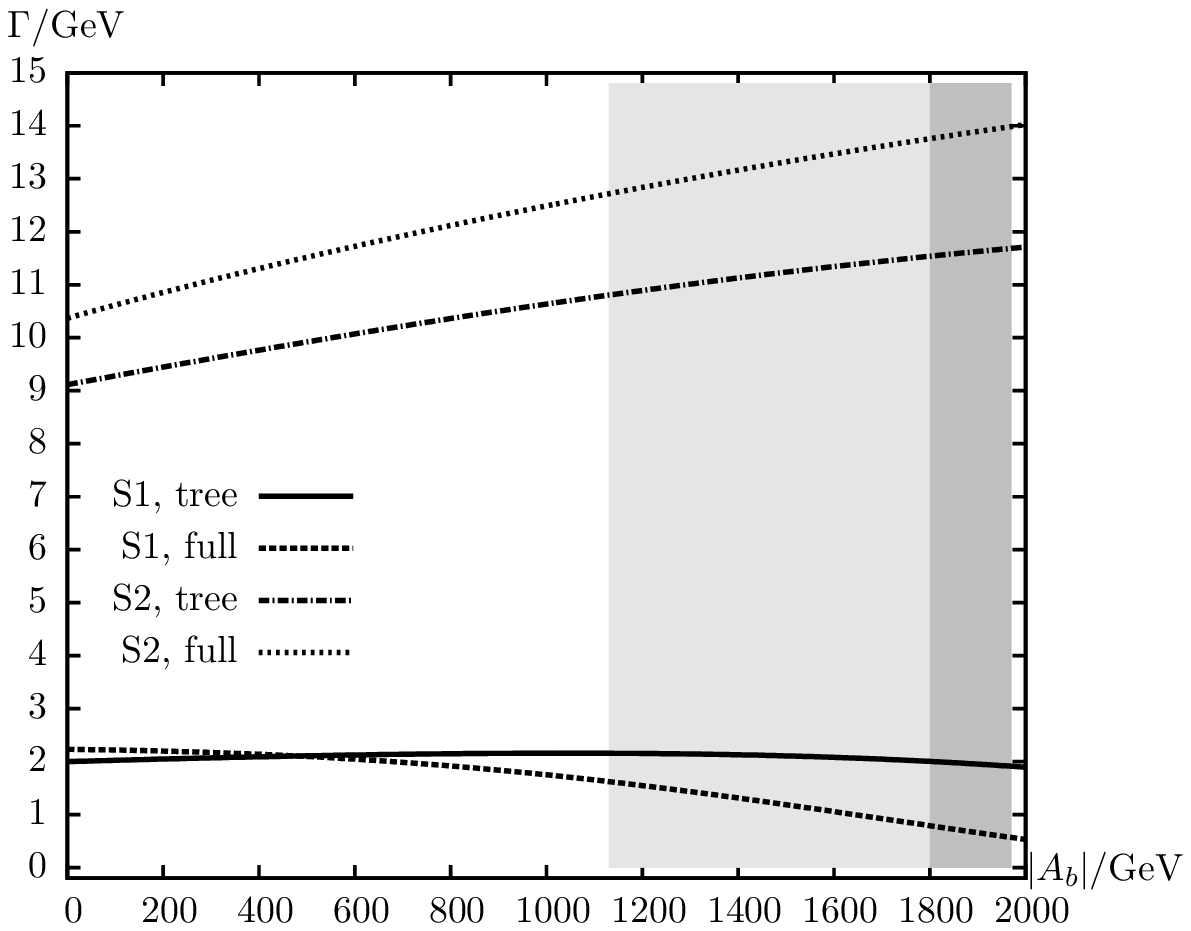} 
\\[4em]
\includegraphics[width=0.49\textwidth,height=7.5cm]{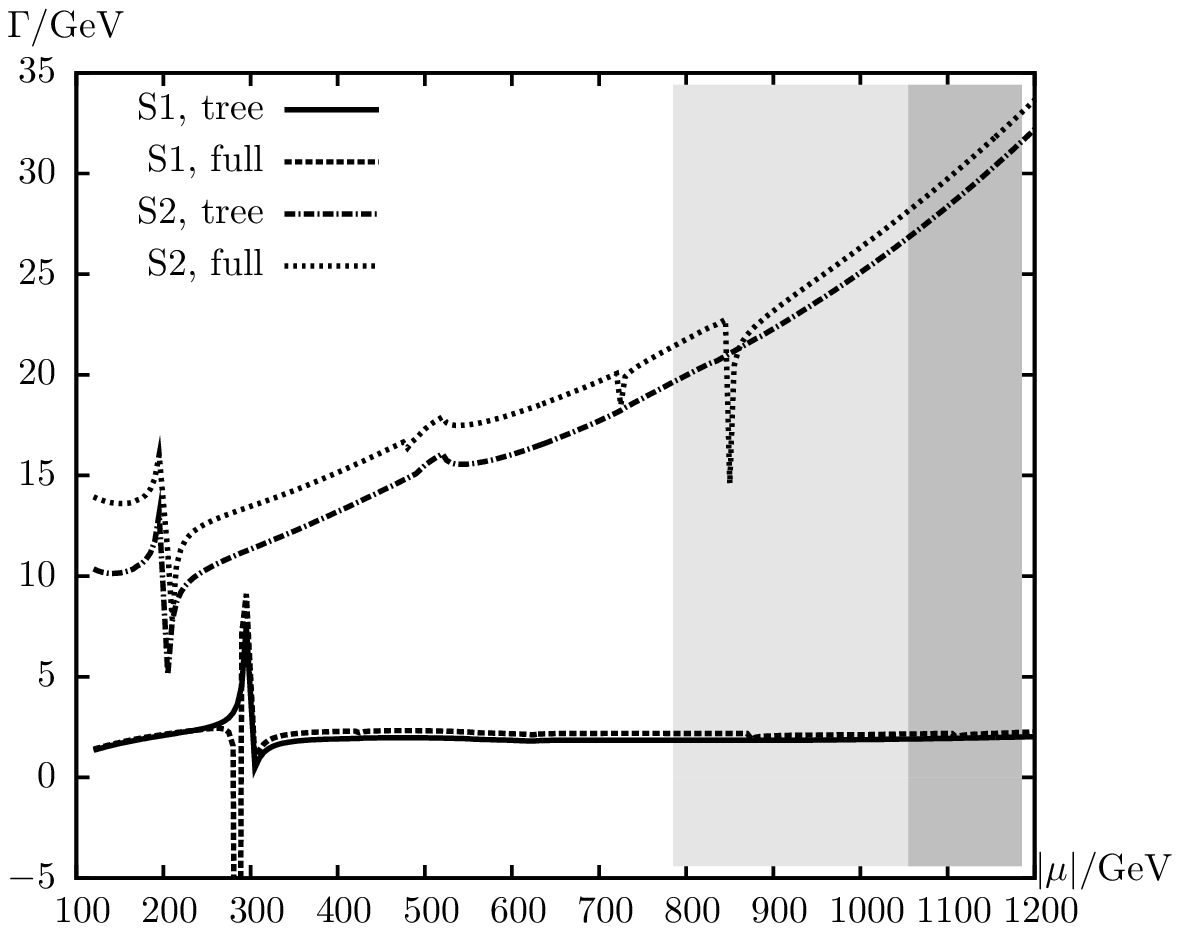}
\hspace{-2mm}
\includegraphics[width=0.49\textwidth,height=7.5cm]{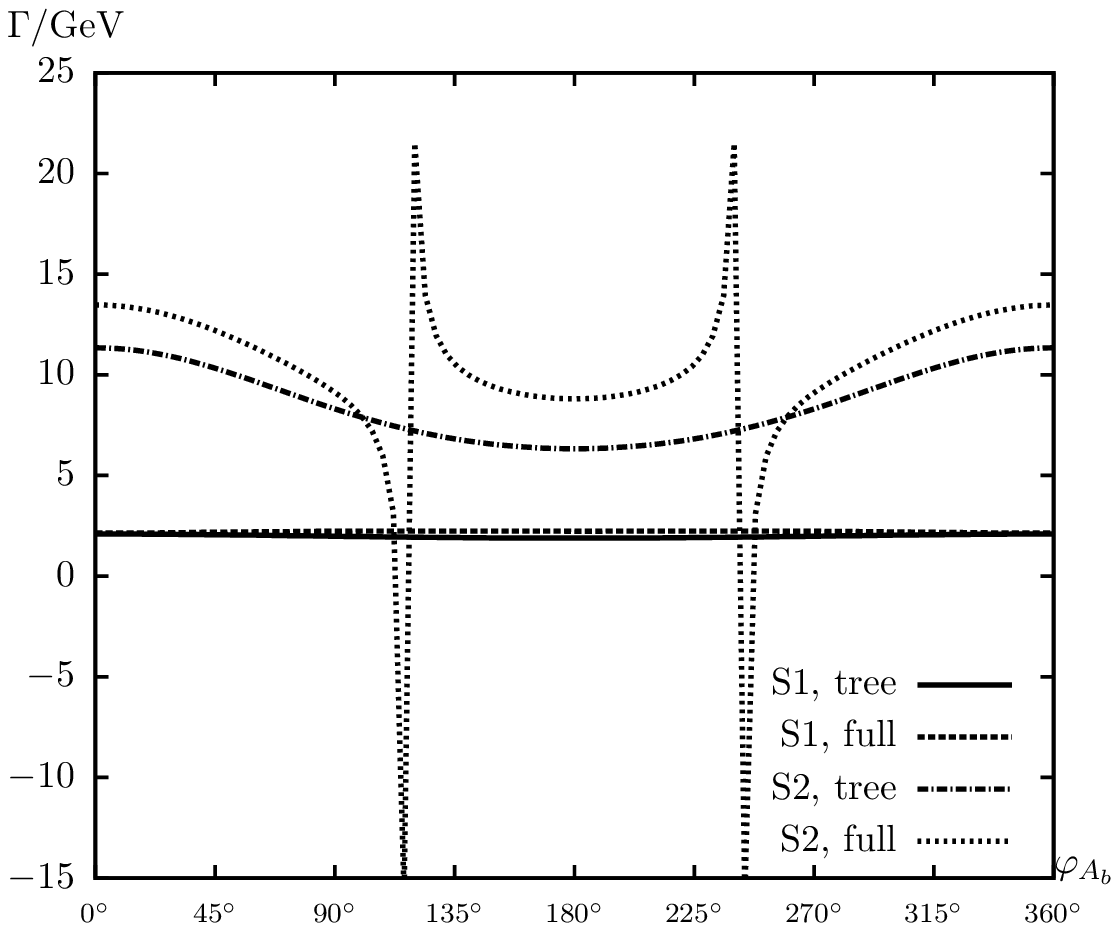}
\end{tabular}
\vspace{2em}
\caption{
$\Ga(\decaySbeW)$. Tree-level and full one-loop corrected partial decay widths 
  for the renormalization scheme RS2. The parameters are chosen
  according to the scenarios \SE\ and \SZ\ 
  (see \refta{tab:para}). 
  For \SE\ the grey region is excluded and for \SZ\ the dark grey region 
  is excluded.
  Upper left plot: $\tb$ varied.
  Upper right plot: $\tb = 20$ and $|\Ab|$ varied.
  Lower left plot: $\tb = 20$ and $|\mu|$ varied.
  Lower right plot: $\tb = 20$ and $\varphi_{\Ab}$ varied.
}
\label{fig:st2sb1W}
\end{center}
\end{figure}

\begin{figure}[htb!]
\begin{center}
\begin{tabular}{c}
\includegraphics[width=0.49\textwidth,height=7.5cm]{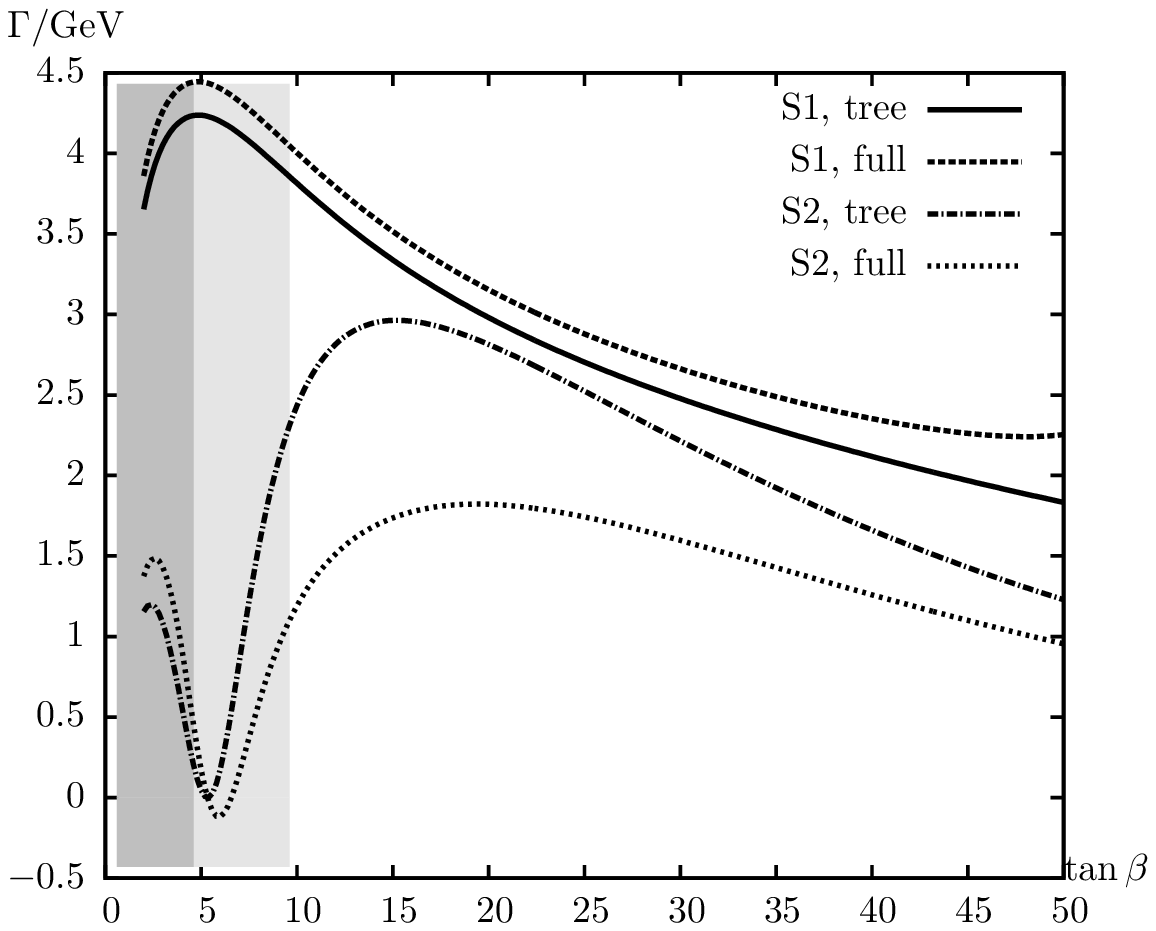}
\hspace{-2mm}
\includegraphics[width=0.49\textwidth,height=7.5cm]{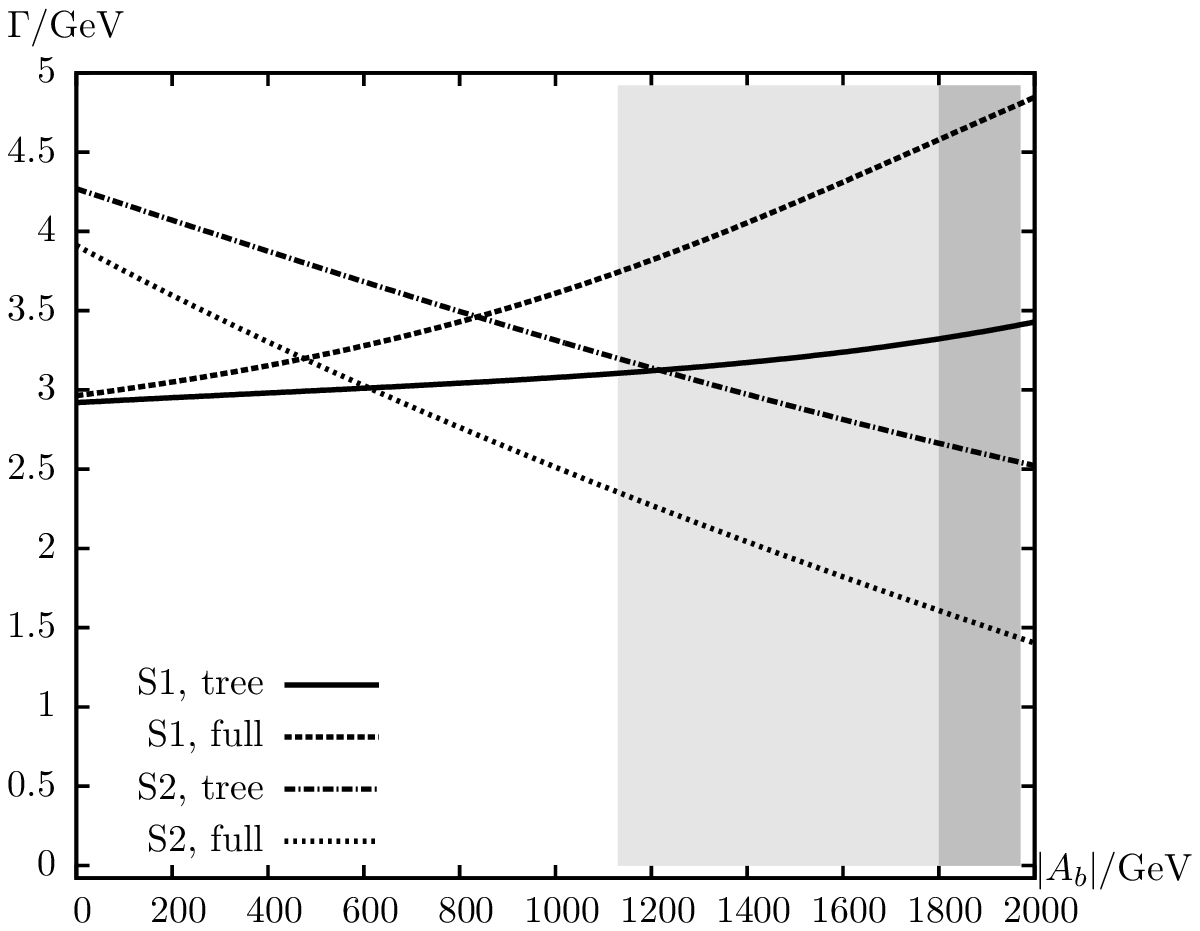} 
\\[4em]
\includegraphics[width=0.49\textwidth,height=7.5cm]{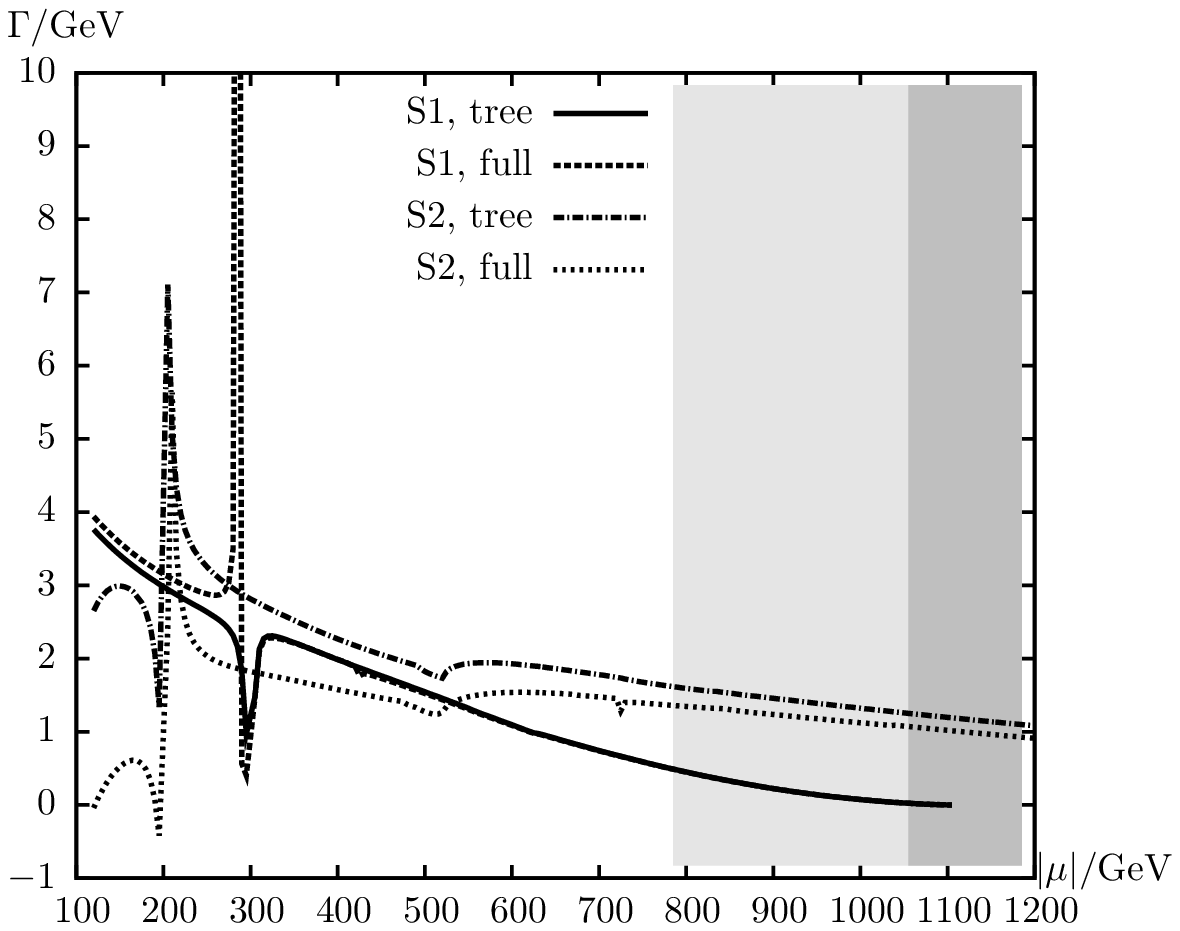}
\hspace{-2mm}
\includegraphics[width=0.49\textwidth,height=7.5cm]{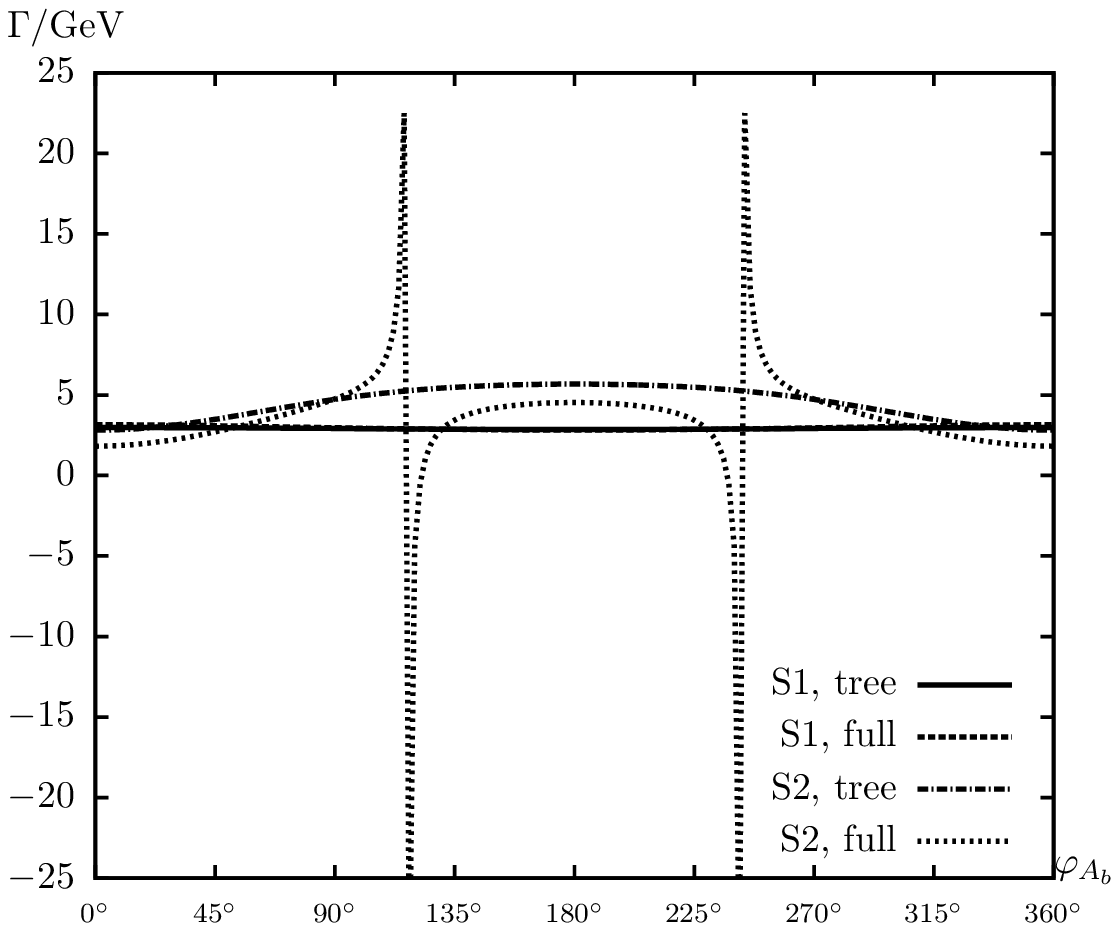}
\end{tabular}
\vspace{2em}
\caption{
$\Ga(\decaySbzW)$. Tree-level and full one-loop corrected partial decay widths 
  for the renormalization scheme RS2. The parameters are chosen
  according to the scenarios \SE\ and \SZ\ 
  (see \refta{tab:para}). 
  For \SE\ the grey region is excluded and for \SZ\ the dark grey region 
  is excluded.
  Upper left plot: $\tb$ varied.
  Upper right plot: $\tb = 20$ and $|\Ab|$ varied.
  Lower left plot: $\tb = 20$ and $|\mu|$ varied.
  Lower right plot: $\tb = 20$ and $\varphi_{\Ab}$ varied.
}
\label{fig:st2sb2W}
\end{center}
\end{figure}

\smallskip
Finally we evaluate the partial decay width of a scalar top quark to a
scalar bottom quark
and a $W$~boson, $\Ga(\decaySbeW)$ and $\Ga(\decaySbzW)$. Since the
$W$~boson is relatively light, also the latter channel is open.
In \reffi{fig:st2sb1W} the results for $\decaySbeW$ are shown, 
in \reffi{fig:st2sb2W} the ones for $\decaySbzW$.
The divergences visible in the various plots are the same ones as found
in the respective plot for $\Ga(\decayH)$. 
An additional (finite) dip is visible in the lower left plot of 
\reffi{fig:st2sb1W} in \SZ\ for $|\mu| \approx 521 \gev$,  
due to an interplay of $t/\cha{2}$ contributions to 
$\Si_{\Sbot_{11}}(\msbe)$, similar to the structure discussed for 
\reffi{fig:dAb}. In this part of the parameter space the results
  calculated within the
  renormalization scheme RS2 have to be discarded.

Overall, the loop corrections to $\Ga(\decaySbeW)$ calculated within the
renormalization scheme RS2 behave similar
to the ones to $\Ga(\decayH)$. The size is relatively small, i.e.
$\lsim 20\%$ and $\lsim 30\%$ of the tree-level results in the upper left
 and  
 in the upper right plot of \reffi{fig:st2sb1W}, respectively, 
for the regions  which are not in conflict with charge- or color
breaking minima  
(for $|\Ab| = 2000 \gev$ a correction of $\sim 70\%$ of the tree-level
result can be observed in 
\SE\ due to the smallness of the tree-level value). 
We find loop corrections of the size of $\lsim 20\%$ of the tree-level
results  in the lower left plot of \reffi{fig:st2sb1W} 
except for very small values of $|\mu|$ and in the lower right plot of
\reffi{fig:st2sb1W}. 
In the latter plot for \SZ\ the known divergences appear at 
$\phiab \approx 117^\circ, 243^\circ$, leading to larger loop
corrections for intermediate values of $\phiab$.
Apart from the latter case the full one-loop corrections to
$\Ga(\decaySbeW)$ are well under control employing the renormalization
scheme RS2.

Similar observations hold for the decay $\decaySbzW$, as shown in
\reffi{fig:st2sb2W}.
In the upper left plot of \reffi{fig:st2sb2W} in the scenario \SZ\ for 
$\tb = |\Ab|/|\mu| \approx 5.3$, the tree-level partial decay width vanishes, 
leading to a ``negative value at the loop level''. 
As discussed above, in this case also $|\cM_{\rm loop}|^2$ would have to
be taken into account, yielding a positive value for $\Ga(\decaySbzW)$.
(A similar situation is found in the lower left plot of
\reffi{fig:st2sb2W} for $|\mu| \approx 200 \gev$.) For somewhat
larger $\tb$ values, loop corrections of $\sim 50\%$ of the tree-level
values are reached, while
in \SE\ they stay below $\sim 23\%$ of the tree-level results.
In the upper right plot of \reffi{fig:st2sb2W} the loop corrections 
are smaller than $\sim 40\%$ of the tree-level values, depending on the
size of $|\Ab|$, see above.
 The loop corrections shown in the lower left plot of \reffi{fig:st2sb2W}
yield  maximal $\sim 9 (37) \%$ of the tree-level results in \SE\
(\SZ), apart from very small $\mu$  
values, where the tree-level partial decay width can become accidentally small.

Finally, looking at the dependence on $\phiab$ in
the lower right plot of \reffi{fig:st2sb2W}, apart from 
the known divergences in \SZ\ around $\phiab \approx 117^\circ, 243^\circ$, 
the loop corrections do not exceed $\sim 6\%$ and $\sim 35\%$ of
the tree-level values
in \SE\ and in \SZ, respectively. Overall, except for the small
parameter regions around $\phiab \approx 117^\circ, 243^\circ$, the full
one-loop corrections to $\Ga(\decaySbzW)$ are well under control employing the
renormalization scheme RS2.

\newpage

\newpage

\newpage

\newpage

\subsection{Comparison with SQCD calculation}
\label{sec:sqcd}

Often QCD corrections to SM or MSSM processes are considered as the
leading higher-order contributions. However, it has also been observed
for SM processes (e.g.\ in the case of $WH$ and $ZH$ production at the
Tevatron and LHC~\cite{WHZH}, for $H + 2$\,jet production at the
LHC~\cite{H2j}, or for the Higgs decay to four fermions in the
SM~\cite{HVV4f}) that the electroweak (EW) corrections can be of similar
size as the QCD corrections. 
Therefore, in the last step of our numerical evaluation, we show the size
of the one-loop effects based on  SUSY QCD (SQCD) only. The size of the
SQCD corrections 
can then be compared to the full calculation presented in the previous
subsection. It should be kept in mind that, following
\refeq{MSbotshift}, also the masses of the scalar bottom quarks depend
on the order of the calculation. Consequently, we do not explicitly compare
SQCD with the full one-loop calculation, but analyze only the size and
the sign of the pure SQCD corrections.

\begin{figure}[htb!]
\vspace{2em}
\begin{center}
\begin{tabular}{c}
\includegraphics[width=0.49\textwidth,height=7.5cm]{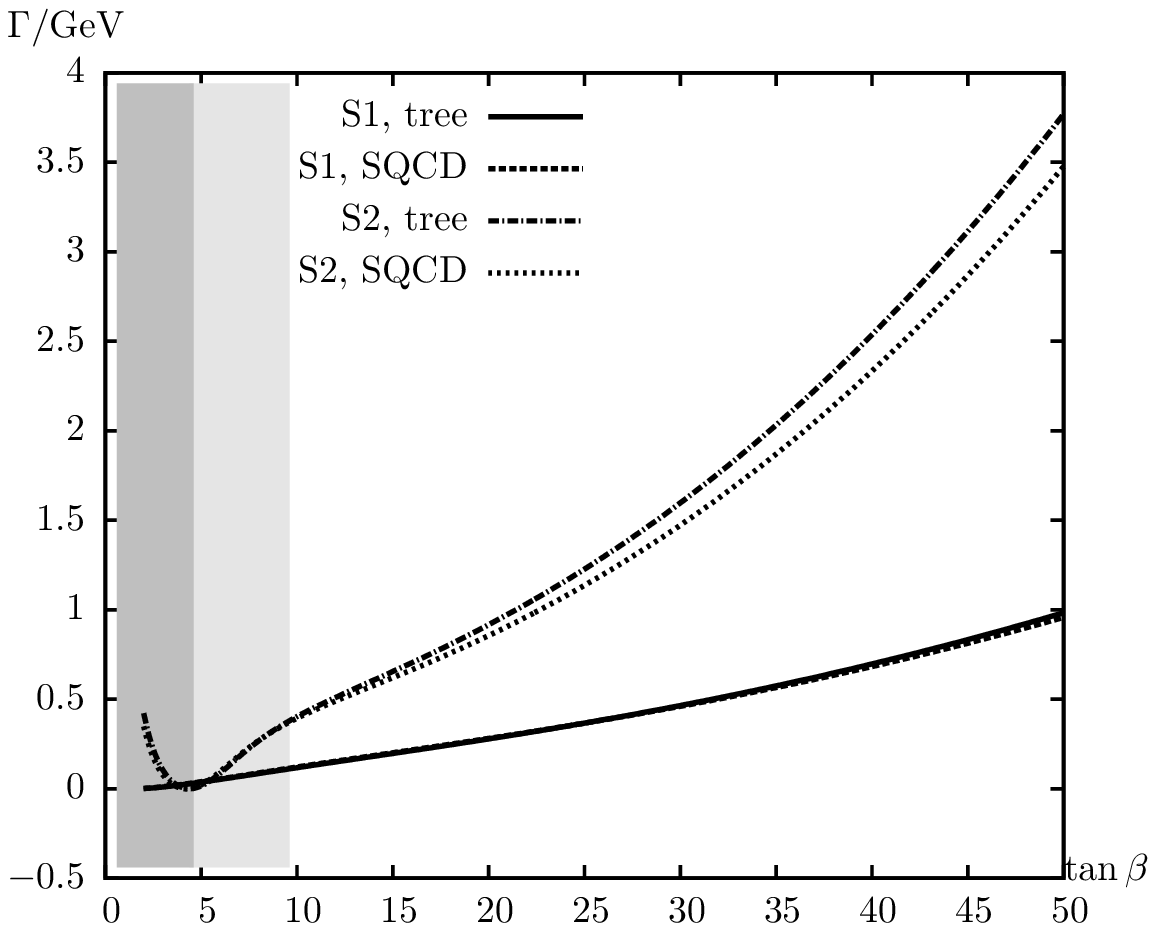}
\hspace{-2mm}
\includegraphics[width=0.49\textwidth,height=7.5cm]{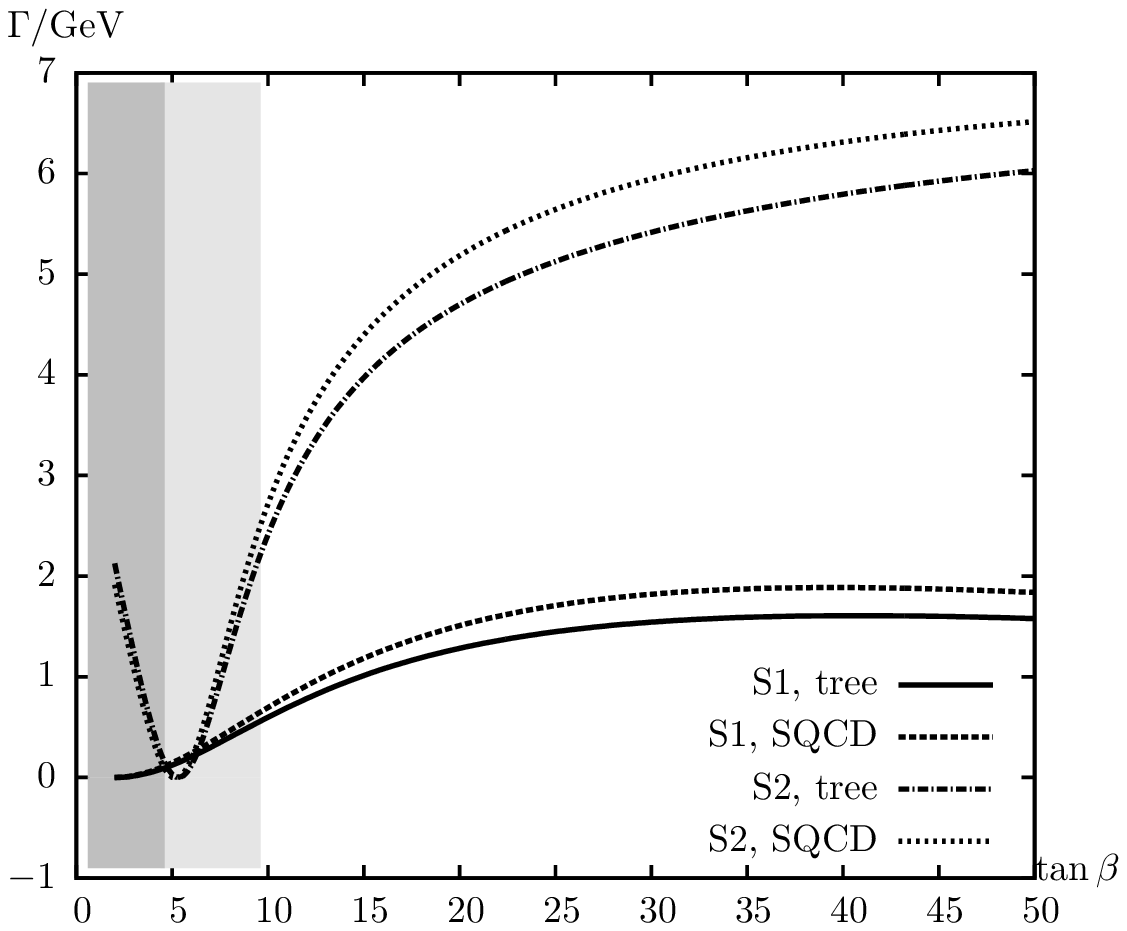} 
\\[4em]
\includegraphics[width=0.49\textwidth,height=7.5cm]{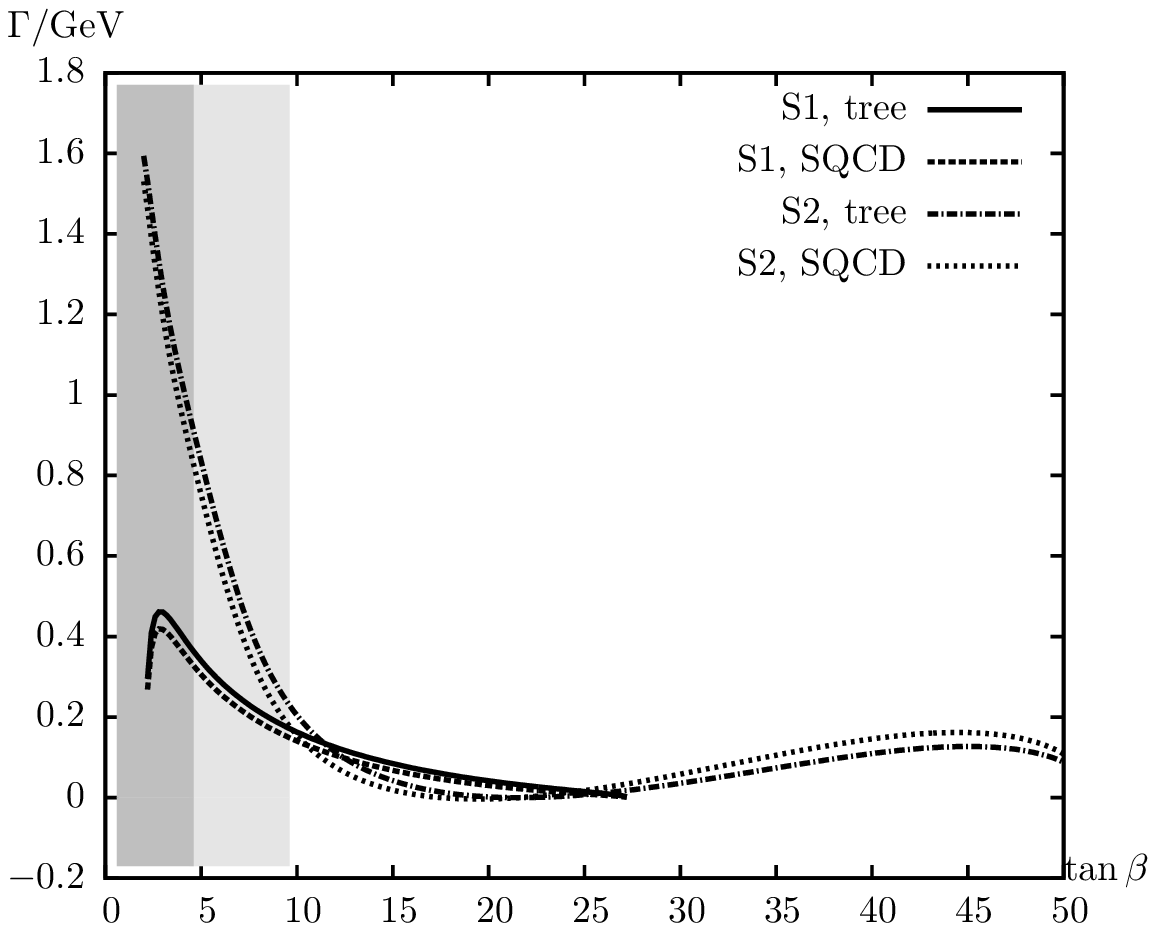}
\hspace{-2mm}
\includegraphics[width=0.49\textwidth,height=7.5cm]{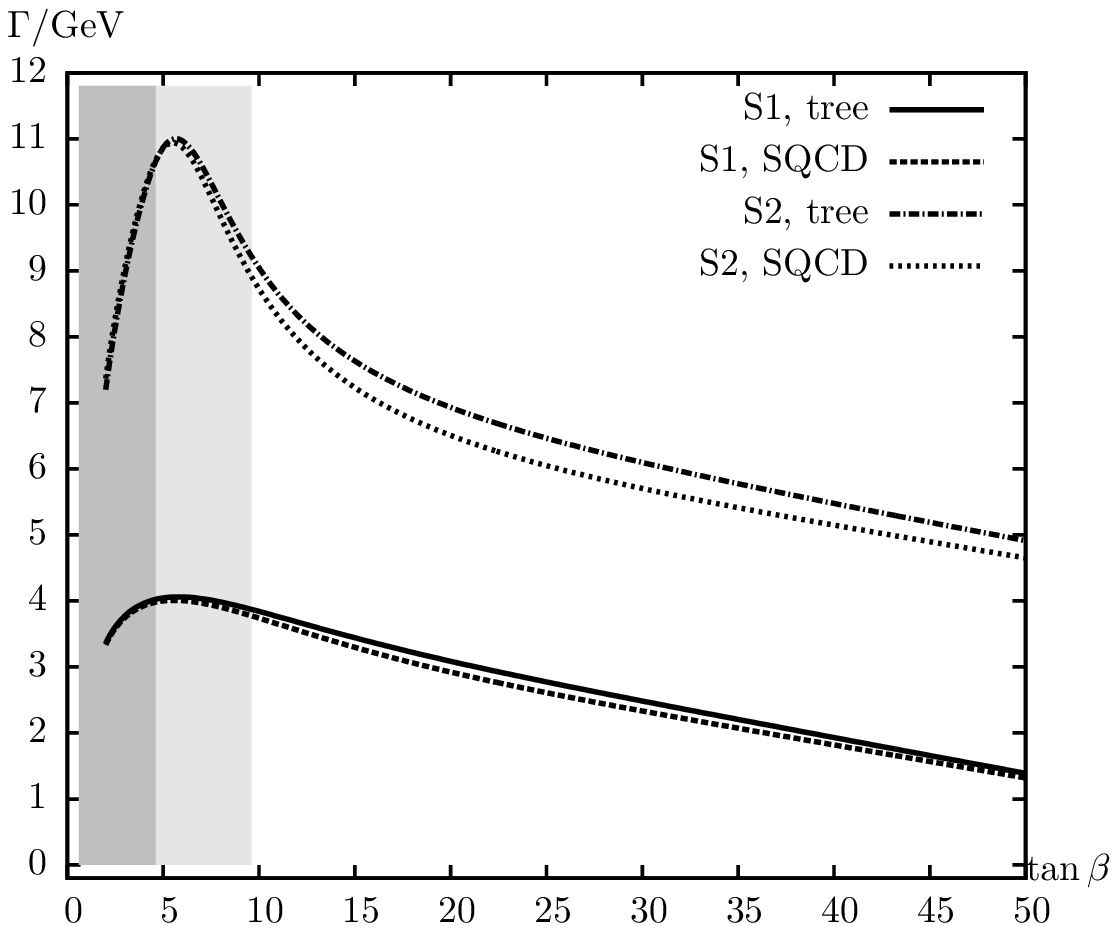}
\end{tabular}
\vspace{2em}
\caption{
  Tree-level and SQCD corrected partial decay widths for the
  renormalization scheme RS2 with $\tb$ varied.  
  The parameters are chosen according to the scenarios \SE\ and \SZ\
  (see \refta{tab:para}). 
  For \SE\ the grey region is excluded and for \SZ\ the dark grey region 
  is excluded.
  Upper left plot: $\Ga(\decayH)$.
  Upper right plot: $\Ga(\decaySbeW)$. 
  Lower left plot: $\Ga(\decaySbzH)$. 
  Lower right plot: $\Ga(\decaySbzW)$. 
}
\label{fig:SQCD}
\end{center}
\end{figure}

In \reffi{fig:SQCD} we show the tree-level values and SQCD one-loop 
corrected partial decay widths for 
$\decayH$, $\decaySbeW$, $\decaySbzH$, $\decaySbzW$,
respectively. The renormalization scheme RS2 is used, 
and hard gluon radiation is taken into account.
The parameters are chosen according to \SE\ and \SZ\ with $\tb$ varied. 
For \SE\ and \SZ\ the grey and the dark grey region 
is excluded via LEP Higgs searches, respectively.
In the lower left plot of \reffi{fig:SQCD} the curves in \SE\ 
end at $\tb \approx 27$ due to the closing of the phase space.
The size of the SQCD one-loop corrections reaches the highest values for
large $\tb$ 
in the case of $\Stopz \to \Sbot_{1,2} H^+$ and for intermediate $\tb$
in the case of $\Stopz \to \Sbot_{1,2} W^+$. 
The relative size in percent of the tree-level values do not exceed 
$-8\%$ in $\Ga(\decayH)$, 
$+18\%$ in $\Ga(\decaySbeW)$,
$-24\%$ in $\Ga(\decaySbzH)$ and
$-6\%$ in $\Ga(\decaySbzW)$. 
The absolute size of the SQCD corrections can be compared with
the upper left plots of \reffis{fig:st2sb1H}--\ref{fig:st2sb2W}, where
the full one-loop corrections are shown. It becomes obvious, especially
in \SZ, that restricting an evaluation to the pure SQCD corrections would
strongly underestimate the full one-loop corrections. (Hard photon radiation
can be as relevant as hard gluon radiation.) Consequently, 
the full set of one-loop corrections must be taken into account to yield
a reliable prediction of the scalar top quark decay width.

\section{Conclusions}

A scalar top quark can decay into a scalar bottom quark 
and a charged Higgs boson or a $W$~boson
if the process is kinematically allowed.
These decay modes can comprise a large part of
the total stop decay width. The decay channels with a charged Higgs
boson in the final state form a potentially important subprocess
of cascade decays 
which are interesting for the search of charged Higgs bosons at the LHC.
In order to arrive at a precise prediction of these scalar top quark
partial decay widths
at least a (full) one-loop calculation has to be performed.
In such a calculation a renormalization procedure has to be applied
that takes into account the top quark/squark as well as the bottom
quark/squark sector in the MSSM. These two sectors are connected via the soft
SUSY-breaking mass parameter $M_{\tilde{Q}_L}$ of the superpartners of the
left-handed quarks, 
which is the same in both sectors due to the $SU(2)_L$ invariance.

Within the MSSM with complex parameters (cMSSM)
we defined six different renormalization schemes for the bottom
quark/squark sector, while 
in the top quark/squark sector we applied a commonly used on-shell
renormalization scheme, which is well suited for processes with external
top and stop quarks. 
In our analysis we focused on the problem that, for
certain parameter sets, an applied renormalization scheme might fail
and cause large counterterm contributions that enhance the loop
corrections to unphysically large values. 
We have analyzed analytically the drawbacks and shortcomings of each of the
six renormalization schemes. Because of the relations between the
parameters that have to be respected also at the one-loop level
 we did not find 
any renormalization scheme that results in reasonably small counterterm
contributions over {\em all}
the cMSSM parameter space we have analyzed (we did not consider
a pure \DRbar~scheme which is not well suited to describe external particles). 
Some renormalization schemes (for instance, the ``on-shell'' scheme
which is defined analogously to the one applied in the
top quark/squark sector) fail over large parts of the parameter
space. Others fail only in relatively small parts where, 
for instance, a divergence due to a vanishing denominator occurs.
The most robust schemes turn out to be the 
``$\mb,\,\Ab$~\DRbar''(RS2) scheme and the
``$\Ab$~vertex, $\re Y_b$~OS''(RS6) scheme.
These renormalization schemes appear to be most suitable for
higher-order corrections 
involving scalar top and bottom quarks.

We performed a detailed numerical analysis for the full one-loop result 
of the partial decay widths corresponding to the four processes $\Stopz
\to \Sbotj H^+/W^+$ ($j = 1,2$) in our ``preferred'' scheme,
``$\mb,\,\Ab$~\DRbar''.  
The higher-order corrections, besides the full set of one-loop diagrams, also
contain soft and hard QED and QCD radiation. 
We evaluated the higher-order predictions of the four partial decay
widths as a function of $\tb$, $\mu$, $\Ab$ and $\phiab$.  
We found mainly modest corrections at the one-loop level.
Larger corrections are mostly found in regions of the parameter space that are
disfavored by experimental constraints and/or charge and color breaking
minima. 
A comparison of the full one-loop calculation with a pure SQCD
calculation showed that the latter one can result in a very poor
approximation of the full result and cannot be used for a reliable
prediction.

A full one-loop calculation of the corresponding branching ratios requires the
calculation of all possible partial decay widths of the scalar top quark
(and consequently a
renormalization of the full cMSSM) and will be presented
elsewhere~\cite{Stop2decay}.


\subsection*{Acknowledgements}

We thank for helpful discussions:
F.~Campanario, 
S.~Dittmaier,
T.~Fritzsche, 
J.~Guasch, 
T.~Hahn, 
W.~Hollik,
L.~Mihaila,  
F.~von der Pahlen, 
T.~Plehn, 
M.~Spira,
D.~St\"ockinger
and 
G.~Weiglein.
The work of S.H.\ was partially supported by CICYT (grant FPA 2007--66387).
Work supported in part by the European Community's Marie-Curie Research
Training Network under contract MRTN-CT-2006-035505
`Tools and Precision Calculations for Physics Discoveries at Colliders'.
H.R.\ acknowledges  support by the Deutsche
Forschungsgemeinschaft via the Sonderforschungsbereich/Transregio
SFB/TR-9 ``Computational Particle Physics'' and the Initiative and
Networking Fund of the Helmholtz Association, contract HA-101 ``Physics
at the Terascale''.


\newpage

\begin{appendix}
\section*{Appendix: \boldmath{$C$}-functions}

As explained in \refse{sec:AbOS_ReYbOS}, 
in RS6 we have to deal with infrared divergent $C$-functions 
(appearing in $\Lambda(p_1^2 = 0,p^2,p^2)$) with vanishing 
Gram-determinants. This case is not implemented in 
{\em LoopTools}~\cite{formcalc}. 
Therefore we follow \citere{cfunc} (and references therein) and 
replace the corresponding $C$-functions by well behaving linear 
combinations of $B$-functions \footnote{FormCalc \cite{formcalc} 
sorts the loop integrals with help of the masses. 
Consequently, any momentum can become zero, 
not only $p_1$. Furthermore {\em LoopTools} uses a different convention 
than \cite{cfunc}: $C_1 = C_{11} - C_{12}$, $C_2 = C_{12}$.}.
For sake of completeness we briefly review our implementation.
The class of $C$-functions with only one external momentum zero, can be
completely reduced to $B$-functions. Having three different masses we
can use partial fraction decomposition:
\begin{align}\label{C0}
C_0(0,p,p,m_1,m_2,m_3) = 
   \frac{B_0(p, m_1, m_3) - B_0(p, m_2, m_3)}{m_1^2 - m_2^2}~.
\end{align}
With only two different masses applying partial differentiation 
(l'Hospital) yields
\begin{align}
\label{dB0}
C_0(0,p,p,m_1,m_1,m_3) &= 
   \frac{\dd B_0(p,m_1,m_3)}{\dd(m_1^2)}~.
\end{align}
We also used symmetry relations and decompositions which can be found 
in \cite{cfunc} and the following short hand notation:
\begin{align}
D_{ij} &= p^4 + m_i^4 + m_j^4 - 2 (p^2 m_i^2 + p^2 m_j^2 + m_i^2 m_j^2)~.
\end{align}
We included the following replacements of $C_i$
functions with $p_1 = 0$:
\begin{align}
C_0(0, p, p, m_1, m_2, m_3) &\to 
   \frac{B_0(p, m_1, m_3) - B_0(p, m_2, m_3)}{m_1^2 - m_2^2}~, \\
C_0(0, p, p, m_1, m_1, m_3) &\to 
   \frac{1}{D_{13}} \Big[ (p^2 + m_3^2 - m_1^2) (2 - B_0(p, m_1, m_3)) \non \\
   &\qquad + (p^2 - m_3^2 - m_1^2) B_0(0, m_1, m_1)
           + 2 m_3^2 B_0(0, m_3, m_3)\Big]~, \\
C_1(0, p, p, m_1, m_2, m_3) &\to 
   \frac{1}{3 (m_1^2 - m_2^2)^2} \Big[ 2 m_1^2 (B_0(p, m_2, m_3) 
   - B_0(p, m_1, m_3)) \non \\
   &\qquad + (m_1^2 - m_2^2) B_0(p, m_2, m_3) - m_1^2 + m_2^2 \non \\
   &\qquad + (3 m_1^2 - 2 m_2^2 - m_3^2 + p^2) B_1(p, m_2, m_3) \non \\
   &\qquad + (m_3^2 - m_1^2 - p^2) B_1(p, m_1, m_3) \Big]~, \\
C_2(0, p, p, m_1, m_2, m_3) &\to 
   \frac{B_1(p, m_1, m_3) - B_1(p, m_2, m_3)}{m_1^2 - m_2^2}~, \\
C_2(0, p, p, m_1, m_1, m_3) &\to 
   \frac{1}{2 p^2} \bigg\{B_0(0, m_1, m_1) - B_0(p, m_1, m_3) \non \\
   &\qquad - \frac{p^2 + m_1^2 - m_3^2}{D_{13}} \Big[ 
           (p^2 - m_1^2 + m_3^2) (2 - B_0(p, m_1, m_3)) \non \\
   &\qquad + (p^2 - m_3^2 - m_1^2) B_0(0, m_1, m_1) 
           + 2 m_3^2 B_0(0, m_3, m_3)\Big] \bigg\}~.
\end{align}
In the case of $p_2 = 0$, we used the following replacements:
\begin{align}
C_0(p, 0, p, m_1, m_2, m_3) &\to 
   \frac{B_0(p, m_1, m_2) - B_0(p, m_1, m_3)}{m_2^2 - m_3^2}~, \\
C_0(p, 0, p, m_1, m_2, m_2) &\to
   \frac{1}{D_{12}} \Big[ (p^2 - m_2^2 + m_1^2) (2 - B_0(p, m_1, m_2)) \non \\
   &\qquad + (p^2 - m_1^2 - m_2^2) B_0(0, m_2, m_2) 
           + 2 m_1^2 B_0(0, m_1, m_1)\Big]~, \\
C_1(p, 0, p, m_1, m_2, m_3) &\to 
   \frac{1}{3 (m_2^2 - m_3^2)^2} \Big[ 2 m_1^2 (B_0(p, m_1, m_2) 
   - B_0(p, m_1, m_3)) \non \\
   &\qquad + (m_2^2 - m_3^2) B_0(0, m_2, m_3) - m_3^2 + m_2^2 \non \\
   &\qquad - (3 m_3^2 - 2 m_2^2 - m_1^2 - p^2) B_1(p, m_1, m_2) \non \\
   &\qquad + (m_3^2 - m_1^2 - p^2) B_1(p, m_1, m_3) \Big]~, \\
C_2(p, 0, p, m_1, m_2, m_3) &\to 
   \frac{1}{3 (m_2^2 - m_3^2)^2} \Big[ 2 m_1^2 (B_0(p, m_1, m_3) 
   - B_0(p, m_1, m_2)) \non \\
   &\qquad - (m_2^2 - m_3^2) B_0(0, m_2, m_3) + m_3^2 - m_2^2 \non \\
   &\qquad - (3 m_2^2 - 2 m_3^2 - m_1^2 - p^2) B_1(p, m_1, m_3) \non \\
   &\qquad + (m_2^2 - m_1^2 - p^2) B_1(p, m_1, m_2) \Big]~.
\end{align}
Finally, for $p_3 = (p_1 + p_2) = 0$ we employed:
\begin{align}
C_0(p, p, 0, m_1, m_2, m_3) &\to 
   \frac{B_0(p, m_1, m_2) - B_0(p, m_2, m_3)}{m_1^2 - m_3^2}~, \\
C_0(p, p, 0, m_1, m_2, m_1) &\to
   \frac{1}{D_{12}} \Big[ (p^2 - m_1^2 + m_2^2) (2 - B_0(p, m_1, m_2)) \non \\
   &\qquad + (p^2 - m_2^2 - m_1^2) B_0(0, m_1, m_1) 
           + 2 m_2^2 B_0(0, m_2, m_2)\Big]~, \\
C_1(p, p, 0, m_1, m_2, m_3) &\to 
   \frac{B_0(p, m_2, m_3) + B_1(p, m_1, m_2) 
   + B_1(p, m_2, m_3)}{m_1^2 - m_3^2}~, \\
C_1(p, p, 0, m_1, m_2, m_1) &\to 
   \frac{1}{2 p^2} \bigg\{ B_0(0, m_1, m_1) - B_0(p, m_1, m_2) \non \\
   &\qquad - \frac{p^2 + m_1^2 - m_2^2}{D_{12}} \Big[ 
           (p^2 - m_1^2 + m_2^2) (2 - B_0(p, m_1, m_2)) \non \\
   &\qquad + (p^2 - m_2^2 - m_1^2) B_0(0, m_1, m_1) 
           + 2 m_2^2 B_0(0, m_2, m_2) \Big] \bigg\}~, \\
C_2(p, p, 0, m_1, m_2, m_3) &\to 
   \frac{1}{3 (m_1^2 - m_3^2)^2} \Big[ 2 m_1^2 (B_0(p, m_2, m_3) 
   - B_0(p, m_1, m_2)) \non \\
   &\qquad - (2 m_1^2 - m_2^2 - m_3^2 + p^2) B_0(p, m_2, m_3) 
           + m_3^2 - m_1^2 \non \\
   &\qquad - (3 m_1^2 - 2 m_3^2 - m_2^2 + p^2) B_1(p, m_2, m_3) \non \\
   &\qquad + (m_2^2 - m_1^2 - p^2) B_1(p, m_1, m_2) \Big]~.
\end{align}


\end{appendix}


\newpage
\pagebreak


\clearpage

\begin{thebibliography}{99} 

\bibitem{mssm} H.P.~Nilles, 
               {\em Phys.\ Rept.} {\bf 110} (1984) 1; \\ 
               H.E.~Haber and G.L.~Kane, 
               {\em Phys.\ Rept.} {\bf 117} (1985) 75; \\  
               R.~Barbieri, 
               {\em Riv.\ Nuovo Cim.} {\bf 11} (1988) 1. 

\bibitem{squark_q_V_als} A.~Bartl, H.~Eberl, K.~Hidaka, S.~Kraml,
        W.~Majerotto, W.~Porod and Y.~Yamada, 
        {\em Phys.\ Lett.} {\bf B 419} (1998) 243
        [arXiv:hep-ph/9710286].

\bibitem{stopsbot_phi_als} A.~Bartl, H.~Eberl, K.~Hidaka, S.~Kraml,
        W.~Majerotto, W.~Porod and Y.~Yamada, 
        {\em Phys.\ Rev.} {\bf D 59} (1999) 115007
        [arXiv:hep-ph/9806299].

\bibitem{sdecay} M.~M\"uhlleitner, A.~Djouadi and Y.~Mambrini,
                 {\em Comput.\ Phys.\ Commun.} {\bf 168} (2005) 46
                 [arXiv:hep-ph/0311167].

\bibitem{sbot_stop_Hpm_altb} L.~Jin and C.~Li,
        {\em Phys.\ Rev.} {\bf D 65} (2002) 035007
        [arXiv:hep-ph/0106253].

\bibitem{A_sferm_sferm_full} C.~Weber, H.~Eberl and W.~Majerotto,
        {\em Phys.\ Lett.\ } {\bf B 572}, 56 (2003)
        [arXiv:hep-ph/0305250];
        {\em Phys.\ Rev.\ }{\bf  D 68} (2003) 093011
        [arXiv:hep-ph/0308146].

\bibitem{H_sferm_sferm_full} C.~Weber, K.~Kovarik, H.~Eberl and W.~Majerotto,
                             {\em Nucl.\ Phys.} {\bf B 776} (2007) 138
                             [arXiv:hep-ph/0701134].

\bibitem{squark_eff} J.~Guasch, S.~Pe\~naranda and R.~Sanchez-Florit,
                     {\em JHEP} {\bf 0904} (2009) 016
                     [arXiv:0812.1114 [hep-ph]].

\bibitem{dissHR} H.~Rzehak, PhD thesis:
                 ``Two-loop contributions in the supersymmetric Higgs
                 sector'', Technische Universit\"at M\"unchen, 2005; 
                 see: {\tt nbn-resolving.de/} \\
                 with {\tt urn}: {\tt nbn:de:bvb:91-diss20050923-0853568146}~.

\bibitem{mhiggsFDalbals} S.~Heinemeyer, W.~Hollik, H.~Rzehak and G.~Weiglein,
                         {\em Eur. Phys. J.} {\bf C 39} (2005) 465
                         [arXiv:hep-ph/0411114].

\bibitem{sbotrenold} A.~Brignole, G.~Degrassi, P.~Slavich and F.~Zwirner,
                     {\em Nucl.\ Phys.} {\bf B 643} (2002) 79
                     [arXiv:hep-ph/0206101].

\bibitem{dissTF} T.~Fritzsche,
                 PhD thesis, Cuvillier Verlag, G\"ottingen 2005,
                 ISBN 3--86537--577--4.

\bibitem{mhcMSSM2L} S.~Heinemeyer, W.~Hollik, H.~Rzehak and G.~Weiglein,
                    {\em Phys.\ Lett.} {\bf B 652} (2007) 300
                    [arXiv:0705.0746 [hep-ph]].

\bibitem{FawziRen} N.~Baro and F.~Boudjema,
                   {\em Phys.\ Rev.} {\bf D 80} (2009) 076010
                   [arXiv:0906.1665 [hep-ph]].

\bibitem{Bartl:2003pd} A.~Bartl, S.~Hesselbach, K.~Hidaka, T.~Kernreiter 
                       and W.~Porod,
                       {\em Phys.\ Rev.} {\bf D 70} (2004) 035003
                       [arXiv:hep-ph/0311338].

\bibitem{Bartl:2004jr} A.~Bartl, E.~Christova, K.~Hohenwarter-Sodek and 
                       T.~Kernreiter,
                       {\em Phys.\ Rev.} {\bf D 70} (2004) 095007
                       [arXiv:hep-ph/0409060].

\bibitem{Ellis:2008hq} J.~Ellis, F.~Moortgat, G.~Moortgat-Pick, 
                       J.~Smillie and J.~Tattersall,
                       {\em Eur.\ Phys.\ J.} {\bf C 60} (2009) 633 
                       [arXiv:0809.1607 [hep-ph]].

\bibitem{Deppisch:2009nj} F.~Deppisch and O.~Kittel,
                          {\em JHEP} {\bf 0909} (2009) 110
                          [Erratum-ibid.\  {\bf 1003} (2010) 091]
                          [arXiv:0905.3088 [hep-ph]].

\bibitem{Deppisch:2010nc} F.~Deppisch and O.~Kittel,
                          arXiv:1003.5186 [hep-ph].

\bibitem{Eberl:2009xe} H.~Eberl, S.~Frank and W.~Majerotto,
                       arXiv:0912.4675 [hep-ph].

\bibitem{feynhiggs} S.~Heinemeyer, W.~Hollik and G.~Weiglein,
                    {\em Comput. Phys. Commun.} {\bf 124} (2000) 76
                    [arXiv:hep-ph/9812320];
                    see {\tt www.feynhiggs.de} .

\bibitem{mhiggslong} S.~Heinemeyer, W.~Hollik and G.~Weiglein,
                     {\em Eur. Phys. J.} {\bf C 9} (1999) 343
                     [arXiv:hep-ph/9812472].

\bibitem{mhiggsAEC} G.~Degrassi, S.~Heinemeyer, W.~Hollik,
                    P.~Slavich and G.~Weiglein, 
                    {\em Eur. Phys. J.} {\bf C 28} (2003) 133
                    [arXiv:hep-ph/0212020].

\bibitem{mhcMSSMlong} M.~Frank, T.~Hahn, S.~Heinemeyer, W.~Hollik, 
                      R.~Rzehak and G.~Weiglein,
                      {\em JHEP} {\bf 0702} (2007) 047
                      [arXiv:hep-ph/0611326].

\bibitem{Stop2decay} T.~Fritzsche, S.~Heinemeyer, H.~Rzehak,
                     C.~Schappacher and G.~Weiglein,
                     {\em in preparation}.

\bibitem{diplTF}  T.~Fritzsche and W.~Hollik,
        {\em Eur.\ Phys.\ J.\ }   {\bf C 24} (2002) 619
        [arXiv:hep-ph/0203159];\\
        T.~Fritzsche,
                 Diploma thesis, Institut f\"ur Theoretische Physik, 
                 Universit\"at Karlsruhe, Germany, Dec. 2000, see:\\
  {\tt www-itp.particle.uni-karlsruhe.de/diplomatheses.de.shtml} .

\bibitem{denner} A.~Denner,
                 {\em Fortsch.\ Phys.} {\bf 41} (1993) 307
                 [arXiv:0709.1075 [hep-ph]].

\bibitem{MSSMrenormierung} W.~Hollik, E.~Kraus, M.~Roth, C.~Rupp,
                           K.~Sibold and D.~St\"ockinger, 
        {\em Nucl.\ Phys.\ } {\bf  B 639} (2002) 3
        [arXiv:hep-ph/0204350].

\bibitem{hr} W. Hollik and H. Rzehak, 
             {\em Eur.\ Phys.\ J.} {\bf C 32} (2003) 127
             [arXiv:hep-ph/0305328].

\bibitem{dr2lA} A.~Djouadi, P.~Gambino, S.~Heinemeyer, W.~Hollik,
                C.~J\"unger and G.~Weiglein,
                {\em Phys. Rev. Lett.} {\bf 78} (1997) 3626
                [arXiv:hep-ph/9612363];
                {\em Phys. Rev.} {\bf D 57} (1998) 4179
                [arXiv:hep-ph/9710438].

\bibitem{sbot_top_cha_alt} J.~Guasch, J.~Sola and W.~Hollik,
        {\em Phys.\ Lett.} {\bf B 437} (1998) 88
        [arXiv:hep-ph/9802329].

\bibitem{sfermprod_alf} H.~Eberl, S.~Kraml and W.~Majerotto,
                        {\em JHEP} {\bf 9905} (1999) 016
                        [arXiv:hep-ph/9903413].

\bibitem{sferm_f_V_full} A.~Arhrib and R.~Benbrik,
        {\em Phys.\ Rev.} {\bf D 71} (2005) 095001
        [arXiv:hep-ph/0412349].

\bibitem{squark_q_chi_full} J.~Guasch, W.~Hollik and J.~Sola,
        {\em Phys.\ Lett.} {\bf B 510} (2001) 211
        [arXiv:hep-ph/0101086];
        {\em JHEP} {\bf 0210} (2002) 040
        [arXiv:hep-ph/0207364].

\bibitem{stop_stop_H_alt} Q.~Li, L.~Jin and C.~Li,
        {\em Phys.\ Rev.} {\bf D 66} (2002) 115008
        [arXiv:hep-ph/0207363].

\bibitem{squark_q_chi_als} S.~Kraml, H.~Eberl, A.~Bartl, W.~Majerotto
        and W.~Porod, 
        {\em Phys.\ Lett.} {\bf B 386} (1996) 175
        [arXiv:hep-ph/9605412];\\
        A.~Djouadi, W.~Hollik and C.~J\"unger,
        {\em Phys.\ Rev.} {\bf D 55} (1997) 6975
        [arXiv:hep-ph/9609419].

\bibitem{squark_q_gl_als} W.~Beenakker, R.~H\"opker and P.~Zerwas,
        {\em Phys.\ Lett.} {\bf B 378} (1996) 159
        [arXiv:hep-ph/9602378].

\bibitem{stop_top_gl_als} W.~Beenakker, R.~H\"opker, T.~Plehn and P.~Zerwas,
         {\em Z.\ Phys.} {\bf C 75} (1997) 349
         [arXiv:hep-ph/9610313].

\bibitem{cfunc} G.~Devaraj and R.~Stuart,
                {\em Nucl.Phys.} {\bf B 519} (1998) 483
                [arXiv:hep-ph/9704308].

\bibitem{pdg} C.~Amsler et al.  [Particle Data Group],
              {\em Phys. Lett.} {\bf B 667} (2008) 1.

\bibitem{RunDec} K.~Chetyrkin, J.~K\"uhn, 
                 {\em Comput. Phys. Commun.} {\bf 133} (2000) 43
                 [arXiv:hep-ph/0004189].

\bibitem{komplexDb} M.~Carena, J.~Ellis, A.~Pilaftsis and C.~Wagner,
                    {\em Nucl.\ Phys.} {\bf B 586} (2000) 92
                    [arXiv:hep-ph/0003180];\\
                    K.~Williams, 
                    PhD thesis, University of Durham, 2008.

\bibitem{deltab1} R.~Hempfling,
                  {\em Phys. Rev.} {\bf D 49} (1994) 6168;\\
                  L.~Hall, R.~Rattazzi and U.~Sarid,
                  {\em Phys. Rev.} {\bf D 50} (1994) 7048
                  [arXiv:hep-ph/9306309];\\
                  M.~Carena, M.~Olechowski, S.~Pokorski and C.~Wagner,
                  {\em Nucl. Phys.} {\bf B 426} (1994) 269
                  [arXiv:hep-ph/9402253].

\bibitem{deltab2} M.~Carena, D.~Garcia, U.~Nierste and C.~Wagner,
                  {\em Nucl. Phys.} {\bf B 577} (2000) 577
                  [arXiv:hep-ph/9912516].

\bibitem{feynarts} J.~K\"ublbeck, M.~B\"ohm and A.~Denner, 
                   {\em Comput. Phys. Commun.} {\bf 60} (1990) 165;\\
                   T.~Hahn, 
                   {\em Comput. Phys. Commun.} {\bf 140} (2001) 418
                   [arXiv:hep-ph/0012260];\\
                   T.~Hahn and C.~Schappacher, 
                   {\em Comput. Phys. Commun.} {\bf 143} (2002) 54
                   [arXiv:hep-ph/0105349].\\
                   The program and the user's guide 
                   are available via {\tt www.feynarts.de} .

\bibitem{formcalc} T.~Hahn and M.~P\'erez-Victoria,
                   {\em Comput. Phys. Commun.} {\bf 118} (1999) 153
                   [arXiv:hep-ph/9807565].

\bibitem{cdr} F.~del Aguila, A.~Culatti, R.~Munoz Tapia and M.~Perez-Victoria,
              {\em Nucl. Phys.} {\bf B 537} (1999) 561
              [arXiv:hep-ph/9806451].

\bibitem{dred} W.~Siegel, 
               {\em Phys. Lett.} {\bf B 84} (1979) 193; \\
               D.~Capper, D.~Jones, and P.~van Nieuwenhuizen,
               {\em Nucl. Phys.} {\bf B 167} (1980) 479. 

\bibitem{dredDS} D.~St\"ockinger,
                 {\em JHEP} {\bf 0503} (2005) 076
                 [arXiv:hep-ph/0503129].

\bibitem{dredDS2} W.~Hollik and D.~St\"ockinger,
                  {\em Phys.\ Lett.} {\bf B 634} (2006) 63
                  [arXiv:hep-ph/0509298].

\bibitem{feynarts-mf} The couplings can be found in the files
                      {\tt MSSM.ps.gz}, {\tt MSSMQCD.ps.gz} and 
                      {\tt HMix.ps.gz} as part of the
                      \fa~package~\cite{feynarts}.

\bibitem{ccb} J.~Frere, D.~Jones and S.~Raby,
              {\em Nucl.\ Phys.} {\bf B 222} (1983) 11;\\
              M.~Claudson, L.~Hall and I.~Hinchliffe,
              {\em Nucl.\ Phys.} {\bf B 228} (1983) 501;\\
              C.~Kounnas, A.~Lahanas, D.~Nanopoulos and M.~Quiros,
              {\em Nucl.\ Phys.} {\bf B 236} (1984) 438;\\
              J.~Gunion, H.~Haber and M.~Sher,
              {\em Nucl.\ Phys.} {\bf B 306} (1988) 1;\\
              J.~Casas, A.~Lleyda and C.~Munoz,
              {\em Nucl.\ Phys.} {\bf B 471} (1996) 3
              [arXiv:hep-ph/9507294];\\
              P.~Langacker and N.~Polonsky,
              {\em Phys.\ Rev.} {\bf D 50} (1994) 2199
              [arXiv:hep-ph/9403306];\\
              A.~Strumia,
              {\em Nucl.\ Phys.} {\bf B 482} (1996) 24
              [arXiv:hep-ph/9604417].


\bibitem{LEPHiggsSM} [LEP Higgs working group],
                     {\em Phys. Lett.} {\bf B 565} (2003) 61
                     [arXiv:hep-ex/0306033].

\bibitem{LEPHiggsMSSM} [LEP Higgs working group],
                       {\em Eur.\ Phys.\ J.} {\bf C 47} (2006) 547
                       [arXiv:hep-ex/0602042].


\bibitem{MSSMcomplphasen} S.~Dimopoulos and S.~Thomas,
                          {\em Nucl.\ Phys.} {\bf  B 465} (1996) 23
                          [arXiv:hep-ph/9510220].

\bibitem{SUSYphases} M.~Dugan, B.~Grinstein and L.~Hall,
                     {\em Nucl.\ Phys.} {\bf B 255} (1985) 413.

\bibitem{EDMDoink} W.~Hollik, J.~Illana, S.~Rigolin and D.~St\"ockinger,
                   {\em Phys. Lett.} {\bf B 416} (1998) 345
                   [arXiv:hep-ph/9707437];
                   {\em Phys. Lett.} {\bf B 425} (1998) 322, 
                   hep-ph/9711322.

\bibitem{EDMrev2} D.~Demir, O.~Lebedev, K.~Olive, M.~Pospelov and A.~Ritz,
                  {\em Nucl. Phys.} {\bf B 680} (2004) 339
                  [arXiv:hep-ph/0311314].

\bibitem{EDMPilaftsis} D.~Chang, W.~Keung and A.~Pilaftsis,
                       {\em Phys. Rev. Lett.} {\bf 82} (1999) 900
                       [Erratum-ibid.\  {\bf 83} (1999) 3972]
                       [arXiv:hep-ph/9811202];\\
                       A.~Pilaftsis,
                       {\em Phys. Lett.} {\bf B 471} (1999) 174
                       [arXiv:hep-ph/9909485].

\bibitem{EDMRitz} O.~Lebedev, K.~Olive, M.~Pospelov and A.~Ritz,
                  {\em Phys. Rev.} {\bf D 70} (2004) 016003
                  [arXiv:hep-ph/0402023].

\bibitem{EDMheavy} P.~Nath,
                   {\em Phys. Rev. Lett.} {\bf 66} (1991) 2565;\\
                   Y.~Kizukuri and N.~Oshimo,
                   {\em Phys. Rev.} {\bf D 46} (1992) 3025.

\bibitem{EDMmiracle} T.~Ibrahim and P.~Nath,
                     {\em Phys. Lett.} {\bf B 418} (1998) 98
                     [arXiv:hep-ph/9707409];
                     {\em Phys. Rev.} {\bf D 57} (1998) 478 
                     [Erratum-ibid.\ {\bf D 58} (1998) 019901] 
                     [Erratum-ibid.\ {\bf D 60} (1998) 079903] 
                     [Erratum-ibid.\ {\bf D 60} (1999) 119901]
                     [arXiv:hep-ph/9708456];\\
                     M.~Brhlik, G.~Good and G.~Kane,
                     {\em Phys. Rev.} {\bf D 59} (1999) 115004
                     [arXiv:hep-ph/9810457].

\bibitem{EDMrev1} S.~Abel, S.~Khalil and O.~Lebedev,
                  {\em Nucl. Phys.} {\bf B 606} (2001) 151
                  [arXiv:hep-ph/0103320].

\bibitem{plehnix} V.~Barger, T.~Falk, T.~Han, J.~Jiang, T.~Li and T.~Plehn,
                  {\em Phys. Rev.} {\bf D 64} (2001) 056007
                  [arXiv:hep-ph/0101106].

\bibitem{EDMrev3} Y.~Li, S.~Profumo and M.~Ramsey-Musolf,
                  arXiv:1006.1440 [hep-ph].

\bibitem{HelmutLL2010} H.~Eberl, 
                       talk given at the {\em Loops \& Legs 2010},
                       W\"orlitz, Germany, April 2010; see:\\ {\tt 
https://indico.desy.de/conferenceOtherViews.py?view=standard\&confId=2200}~.

\bibitem{lepewwgNEW} ALEPH Collaboration, CDF Collaboration, D0
  Collaboration, DELPHI Collaboration, L3 Collaboration, OPAL
  Collaboration, SLD Collaboration, LEP Electroweak Working Group,
  Tevatron Electroweak Working Group, SLD electroweak heavy flavour
  groups, 
                     arXiv:0911.2604 [hep-ex].

\bibitem{WHZH} M.~Ciccolini, S.~Dittmaier and M.~Kr\"amer,
               {\em Phys.\ Rev.} {\bf D 68} (2003) 073003
               [arXiv:hep-ph/0306234].

\bibitem{H2j} M.~Ciccolini, A.~Denner and S.~Dittmaier,
              {\em Phys.\ Rev.\ Lett.} {\bf 99} (2007) 161803
              [arXiv:0707.0381 [hep-ph]];
              {\em Phys.\ Rev.} {\bf D 77} (2008) 013002
              [arXiv:0710.4749 [hep-ph]].

\bibitem{HVV4f} A.~Bredenstein, A.~Denner, S.~Dittmaier and M.~Weber,
                {\em JHEP} {\bf 0702} (2007) 080
                [arXiv:hep-ph/0611234].


\end{thebibliography}
\end{document}